\documentclass[aps,twocolumn,showpacs,floatfix]{revtex4}

\usepackage{dcolumn,multirow}
\usepackage{graphicx}
\usepackage{amsfonts,amsmath,bm,amssymb}
\usepackage{comment}
\usepackage[usenames]{color}
\usepackage[para]{threeparttable}
\usepackage{etoolbox}
\usepackage{titlesec}

\begin{document}

\title{ Structural and  metal-insulator transitions in rhenium based double perovskites via orbital ordering }

\author{Alex Taekyung Lee}
\affiliation{Department of Applied Physics and Applied Mathematics, Columbia University, New York, New York 10027, USA }

\author{Chris A. Marianetti}
\affiliation{Department of Applied Physics and Applied Mathematics, Columbia University, New York, New York 10027, USA }

\date{\today }

\begin{abstract}
Re-based double perovskites (DPs) have garnered substantial attention due to their high
Curie temperatures ($T_C$) and display of complex interplay of structural and
metal-insulator transitions (MIT).  Here we systematically study the ground state
electronic and structural properties for a family of Re-based DPs $A_2B$ReO$_6$ ($A$=Sr,
Ca and $B$=Cr, Fe), which are related by a common low energy Hamiltonian, using density
functional theory + $U$ calculations.  We show that the on-site interaction $U$ of Re
induces orbital ordering (denoted C-OO), with each Re site having an occupied $d_{xy}$
orbital and a C-type alternation among $d_{xz}/d_{yz}$, resulting in an insulating state
consistent with experimentally determined insulators Sr$_2$CrReO$_6$, Ca$_2$CrReO$_6$, and
Ca$_2$FeReO$_6$.  The threshold value of $U_{\textnormal{Re}}$ for orbital ordering is
reduced by inducing $E_g$ octahedral distortions of the same C-type wavelength (denoted
C-OD), which serves as a structural signature of the orbital ordering;  octahedral tilting
also reduces the threshold.  The C-OO, and the concomitant C-OD, are a spontaneously
broken symmetry for the Sr based materials (i.e. $a^0a^0c^-$ tilt pattern),  while not for
the Ca based systems (i.e. $a^-a^-b^+$ tilt pattern).  Spin-orbit coupling does not
qualitatively change the physics of the C-OO/C-OD, but can induce relevant  quantitative
changes.  We prove that a single set of
$U_{\textnormal{Cr}},U_{\textnormal{Fe}},U_{\textnormal{Re}}$ capture the experimentally
observed metallic state in Sr$_2$FeReO$_6$ and insulating states in other three systems.
We predict that the C-OO is the origin of the insulating state in Sr$_2$CrReO$_6$, and
that the concomitant C-OD may be experimentally observed at sufficiently low temperatures
(i.e. space group $P4_2/m$) in pure samples.  Additionally, given our prescribed values of
$U$, we show that the C-OO induced insulating state in Ca$_2$CrReO$_6$ will survive even
if the C-OD amplitude is suppressed (e.g. due to thermal fluctuations).  The role of the
C-OO/C-OD in the discontinuous, temperature driven MIT in Ca$_2$FeReO$_6$ is discussed.

\end{abstract}


\maketitle

\section{Introduction}
\subsection{General Background}

There is a huge phase space of possibilities for perovskite based transition metal oxides
with more than one type of transition metal which nominally bears $d$ electrons, and
experimental efforts are continuing to expand in this direction; including chemical
synthesis \cite{Re-Kato,Teresa-SrCrReO,Hauser-SrCrRe,review-Ibarra,Vasala-review} and
layer-by-layer growth by pulsed laser deposition
\cite{Singh-LNMO,Philipp-SrCrWO,review-Ibarra}.  Given that many of these materials will
exhibit strongly correlated electron behavior, it will be critical to have appropriate
first-principles based approaches which can be applied to this vast phase space in order
to guide experimental efforts; allowing for the development of novel, functional
materials.  Nearly two decades ago, room-temperature ferrimagnetism (sometimes loosely
referred to as ferromagnetism) was discovered in the double-perovskite (DP) transition
metal oxides (TMO)  Sr$_2$FeMoO$_6$  \cite{SrFeMo-Kobayashi}, attracting much attention to
DP TMO's due to their rich physics and potential for spintronic applications
\cite{review-Ibarra}.  Recent first-principles efforts have shown promise in identifying
new, novel materials in this phase
space\cite{ChenPRL2013,KarolakPRB2015,KimPRB2015,Puggioni2015087202}.

Among the various double perovskites, Re-based DPs are a particularly intriguing class;
and the small set $A_2B$ReO$_6$ ($A$=Sr, Ca and $B$=Cr, Fe) already contains a wealth of
interesting physics and impressive metrics.  Moreover, this particular set of Re-based DPs
materials forms a sort of family which descends from the same low energy Hamiltonian of Re
dominated orbitals, despite the fact that Cr and Fe have different numbers of electrons;
and this can be deduced from nominal charge counting along with some amount of post facto
knowledge (see Section \ref{sec:general_aspects} for a more detailed explanation).  Given
that Sr and Ca are isovalent (i.e. nominally $2+$), these two cations serve as binary
parameter to modify the degree and type of octahedral tilting, changing the bandwidth of
the system. 
Switching between Cr and  Fe changes the valence by two electrons and alters
the $B$ site energy. However, Cr and Fe are totally analogous in the sense that both yield
a filled spin shell given a predominant octahedral crystal field and a high spin
configuration (i.e. $t_{2g,\uparrow}^3$ and $t_{2g,\uparrow}^3e_{g,\uparrow}^2$, respectively). 

Experiment dictates that the resulting four permutations of $A_2B$ReO$_6$ yield both
metallic and insulating ground states, insulator to metal transitions as a function of
temperature (for reasonable temperature scales), structural transitions as a function of
temperature, and in some cases very high ferrimagnetic to paramagnetic transition
temperatures.  Moreover, this Re-based family of DP contains unexplained phenomena, such
as the discontinuous,  isostructural phase transition in Ca$_{2}$FeReO$_{6}$.  Therefore,
there are a variety of phenomenological, qualitative, and quantitative challenges which
need to be addressed in this family. 

Given that all of these compounds are strongly magnetically ordered at low temperatures,
it is reasonable to expect that DFT+$U$ might provide an overarching, qualitative view of
the physics; perhaps even quantitative.  In this work, we use DFT+$U$ calculations to
investigate the electronic and structural aspects of $A_{2}B$ReO$_{6}$ ($A$=Sr, Ca, and
$B$=Cr, Fe),  systematically accounting for the effects of octahedral distortions and
rotations; in addition to carefully exploring the effect of the Hubbard $U$ for both the
$B$ sites and Re.  We show that a single set of
$U_{\textnormal{Re}},U_{\textnormal{Fe}},U_{\textnormal{Cr}}$  can obtain qualitative
agreement with known experiments of all four compounds.  Particular attention is payed to
isolating the effects of the Hubbard $U$ by additionally considering cubic reference
structures in the absence of any octahedral distortions or tilting.  Finally, we explore
the effect of spin-orbit coupling, demonstrating that it can perturb the C-OO and the
resulting C-OD, but the qualitative trends hold.

The rest of the paper is organized as follows.  Sections \ref{lit_review} and
\ref{sec:orbital_ordering} address the previous literature of the Re-based double
perovskites and orbital ordering physics in other perovskites, respectively.  Section
\ref{comp_method} details the computational methods and provides a brief discussion on the
value of $U$, while a detailed analysis of the optimal $U$ values is given in Section
\ref{U_of_Re}.  Section \ref{sec:general_aspects} provides a minimal analysis of the
various physical mechanisms at play in this family of materials, highlighting the key
findings in our paper; while detailed calculations which shape our conclusions can be
found in Sections \ref{sec:electronic_struct} and \ref{sec:SOC}.  Section
\ref{sec:exp_chal} discusses future experiments which could test our predictions, and
Section \ref{sec:summary} presents the summary of the paper.

\subsection{Literature review of $A_2B$ReO$_6$ ($A$=Sr, Ca and $B$=Cr,Fe)}
\label{lit_review}

\begin{figure}
\begin{center}
\includegraphics[width=0.45\textwidth, angle=0]{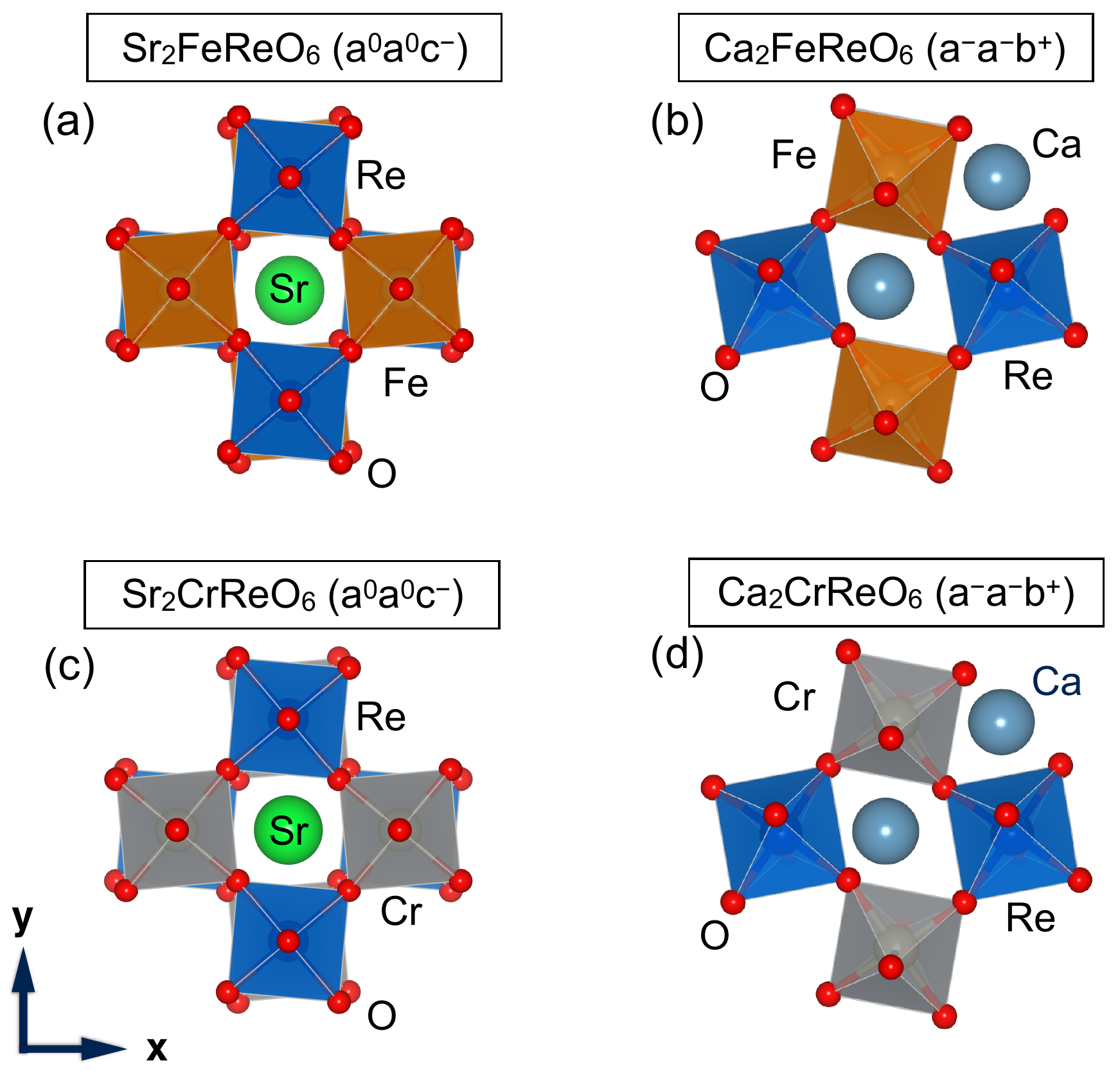}
\caption{ 
    Orthographic view of the crystal structures of (a) Sr$_2$FeReO$_6$, (b) Ca$_2$FeReO$_6$, (c) Sr$_2$CrReO$_6$, 
    and (d) Ca$_2$CrReO$_6$. The octahedral tilt pattern  in listed in each case.
}
\label{FeRe-str}
\end{center}
\end{figure}

\begin{figure}
\begin{center}
\includegraphics[width=0.5\textwidth, angle=0]{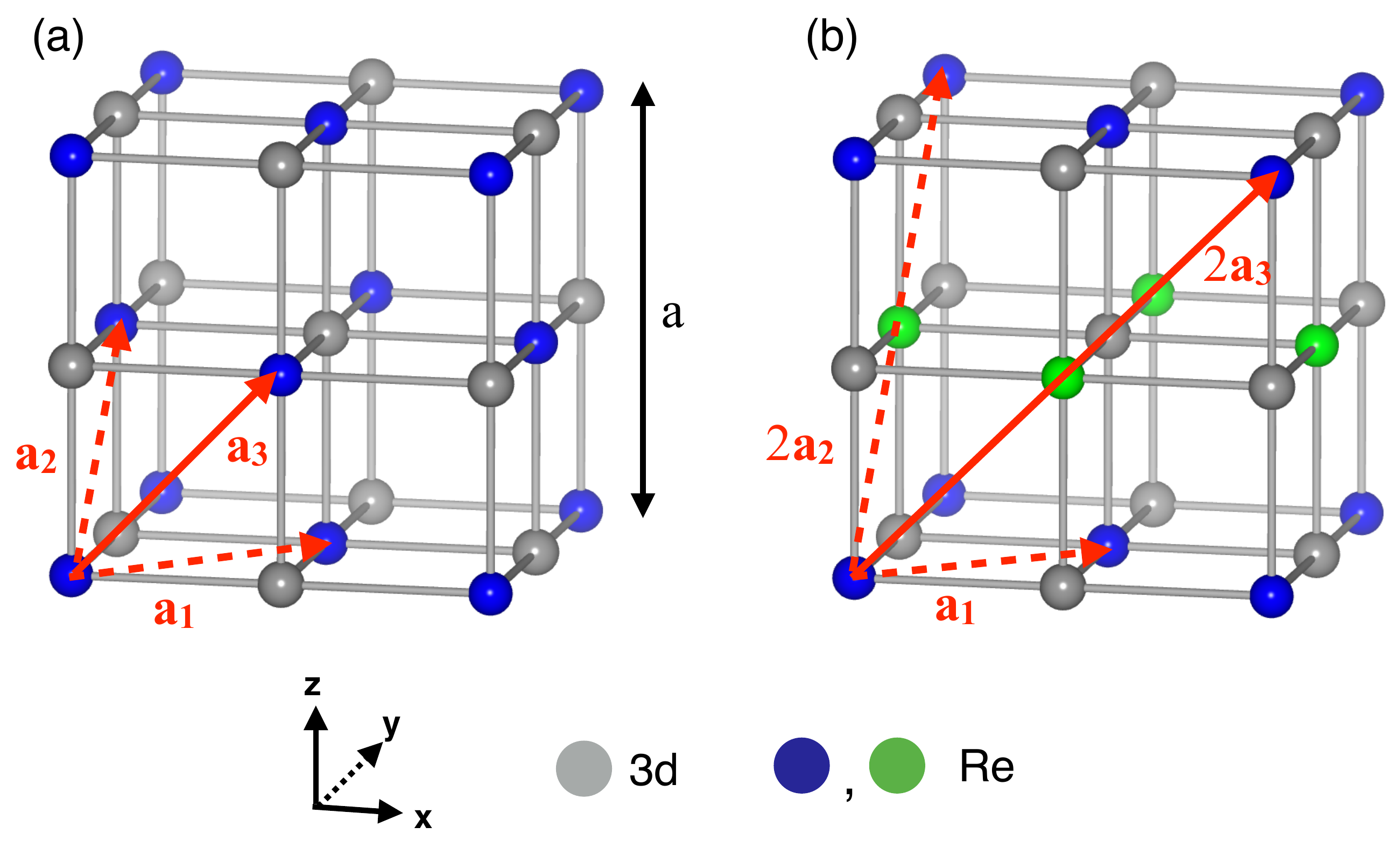}
\caption{ Schematic of the Face-centered cubic (FCC) lattice of Re-based DP. Oxygen atoms
are positioned at the midpoint between the Re and 3$d$ site; though not pictured.  Panel
(a) defines the choice of primitive lattice vectors in cubic phase, corresponding to 
  $\bm{a_1} = a/2 (\hat{\imath} +\hat{\jmath} )$, 
  $\bm{a_2} = a/2 (\hat{\jmath} + \hat{k}     )$, 
 $\bm{a_3} = a/2  (\hat{\imath} + \hat{k}     )$.
  Panel (b) shows the  $\bm{q_\textnormal{fcc}} = \left(0,1/2,1/2\right)$ phase modulation
  of the Re atoms (i.e. green/blue color), along with the choice of supercell lattice
  vectors which accommodate this motif.
}
\label{fcc_lattice}
\end{center}
\end{figure}

Here we review the experimental literature, in addition to some of the theoretical
literature, on our Re-based compounds of interest: $A_{2}B$ReO$_{6}$ ($A$=Sr, Ca, and
$B$=Cr, Fe). All four compounds form a perovskite structure with the Re/$B$ atoms ordering
in a $\bm{q_\textnormal{sc}} = \left( \frac{1}{2},\frac{1}{2},\frac{1}{2}\right)$ motif
with respect to the primitive simple cubic perovskite lattice vectors (see Figure
\ref{fcc_lattice}).  All systems are ferrimagnetically ordered below room temperature with
the Re and $B$ atoms having opposite spins.  We begin by presenting an experimental table
of the crystal structures for the ground state and at temperatures above the structural
transition; except for Ca$_{2}$CrReO$_{6}$, which is not known to have a transition near
room temperature (see Table \ref{str_temp}). Additionally, we tabulate the transition
temperatures and the nature of the ground state (i.e. metal vs. insulator). We will also
discuss other experimental viewpoints from the literature, some with dissenting views,
that are not represented in this table.

Bulk Sr$_{2}$FeReO$_{6}$ is tetragonal at 5K ($I4/m$, space group 87) as shown in Fig.
\ref{FeRe-str}(a), metallic (even in well ordered
samples)\cite{Teresa-SrCrReO,Re-Kato,Auth,Teresa-FeRe}, has $a^{0}a^{0}c^{-}$ octahedral
tilting, and in-plane and out-of-plane $\angle$Fe-O-Re are 171.9 and 180$^\circ$,
respectively \cite{Teresa-SrCrReO}.  Upon increasing temperature, it undergoes a
tetragonal-to-cubic phase transition at  $T_t=490K$ to space group 225 ($Fm\bar{3}m$),
removing octahedral tilting.

Ca$_{2}$FeReO$_{6}$ is monoclinic at 7K ($P2_{1}/n$, space group 14-2), and has
$a^{-}a^{-}b^{+}$ octahedral tilting (see Fig. \ref{FeRe-str}(b)).  It is generally known
as insulator at low temperature \cite{Iwasawa,Oikawa,Re-Kato,Teresa-FeRe}, though Fisher
\emph{et al.} suggested that it may be a bad metal \cite{Fisher}.  With increasing
temperature, Ca$_{2}$FeReO$_{6}$ undergoes a concomitant structural and metal-insulator
transition (MIT) at 140K \cite{Iwasawa,Oikawa}.  Interestingly, the structures above and
below the transition have the same space group symmetry, and octahedral tilting, but the
structural parameters are slightly different \cite{Oikawa}.  Based on the experimental
results, we infer that the predominant structural change at the phase transition is the
enhancement and reorientation of a local axial octahedral distortion of the Re-O
octahedron (i.e. linear combinations $A_{1g}+E_g$ octahedral modes, as defined from the
cubic reference) which order in an C-type antiferro (i.e. $\bm{q_{\textnormal{fcc}}} =
\left(0,\frac{1}{2},\frac{1}{2}\right)$) manner with respect to the primitive
face-centered cubic lattice vectors of the double perovskite (see Figure \ref{fcc_lattice}).
 We refer to this as a C-type octahedral distortion (C-OD) (see Table
 \ref{CaFeReO-Re_mode} for projections onto the octahedral mode amplitudes).

The C-OD will be demonstrated to be a signature of orbital ordering of the Re electrons;
which we will prove  to be the common mechanism of the MIT in this entire family of
materials.  The C-OD is not a spontaneously broken symmetry in the space group of
Ca$_{2}$FeReO$_{6}$ (e.g. there is a small non-zero amplitude in the high temperature
phase), and there are two symmetry inequivalent variants (i.e. C-OD$^+$ and C-OD$^-$, see
Fig. \ref{COD_orientation}) which represent the low and high temperature structures,
respectively.  Incidentally, the C-OD is a spontaneously broken symmetry in the Sr-based
crystals (due to the $a^0a^0c^-$ tilt pattern), whereby C-OD$^+$ and C-OD$^-$ are
identical by symmetry.

In order to clearly characterize the C-OD, the bond lengths of the  Re-O octahedron  from
the experimental structures are summarized in Fig.  \ref{CaFeReO-bonds}.  Above the
structural transition, the Re-O bonds are split into three sets of two equal bond lengths
(where the equal bonds arise from the inversion symmetry at the Re site), but two of the
three sets are very similar.  Specifically, at $T$=300K
$d_{\textnormal{Re1-O1}}=1.959\AA$, $d_{\textnormal{Re1-O2}}=1.954\AA$, and
$d_{\textnormal{Re1-O3}}=1.939\AA$, where Re1-O1 and Re1-O2 are approximately within the
$a-b$ plane and Re1-O3 is approximately along the $c$-axis (see Fig. \ref{COD_and_OO}). In
order to quantify relevant aspects of the octahedral distortions, we will define a
parameter $d_{|x-y|}=|d_{\textnormal{Re-O1}}-d_{\textnormal{Re-O2}}|$, which is small in
the high temperature phase (i.e. $d_{|x-y|} =0.005\AA$ at $T$=300K); and $d_{|x-y|}$ is
precisely the amplitude of the $E_g^{(0)}$ octahedral mode in the unrotated local
coordinate system. For the symmetry equivalent Re within the unit cell (i.e. Re2), the
nearly equivalent O1 and O2 bond lengths are swapped (i.e.  $d_{|x-y|} $ is identical but
the direction of the long/short bonds have reversed); while Re-O3 is identical.  

Upon
changing to the low temperature phase, there is a modest change whereby the Re-O3 bond
length shifts up by 0.006$\AA$ (i.e. equivalently in both Re), and a more dramatic change
whereby the splitting between Re-O1 and Re-O2 becomes substantially larger (i.e.
$d_{|x-y|} =0.014\AA$).  As in the high temperature structure, symmetry dictates that the
direction of $d_{|x-y|} $ alternates between the two Re sites.  The main difference is
that $d_{|x-y|}$ acquires an appreciable value in the low temperature phase, and the C-OD
switches between C-OD$^+$ and C-OD$^-$ (see Section \ref{subsubsec_cafe} for a more
detailed discussion).

Interestingly, Granado \emph{et al.} suggested that there is phase separation between 10K
and 650K, with all three phases being monoclinic \cite{Granado}.  More specifically, the
most abundant phases are found to be the M1 and M2 phases, with fraction of 55\% and 45\%,
respectively, and the main differences between the two phases are the $b$-lattice
parameter and angle $\beta$.  Similarly, Westerburg \emph{et al.} also observed two
different phases below 300K \cite{Westerburg}.  We note that the M1 and M2 phases in
Granado \emph{et al.}'s results \cite{Granado} are similar to the low-$T$ and high-$T$
phases reported by Oikawa \emph{et al.}, where the separation was not detected
\cite{Oikawa}.  M1, which has the largest portion at low temperature, has a $b$ lattice
parameter which is $\sim$0.015\AA\ smaller and a $\beta$ which is $\sim$0.1$^\circ$ larger
than those of the M2 phase, which constitutes $\sim$90\%  of the high T phase
\cite{Granado}.  Similarly, at 140K, the low-$T$ phase has smaller $b$ and larger $\beta$
than the high-$T$ phase in Oikawa \emph{et al.}'s report \cite{Oikawa}.

Having clarified the nature of the experimentally measured structural distortions in
Ca$_2$FeReO$_6$, we return to the issue of the MIT as addressed in the literature.  Since
there are nominally only Re $t_{2g}$ states near the Fermi level, the MIT is a gapping of
these states.  Oikawa \emph{et al.} suggested that the $d_{xy}$+$d_{yz}$ and
$d_{xy}$+$d_{zx}$ orbitals are randomly arranged at Re sites in the metallic phase,
whereas the $d_{yz}$+$d_{zx}$ orbitals are preferentially occupied in the insulating
phase; and the splitting between $d_{yz}$+$d_{zx}$ and $d_{xy}$ orbitals produce the
energy gap \cite{Oikawa}.  Previous local spin density functional theory (LSDA) studies
showed that Ca$_2$FeReO$_6$ is metallic without considering on-site Coulomb repulsion $U$
term for Re ($U_{\textnormal{Re}}$) \cite{CaFeRe-Wu,Szotek} and a gap is opened with the
large value $U$=3$-$4 eV \cite{Iwasawa,Jeon}. 

Gong \emph{et al.} concluded that the Re $t_{2g}$ states order into a $d_{xy}$+$d_{zx}$
configuration using the modified Becke-Johnson (mBJ) exchange-correlation potential, and
showed that Ca$_2$FeReO$_6$ is insulating; though this study did not explicitly identify
the C-type orbital ordering that drives this insulating state.  A noteworthy approximation
made in their work is that the atomic coordinates are relaxed within GGA, where
$d_{|x-y|}$ is only 0.004\AA, which is far smaller than low temperature experimental value
in the insulating state.  The importance of this amplified C-OD amplitude will be clearly
demonstrated within our work. 

Antonov \emph{et al.}\cite{Ernst} reported the electronic structure of Ca$_2$FeReO$_6$
using LSDA+$U$+spin-orbit-coupling calculations.  Using structures obtained from
experiment at  different temperatures, which encompasses the discontinuous phase
transition at $T$=140K \cite{Oikawa}, they showed that spin and orbital moments also have
abrupt changes across the transition, while both change linearly with structures from
temperatures below and above the MIT \cite{Ernst}.

\begin{table}
\begin{threeparttable}
\begin{ruledtabular}
\caption{ Space group ($sym$), amplitude of the $E_g^{(0)}$ octahedral mode ($d_{|x-y|}$,
in units of $\AA$)  for Re, and metallic/insulating ($M$/$I$) nature for both low and high
temperatures of Re-based DPs. Magnetic transition temperatures ($T_{\textnormal{C}}$) and
structural phase transition temperature ($T_t$) are also tabulated. Question marks
indicate unknown or uncertain data.  }
\label{str_temp}
\renewcommand{\arraystretch}{1.4}
\begin{tabular}{c c c c c c c c}
Materials				& $T_{\textnormal{C}}$		& $T_t$	 & $T$	& $sym$	& $d_{|x-y|}$	& $M/I$	 	 \\
\hline
\multirow{2}{*}{Sr$_2$FeReO$_6$}		
& \multirow{2}{*}{420K}\tnote{a}		&\multirow{2}{*}{490K}\tnote{a} 		& 5K		& $I4/m$\tnote{a}	& 0\tnote{a}	& $M$\tnote{a} \\
& 							&							& 500K 	& $Fm\bar{3}m$	& 0			& $M$ \\				
\hline
\multirow{2}{*}{Sr$_2$CrReO$_6$}		
& \multirow{2}{*}{620K}\tnote{a}		& \multirow{2}{*}{260K} \tnote{a}	& 2K		& $I4/m$?\tnote{*}		& ?\tnote{*}	& $I$\tnote{b} \\
& 							&							& 300K	& $Fm\bar{3}m$\tnote{a}	& 0\tnote{a}	& $M$\tnote{a,b} \\					
\hline
\multirow{2}{*}{Ca$_2$FeReO$_6$}		
& \multirow{2}{*}{540K}\tnote{c}		& \multirow{2}{*}{140K} \tnote{c}	& 7K		& $P2_{1}/n$\tnote{c}	& 0.014\tnote{c}	& $I$\tnote{c} \\
& 							&							& 300K	& $P2_{1}/n$		& 0.005			& $M$ \\		
\hline
\multirow{2}{*}{Ca$_2$CrReO$_6$}		
& \multirow{2}{*}{360K}\tnote{d}		& \multirow{2}{*}{?}			& 4.2K	& $P2_{1}/n$\tnote{d}	& ?				& $I$\tnote{d}	 \\
& 							&						& 300K 	& $P2_{1}/n$			& 0.005\tnote{d}	& $I$	 \\		
  \end{tabular}
\end{ruledtabular} 
\begin{tablenotes}[normal,flushleft]
\item[*] a recent experiment finds an insulating state, but the structural parameters have not been measured \cite{Hauser-SrCrRe} \\
\item[a] ref. \cite{Teresa-SrCrReO}   
\item[b] ref. \cite{Hauser-SrCrRe}  
\item[c] ref. \cite{Oikawa}
\item[d] ref. \cite{Re-Kato}      

\end{tablenotes}
\end{threeparttable}
\end{table}

Sr$_{2}$CrReO$_{6}$ has been determined to be tetragonal with an $a^{0}a^{0}c^{-}$
octahedral tilt pattern  (i.e. space group $I4/m$, see Fig.
\ref{FeRe-str}(c))\cite{Re-Kato,Teresa-SrCrReO}.  Teresa \emph{et al.} reported a
structural transition at $T=260K$, going from $I4/m$ to $Fm\bar 3 m$ (with increasing
temperature) whereby the octahedral tilts and the tetragonality are disordered.
Alternatively, Kato \emph{et al.} found that Sr$_{2}$CrReO$_{6}$ is still
$I4/m$\cite{Re-Kato,Kato-SrCrReO} at room temperature, implying that the transition
temperature was even higher in these particular samples; while Winkler \emph{et al.} found
that it was cubic ($Fm\bar{3}m$) at room temperature\cite{Winkler}.  This is likely a
minor discrepancy given that the structure measured by Kato \emph{et al.} at room
temperature (300K) only has small deviations from $Fm\bar{3}m$: the in-plane Cr-O-Re angle
in the tetragonal structure is 179.7$^\circ$, close to 180$^\circ$, and the lattice
parameters are nearly cubic with $\sqrt{2}a$=7.817 and $c$=7.809 \AA\cite{Re-Kato}.

Recent experiments have found Sr$_{2}$CrReO$_{6}$ to be insulating at low temperatures, in
contrast with earlier work which found metallic states.  Specifically, Hauser \emph{et
al.} found that a Sr$_{2}$CrReO$_{6}$ film grown on STO, where the strain is less than
0.05\%, is insulating at 2K with a 0.21eV energy gap \cite{Hauser-SrCrRe}.  Alternatively,
numerous samples obtained from chemically synthesis were all found to be  metallic
\cite{Re-Kato,Kato-SrCrReO,Teresa-SrCrReO,Winkler}, in addition to previous thin film
samples\cite{Asano-SrCrRe}.  It should be noted that Kato \emph{et al.} emphasized that
Sr$_{2}$CrReO$_{6}$ is a very bad metal, and lies at the vicinity of a Mott-insulating
state \cite{Kato-SrCrReO}.  Moreover, Hauser \emph{et al.} suggested that oxygen vacancies
are the reason why Sr$_{2}$CrReO$_{6}$ samples reported in previous studies were
metallic \cite{Hauser-Vo,Lucy-Vo}.  Indeed, previously reported metallic
Sr$_{2}$CrReO$_{6}$ samples have a large amount of defects, such as Cr/Re anti-site
defect: 9\%\cite{Majewski}, 15\% \cite{Michalik}, 10$-$12\% \cite{Teresa-SrCrReO}, and
23.3\% \cite{Re-Kato}.  However, there is not yet theoretical justification for why
Sr$_{2}$CrReO$_{6}$ might be an insulator.  Unfortunately, full structural parameters have
not yet been extracted from the insulating film at low temperatures\cite{Hauser-SrCrRe},
which could reveal signatures of an orbitally ordered insulator which we predict in our
analysis (see Section \ref{subsubsec_srcr}).

To our knowledge, there are only few experiments on Ca$_{2}$CrReO$_{6}$
\cite{Re-Kato,Kato-SrCrReO}; finding a monoclinic crystal structure (space group
$P2_{1}/n$, see Fig. \ref{FeRe-str}(d)) and an insulating ground state.  The energy gap is
not reported yet, though the reflectivity spectra and optical conductivity were measured
\cite{Kato-SrCrReO}.  Theoretically, the recent mBJ study of Gong \emph{et al.} suggested
that the energy gap is 0.38eV, much larger than that of Ca$_{2}$FeReO$_{6}$.  The
resistivity curve suggests that it is still insulating at room temperature, but the
resistivity will have an error given the 12-13.7\% of $B$-site disorder
\cite{Re-Kato,Kato-SrCrReO}, similar to the case of Sr$_{2}$CrReO$_{6}$.  In addition,
structural parameters as a function of temperature have not yet been reported, which will
be relevant to testing the predictions in our study.

In the existing literature, the effect of electronic correlation, orbital ordering, and
octahedral distortions have not been sufficiently isolated to give a universal
understanding of this family.  Most importantly, the origin of the MIT and it's relation
to orbital ordering and the concomitant C-OD have not been elucidated.  Theory and
computation will be critical to separating cause from effect.

\subsection{Orbital ordering }
\label{sec:orbital_ordering}

Orbital ordering is a well known phenomena in transition metal
oxides \cite{Kugel,Tokura2000462,Kugel1982231}, and it can drive a material into an insulating ground state.  
Two main mechanisms which
drive orbital ordering are the electron-lattice ($e$-$l$) coupling, with a very relevant scenario being the well known Jahn-Teller (JT) Effect,
and electron-electron ($e$-$e$) interactions. 
Disentangling these two effects in a real system can be challenging, as both mechanisms result in orbital ordering and a concomitant lattice distortion;
though the latter could be vanishingly small in the case of $e$-$e$ driven orbital ordering. 
A
complicating factor in both theory and experiment is that preexisting
structural distortions (e.g. octahedral tilting) may preclude the orbital
ordering from being a spontaneously broken symmetry; meaning that orbital
ordering is always present and the only question is a matter of degree.
In the event that the $e$-$e$ interactions are driving the ordering, a further question is if orbital ordering is critical to realizing the insulating
state (ie. Slater-like $e$-$e$ driven orbital ordered insulator) or if Mott physics generates the insulating state (ie. the system remains insulating even
if the orbitals are thermally disordered). This latter question can also be cumbersome to disentangle.

In the context of DFT+$U$ calculations, the Hubbard $U$ captures a very relevant portion
of the $e$-$e$ interactions which drive orbital-ordering; similar to the $U$ in a model Hamiltonian which gives rise to superexchange\cite{Kugel}. 
The $e$-$l$ coupling is accounted for in the DFT portion of the calculation (assuming a local
or semi-local approximation to the DFT functional). If experimentally deduced orbital ordering is accounted for at the level
of DFT (ie. $U$=0), then $e$-$l$ couplings are likely playing a dominant role; 
while if DFT does not predominantly capture the orbital ordering, 
then the $e$-$e$ interactions are likely playing a dominant role. In this case of dominant $e$-$e$ interactions, if a particular spatial ordering is a necessary condition to 
drive an insulating state within DFT+$U$ (for a physical value of $U$), then the resulting insulating
state could be labeled as Slater-like. If an insulating state is achieved for an arbitrary ordering of the orbitals, then the system would be considered Mott-like.

%An appreciable value of $E_g^{(0)}$ indicates orbital ordering, making it a useful metric; though it is not a necessary condition for orbital ordering. 
%For example, if the $e$-$e$ is dominant and the $e$-$l$ is weak, the orbitals may order while the  $E_g^{(0)}$ octahedral distortion may be thermally quenched
%with relatively small temperatures. Alternatively, reversing the strenghts of the mechanism would require the $E_g^{(0)}$ octahedral distortion as a necessary condition.

\begin{table}
\begin{threeparttable}
\begin{ruledtabular}
  \caption{ Space group ($sym$), amplitude of the $E_g^{(0)}$ octahedral mode
  ($d_{|x-y|}$)   for orbitally active transition metal, and
  metallic/insulating ($M$/$I$) nature for both low and high temperatures of
  various perovskites which have orbital ordering. Magnetic transition
  temperatures ($T_{\textnormal{mag}}$) are also tabulated, where
  $T_{\textnormal{mag}}=T_N$, except for YTiO$_3$, and Ba$_2$NaOsO$_6$, where
  $T_{\textnormal{mag}}=T_C$. Question marks indicate unknown or uncertain
  data.  }
\label{mott_slater_table}
\renewcommand{\arraystretch}{1.35}
\begin{tabular}{c c c c c c c c}
Materials		& $T_{\textnormal{mag}}$		& $T$	& $sym$	& $d_{|x-y|}$ (\AA)	& $M/I$	& ref 	 \\
	\hline
\multirow{2}{*}{LaMnO$_3$}	%T_N
& \multirow{2}{*}{140K}	& 300K	& $Pbnm$		& 0.271\tnote{a}	& $I$		
& \multirow{2}{*}{\shortstack[l]{\cite{LMO-PRB1998,MandalPRB2001,Yin-LaMnO3,Moussa-LaMnO3,Sanchez-LaMnO3}}} \\
&	 				& 798K	& $Pbnm$ & 0.047			& $M$\tnote{b}	& \\
\hline
\multirow{2}{*}{KCuF$_3$}	%T_N
& \multirow{2}{*}{38K}	& 300K	& $I4/mcm$	& 0.372\tnote{c}	& $I$		
& \multirow{2}{*}{\shortstack[l]{\cite{Marshall-KCuF3,LDA+U1,Pavarini_KCuF_2008,Leonov2008096405,Hutchings-KCuF3,Paolasini-KCuF3}}}\\
&					& 900K	& $I4/mcm$	& 0.453			& $I$?	& \\
\hline
\multirow{2}{*}{LaTiO$_3$}	%T_N
& \multirow{2}{*}{146K}	& 8K		& $Pbnm$		& 0.021\tnote{d}	& $I$	 
& \multirow{2}{*}{\shortstack[l]{\cite{Cwik-LTO,Keimer-LTO,Hemberger2003066403,Iliev2004172301,Fritsch2002212405,
Pavarini_LTO_2004,Pavarini_LTO_2005,Mochizuki20041833,Mochizuki2004154,Arima199317006,Cheng-LTO}}} \\
&					& 293K 	& $Pbnm$		& 0.026			& $I$		& \\
\hline						
\multirow{2}{*}{YTiO$_3$}		%T_C
& \multirow{2}{*}{30K}	& 2K		& $Pbnm$		& 0.054\tnote{e}	& $I$	 
& \multirow{2}{*}{\shortstack[l]{\cite{Arima199317006,Cheng-LTO,KomarekPRB2007,Pavarini_LTO_2004,Pavarini_LTO_2005,Ament-YTO,
Mochizuki20041833,Mochizuki2004154,Akimitsu20013475,Itoh19992783,Kiyama20051123,Iga2004257207,Nakao2002184419,
Taguchi1993511,Okimoto19959581}}}\\
& 					& 290K	& $Pbnm$		& 0.051			& $I$		& \\
\hline						
\multirow{3}{*}{LaVO$_3$}   %T_N
& \multirow{3}{*}{143K}	& 10K 	& $P2_{1}/n$		& 0.061\tnote{f}			& $I$		
& \multirow{3}{*}{\shortstack[l]{\cite{Bordet1993,ZhouPRL2008,RenPRB2003,MiyasakaPRL2000,MiyasakaPRB2003,        
RaychaudhuryPRL2007,FangPRL2004,MiyasakaJPSJ2002}}} \\	
&					& 150K	& $Pbnm$		& 0.013		& $I$	 	& \\
& 					& 295K 	& $Pbnm$		& 0.001			& $I$		& \\						
\hline
\multirow{2}{*}{YVO$_3$\tnote{*}}		%T_N
& \multirow{2}{*}{116K}	& 5K		& $Pbnm$		& 0.050\tnote{g}	& $I$	 
& \multirow{2}{*}{\shortstack[l]{\cite{ReehuisPRB2006,BenckiserPRB2013,KawanoJPSJ1994,NoguchiPRB2000,
BlakePRB2002,UlrichPRL2003,BlakePRL2001,MiyasakaJPSJ2002,MiyasakaPRB2003,FangPRL2004,RaychaudhuryPRL2007,BenckiserNJP2008}}}	\\
& 					& 295K 	& $Pbnm$		& 0.014			& $I$		& \\		
\hline
\multirow{2}{*}{Ba$_2$NaOsO$_6$}		%T_C
& \multirow{2}{*}{7K}	& 5K			& $Fm\bar{3}m$	& 0\tnote{h}	& $I$	 
& \multirow{2}{*}{\shortstack[l]{\cite{Erickson2007,Stitzer2002,Steele2011,Xiang2007,Balents2010,Jackeli2017,KWLee2007,Pickett2015,Pickett2016}}} \\
& 					& high $T$	& $Fm\bar{3}m$	& 0			& ?	& \\								
\hline
\multirow{2}{*}{Sr$_2$CeIrO$_6$}	%T_N %d=0.01069 at 30K
& \multirow{2}{*}{21K}	& 2K		& $P2_{1}/n$		& 0.008\tnote{i}		& $I$	 
& \multirow{2}{*}{\shortstack[l]{\cite{HaradaJPCM2000,Felser-CeIr}}}\\
& 					& 300K	& $P2_{1}/n$		& 0.043				& $I$		& \\															
\end{tabular}
\end{ruledtabular} 
\begin{tablenotes}[para,flushleft]
\small
\item[a] ref. \cite{LMO-PRB1998}   
\item[b] ref. \cite{MandalPRB2001}   
\item[c] ref. \cite{Marshall-KCuF3}   
\item[d] ref. \cite{Cwik-LTO}   
\item[e] ref. \cite{KomarekPRB2007}   
\item[f] ref. \cite{Bordet1993}   
\item[g] ref. \cite{ReehuisPRB2006}  
\item[h] ref. \cite{Steele2011}  
\item[i] ref. \cite{HaradaJPCM2000}   \\
\item[*] YVO$_3$ is $P2_1/n$ between 77K and 200K.
\end{tablenotes}
\end{threeparttable}
\end{table}

Classic examples of perovskites which display antiferro orbital ordering, and are insulators, include 
the 3$d^4$  LaMnO$_3$ \cite{LMO-PRB1998,MandalPRB2001,Yin-LaMnO3,Moussa-LaMnO3,Sanchez-LaMnO3} and
the 3$d^9$ KCuF$_3$\cite{Marshall-KCuF3,LDA+U1,Pavarini_KCuF_2008,Leonov2008096405,Hutchings-KCuF3,Paolasini-KCuF3}
which have ordering of $e_g$ electrons;
the 3$d^1$ ($t_{2g}$) materials LaTiO$_3$ and YTiO$_3$ 
\cite{Cwik-LTO,Keimer-LTO,Hemberger2003066403,Iliev2004172301,Fritsch2002212405,
Pavarini_LTO_2004,Pavarini_LTO_2005,Mochizuki20041833,Mochizuki2004154,Arima199317006,Cheng-LTO,KomarekPRB2007,Ament-YTO,
Akimitsu20013475,Itoh19992783,Kiyama20051123,Iga2004257207,Nakao2002184419,
Taguchi1993511,Okimoto19959581}; and
the 3$d^2$  ($t_{2g}$)  perovskites LaVO$_3$ and YVO$_3$ 
\cite{Bordet1993,ZhouPRL2008,RenPRB2003,MiyasakaPRL2000,MiyasakaPRB2003,        
RaychaudhuryPRL2007,FangPRL2004,MiyasakaJPSJ2002,ReehuisPRB2006,BenckiserPRB2013,KawanoJPSJ1994,NoguchiPRB2000,
BlakePRB2002,UlrichPRL2003,BlakePRL2001,BenckiserNJP2008}.
It is useful to make some empirical characterization of these classic examples 
to provide context for the orbital ordering we identify in this study (see Table \ref{mott_slater_table}). 
All of these systems are insulators until relatively high temperatures. 

%Additionally, magnetism is a relatively low energy detail
%in most cases, given that the magnetic ordering temperature typically far below the onset of orbital ordering; and always below the metal-insulator transition.

All of the aforementioned examples have the GdFeO$_3$ tilt pattern ($a^{-}a^{-}b^{+}$)
except KCuF$_3$, and thus there is only orbital degeneracy in KCuF$_3$ (assuming a
reference state where the $E_g^{(1)}$ strain mode is zero).  Therefore, antiferro orbital
ordering could be a spontaneously broken symmetry for KCuF$_3$, while the other systems
will always display some degree of orbital polarization and octahedral distortion.  In all
cases, the most relevant lattice distortion is an $E_g^{(0)}$ distortion, 
driven by both $e$-$l$ and $e$-$e$
interactions.  The $e$-$l$ coupling is generally much larger for scenarios involving $e_g$
electrons as compared to $t_{2g}$.

DFT+$U$ calculations can be helpful in disentangling the effects of $e$-$e$ interactions and $e$-$l$ coupling.
In the aforementioned classic examples of orbital ordering
involving $e_{g}$ electrons, important contributions are realized  from both
$e$-$e$ interactions and $e$-$l$ coupling.  In LaMnO$_3$, an antiferro
$E_g^{(0)}$ Jahn-Teller distortion, and corresponding orbital ordering, is found even at
the level of GGA (ie. $U_{\textnormal{Mn}}$=0): the $e$-$l$ coupling is strong
enough to recover 0.8 of the experimentally observed Jahn-Teller
distortion\cite{Yin-LaMnO3}. However, a non-zero $U_{\textnormal{Mn}}$ is
needed to properly capture the energy stabilization, insulating ground state,
and full magnitude  of the  Jahn-Teller distorted, orbitally ordered state.  
In
KCuF$_3$, pure GGA is sufficient to spontaneously break symmetry and obtain the
antiferro $E_g^{(0)}$ Jahn-Teller distortion that is observed in experiment,
though the stabilization energy is grossly underestimated and the distortion
magnitude is too small\cite{Leonov2008096405}.  Including the on-site $U$ gives
reasonable agreement with experiment (both within DFT+$U$ and
DFT+DMFT)\cite{LDA+U1,Pavarini_KCuF_2008,Leonov2008096405}.  Alternatively, if
one remains in the cubic reference structure, preventing coupling with the
lattice, an on-site $U$ of 7eV can drive the orbitally ordered insulator with a
corresponding transition temperature of roughly 350K\cite{Pavarini_KCuF_2008}.
Therefore, both mechanism can drive the same instability, but in isolation the
on-site $U$ recovers a larger component of the stabilization energy;
though both ingredients are necessary to quantitatively describe experiment.
In both LaMnO$_3$ and KCuF$_3$, $e$-$e$ interactions and $e$-$l$ coupling
both play a direct, relevant role.

In the $t_{2g}$-based systems, the $e$-$l$ coupling is expected to be smaller.  DFT+$U$
studies for LaTiO$_3$ show that at $U$=0, the system is metallic and has a very small
$E_g^{(0)}$ distortion of $d_{|x-y|}$=0.004$-$0.005$\AA$ \cite{AhnJKPS2006,HePRB2012}; in
contrast to experiment which yields an insulator with $d_{|x-y|}$=0.021$\AA$. Increasing
the $e$-$e$ interactions to $U$=3.2eV/$J$=0.9eV \cite{AhnJKPS2006}, an insulator is obtained
and $d_{|x-y|}$=0.018$\AA$, in much better agreement with experiment (see Table
\ref{mott_slater_table}).  DFT will always have a small value of $d_{|x-y|}$ due to the
broken symmetry caused by the octahedral tilting, and the $e$-$l$ coupling within DFT
provides no strong enhancement of this distortion.  Applying the Hubbard $U$ both orders
the orbitals and induces an appreciable value of $d_{|x-y|}$.  This concomitant
$d_{|x-y|}$ distortion may increase the potency of the Hubbard $U$, such as increasing the
resulting band gap (See results of Re-based family in Section \ref{results_discuss}).  
The vanadates 
behave in a similar fashion.  DFT (ie. $U$=0) calculations for LaVO$_3$, for the
low-temperature phase $P2_1/n$ (or alternatively, $P2_1/a$ or $P2_1/b$), showed that
LaVO$_3$ is metallic and $d_{|x-y|}$=0.001-0.002$\AA$\cite{HePRB2012}, in stark
disagreement with experiment which shows insulating behavior and $d_{|x-y|}$=0.061$\AA$
(see Table \ref{mott_slater_table}). Hybrid functional calculations, which are very
similar in nature to DFT+$U$, recover the insulating state and a appreciable $d_{|x-y|}$
amplitude.  
In these $t_{2g}$-based systems, DFT gets
$d_{|x-y|}$ wrong by a factor of approximately 4-5 and 30-60 for the titanates and the
vanadates, respectively; a much more dramatic failure than in the e$_g$-based materials.

Within experiment, one cannot easily isolate different terms in the Hamiltonian, though it
may be possible to thermally quench the lattice distortion and determine if the orbital
polarization persists. If so, this would strongly indicate that $e$-$e$ interaction are
dominant in driving orbital ordering. Furthermore, experiment could possibly determine if
the system remains gapped upon thermally disordering the orbitals. As mentioned above,
octahedral tilting is often a higher energy scale which already breaks symmetry, and it
will not be totally clear what the quenched value of the distortion and or the orbital
polarization should be.  Below we tabulate the amplitude of the $E_g^{(0)}$ distortion and
the metallic/insulating nature at a low temperature (i.e. a temperature below the orbital
ordering) and a high temperature (i.e. either above the orbital ordering or the highest
temperature measured); the magnetic transition temperatures are also included. 

Three scenarios can be identified. First, the $E_g^{(0)}$ distortion may be essentially
unchanged as a function of temperature, or even enhanced, while the material remains
insulating (i.e. KCuF$_3$, LaTiO$_3$, and YTiO$_3$). Second, the $E_g^{(0)}$ distortion
may be largely quenched via temperature and the system concomitantly becomes metallic
(i.e. LaMnO$_3$). Third, the $E_g^{(0)}$ distortion may be largely quenched and the system
remains insulating (i.e. LaVO$_3$ and YVO$_3$). In the first two scenarios, little can be
deduced without further analysis: the orbital ordering and structural distortion are
either frozen in or are simultaneously washed out. For the vanadates, we learn that the
lattice distortion is irrelevant for attaining the insulating state: $e$-$e$ interactions
drive the orbital ordering. Further analysis would be needed to know if the insulator is
Slater-like or Mott-like.

The orbital ordering which we identify in the 3$d$-5$d$ Re-based DP's of this
study has a number of distinct circumstances as compared to these classic 3$d$
single perovskites.  First, in the Re-DP's the magnetic transition temperatures
($T_{mag}$) are a  much larger energy scale (see Table \ref{str_temp}), which means that the spins are strongly ordered 
well before orbital physics comes into play;  whereas the reverse is true in the single 3$d$ perovskites. 
%In LaTiO$_3$, the N\'eel temperature $T_N$ is 146K and it is still
%insulating at 300K \cite{Arima199317006,Cwik-LTO}.
%In YTiO$_3$, the Curie temperature is $T_C$=30K and it remains insulating at room temperature 
%\cite{Iga2004257207,Taguchi1993511,Okimoto19959581}.  
%XX
%In KCuF$_3$, $T_N$ is 38K \cite{Hutchings-KCuF3} while the JT distortion and orbital ordering is stable below 800K \cite{Paolasini-KCuF3}.
%Recent neutron powder diffraction experiments showed that the large JT distortion even exists at 900K \cite{Marshall-KCuF3}. 
%Similarly, $T_N$ of LaVO$_3$ and YVO$_3$ are 143K and 116K, respectively, while they are still insulating at room temperature 
%\cite{MiyasakaPRL2000,BenckiserNJP2008}.
%In LaMnO$_3$, $T_N$ is 140K in \cite{Moussa-LaMnO3} while the JT transition and orbital ordering is stable until
%750K\cite{Sanchez-LaMnO3}.
%Alternatively, in Ca$_2$FeReO$_6$ the ferrimagnetic transition temperature $T_C$=520K, 
%which  is much higher than $T_{\textnormal{MIT}}$=140K, opposite to the classic 3$d$ perovskites.
Another difference is that the electronic structure of the 3$d$-5$d$ DPs are
generally governed by 5$d$ orbitals, which may have a non-trivial spin-orbit
interaction\cite{Balents2010}. While 5$d$ orbitals are more delocalized than
3$d$ orbitals, it should be kept in mind that that the rock salt ordering of
the DP's results in relatively small effective Re bandwidths (see Section
\ref{sec:general_aspects}).

A well studied class of DP's where orbital ordering may be relevant is the $A_2BB'$O$_6$
double perovskites, where $B$ has fully filled or empty $d$ orbital and $B'$ is a 5$d$
transition metal.  One well studied type of family is the $B'$=5$d^1$ Mott insulators,
such as Ba$_2B$OsO$_6$ ($B$=Li, Na)
\cite{Erickson2007,Stitzer2002,Steele2011,Xiang2007,Balents2010,Jackeli2017,KWLee2007,Pickett2016,Pickett2015}
and Ba$_2B$MoO$_6$ ($B$=Y, Lu)  \cite{Steele2011} (see Table \ref{mott_slater_table}).
These materials have very weak magnetic exchange interactions (e.g. $T_C$ of
Ba$_2$NaOsO$_6$ is 6.8-8 K \cite{Erickson2007,Stitzer2002,Steele2011}), and exotic phases
have been proposed such as quantum-spin-liquids, valence-bond solids, or spin-orbit dimer
phases\cite{Balents2010,Jackeli2017}.  Xiang \emph{et al.} \cite{Xiang2007} studied
Ba$_2$NaOsO$_6$ using the first-principles calculations, and suggested that an insulating
phase cannot be obtained within GGA+$U$ up to $U-J$=0.5 Ryd: orbital ordering is not
observed in their electronic band structure within GGA+$U$.  They also show that
Ba$_2$NaOsO$_6$ is insulating within GGA+$U$ when including spin-orbit coupling (SOC) with
$U-J$=0.2Ryd and the [111] magnetization axis.  Gangopadhyay \emph{et al.}
\cite{Pickett2015,Pickett2016} also proposed that SOC is essential to obtain a nonzero
band gap, using hybrid functional + SOC calculations.  Based on experiment, Erickson
\emph{et al.} proposed that Ba$_2$NaOsO$_6$ has orbital ordering with a non-zero
wavevector, deduced in part from the small negative Weiss temperature from magnetic
susceptibility measurements \cite{Erickson2007}.

Another analogous example is the Ir-based double perovskite Sr$_2$CeIrO$_6$ (see Table \ref{mott_slater_table}), 
where Ce has a filled shell and the Ir 5$d$ nominally have 5
electrons (or one hole) in the $e_g^\pi + a_{1g}$ orbitals (i.e. descendants of
$t_{2g}$)\cite{Felser-CeIr}, and this results in  weak antiferromagnetic coupling (i.e.
$T_N$=21K).  Additionally, orbital ordering has been identified in this material, where
the hole orders in the $e_g^\pi$ shell among the $d_{xz}$ and $d_{yz}$ orbitals with an
antiferro modulation. The orbital ordering is accompanied by a $E_g^{(0)}$ structural
distortion, though the experimental temperature dependence is rather unusual.
At 2K and 300K, $d_{|x-y|}$=0.008$\AA$ and $d_{|x-y|}$=0.043$\AA$, respectively (see Table \ref{mott_slater_table}),
showing a strong increase in amplitude with increasing temperature\cite{HaradaJPCM2000};
while $d_{|x-y|}$=0.049$\AA$ is obtained within GGA+$U$ ($U$=4eV and $J$=1eV) \cite{Felser-CeIr}.
The authors
attribute the orbital ordering to the Jahn-Teller effect, though they demonstrate that
$U_{\textnormal{Ir}}$ is a necessary condition for opening a band gap\cite{Felser-CeIr}.
It should be noted that the wavevector of the antiferro orbital and structural ordering in
this system is the same as what we identify in the Re-based family in the present work.

Re-based double perovskites are quiet distinct from the aforementioned double perovskites
with empty or fully filled $d$ shell on the $B$ ion.  Unlike these latter materials,
Re-based DPs have nonzero magnetic spin for $B$ (e.g., Cr has spin 3/2 and Fe has spin
5/2), and thus have a strong antiferromagnetic exchange interaction between $B$ and $B'$;
resulting in a $T_C$ that is  much higher than room temperature (e.g. $T_C$ in
Sr$_2$CrReO$_6$ is 620K, see Table \ref{str_temp}). Therefore, the spin degrees of freedom
are locked in until relatively high temperatures, creating an ideal testbed to probe
orbital physics. The family of Re-based DP's evaluated in this study are ideally
distributed in parameter space about the orbital ordering phase transition.

\section{Computation Details}
\label{comp_method}
We used the projector augmented wave (PAW) method \cite{PAW,Kresse19991758} in order to numerically solve the Kohn-Sham equations,
as implemented in the VASP code \cite{VASP}. 
The exchange-correlation functional was approximated using the revised version of the generalized gradient
approximation (GGA) proposed by Perdew \emph{et al.} (PBEsol) \cite{PBEsol}.
%Selective testing is also done using the local density approximation (LDA) \cite{CA,Perdew}.
In all cases, the spin-dependent version of the exchange correlation functional is employed; both with and without
spin-orbit coupling (SOC).
A plane wave basis with a kinetic energy cutoff of 500 eV was employed. We used a $\Gamma$-centered \textbf{k}-point 
mesh of 9$\times$9$\times$7 (11$\times$11$\times$9 for density of states).
Wigner-Seitz radii of 1.323, 1.164, and 1.434 \AA~ were used for site 
projections on Cr, Fe and Re atoms, respectively, as implemented in the VASP-PAW projectors.

The GGA+$U$ scheme within the rotationally invariant formalism and the fully localized
limit double-counting formula \cite{LDA+U1} is used to study the effect of electron
correlation.  The electronic and structural properties critically depend on
$U_{\textnormal{Re}}$, and  therefore we carefully explore a range of values.  We also
explore how the results depend on $U_{\textnormal{Cr}}$ and $U_{\textnormal{Fe}}$, which
play a secondary but relevant role in the physics of these materials.  We do not employ an
on-site exchange interaction $J$ for any species, as this is already accounted for within
the spin-dependent exchange-correlation potential \cite{Hyowon,Chen2015}.  

A post facto
analysis of our results demonstrate that a single set of values (which are reasonable as
compared to naive expectations and previous work) can account for the electronic and
crystal structure of this family (see Section \ref{U_of_Re}), and it is useful to provide
this information at the outset for clarity. In the absence of spin-orbit coupling, values
of $U_{\textnormal{Fe}}$=4 eV, $U_{\textnormal{Cr}}$=2.5 eV, and $U_{\textnormal{Re}}$=2
eV are found; including spin-orbit coupling requires $U_{\textnormal{Re}}$ to be slightly
decreased to 1.9 eV in order to maintain the proper physics.  In subsequent discussions,
the units of $U$ will always be in electron volts (eV), and this may be suppressed for
brevity.

We used experimental lattice parameters throughout (see Table \ref{str_temp}), and the
reference temperature is 300K  unless otherwise specified.  Atomic positions within the
unit cell were relaxed until the residual forces were less than 0.01 eV/\AA.  In select
cases we do relax the lattice parameters as well to ensure no qualitative changes occur,
and indeed the changes are small and inconsequential in all cases tested.

\section{Results and discussion}
\label{results_discuss}

\subsection{General Aspects of the Electronic Structure}
\label{sec:general_aspects}

\begin{figure}
\begin{center}
\includegraphics[width=0.50\textwidth, angle=0]{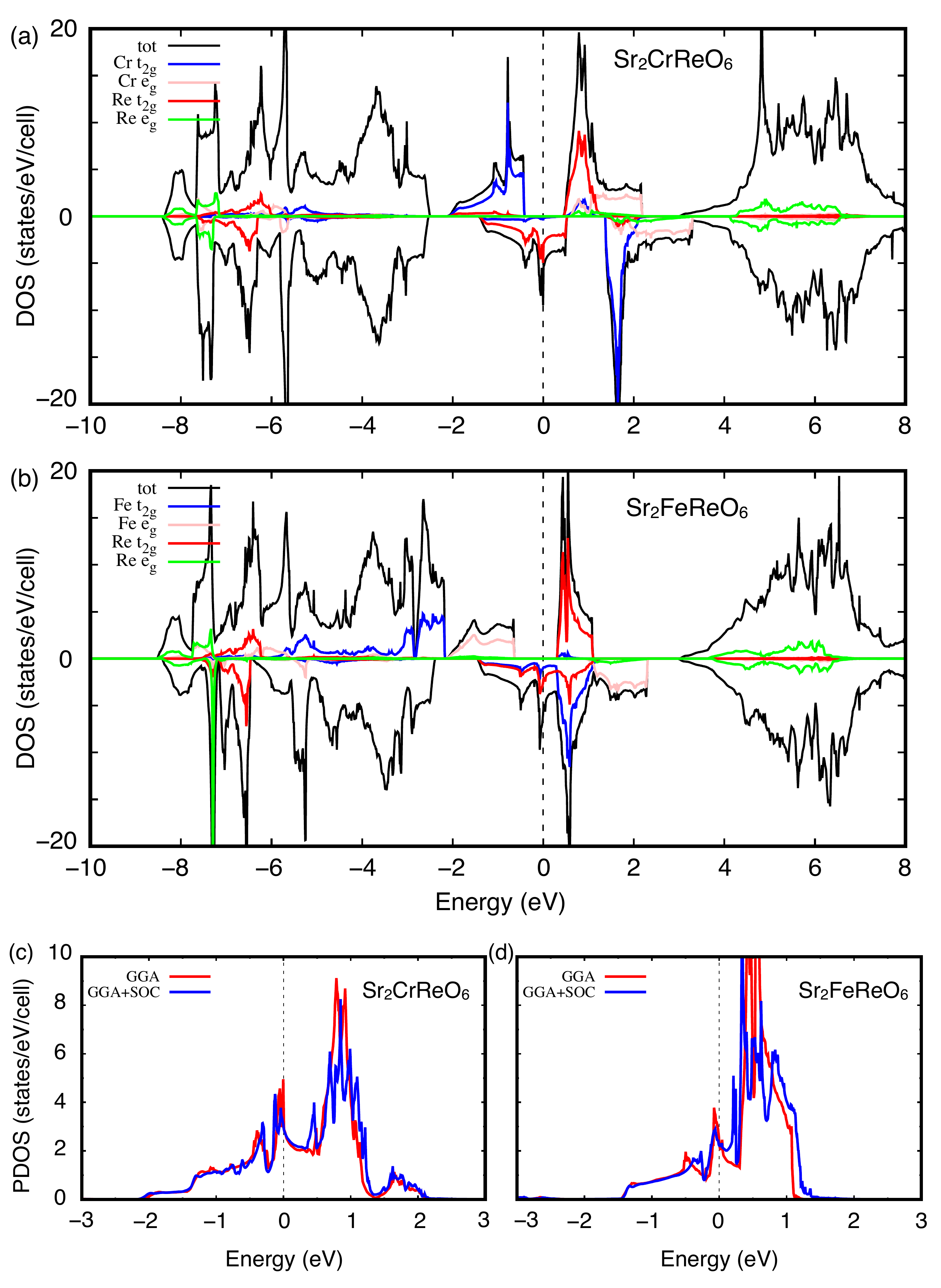}
\caption{ (a),(b)  Total and atom/orbital projected spin-resolved  density of states (DOS)
from DFT for (a) Sr$_2$CrReO$_6$ and  (b) Sr$_2$FeReO$_6$.  The majority spin are shown as
a positive DOS while the minority are negative.  Orbital projections are given for
$t_{2g}$ and $e_g$ states for Re, Fe, and Cr.  (c),(d) Illustrating the effect of
spin-orbit coupling in the Re Projected density of states, comparing GGA (red solid line)
and GGA+SOC (blue solid line) for (c) Sr$_2$CrReO$_6$ and  (d) Sr$_2$FeReO$_6$.  For GGA,
majority and minority spins are summed for comparison.  The Fermi energy is zero in all
panels, and Fm$\bar 3$m is used throughout.  }
\label{electronic_struct}
\end{center}
\end{figure}

\begin{figure}
\begin{center}
\includegraphics[width=0.5\textwidth, angle=0]{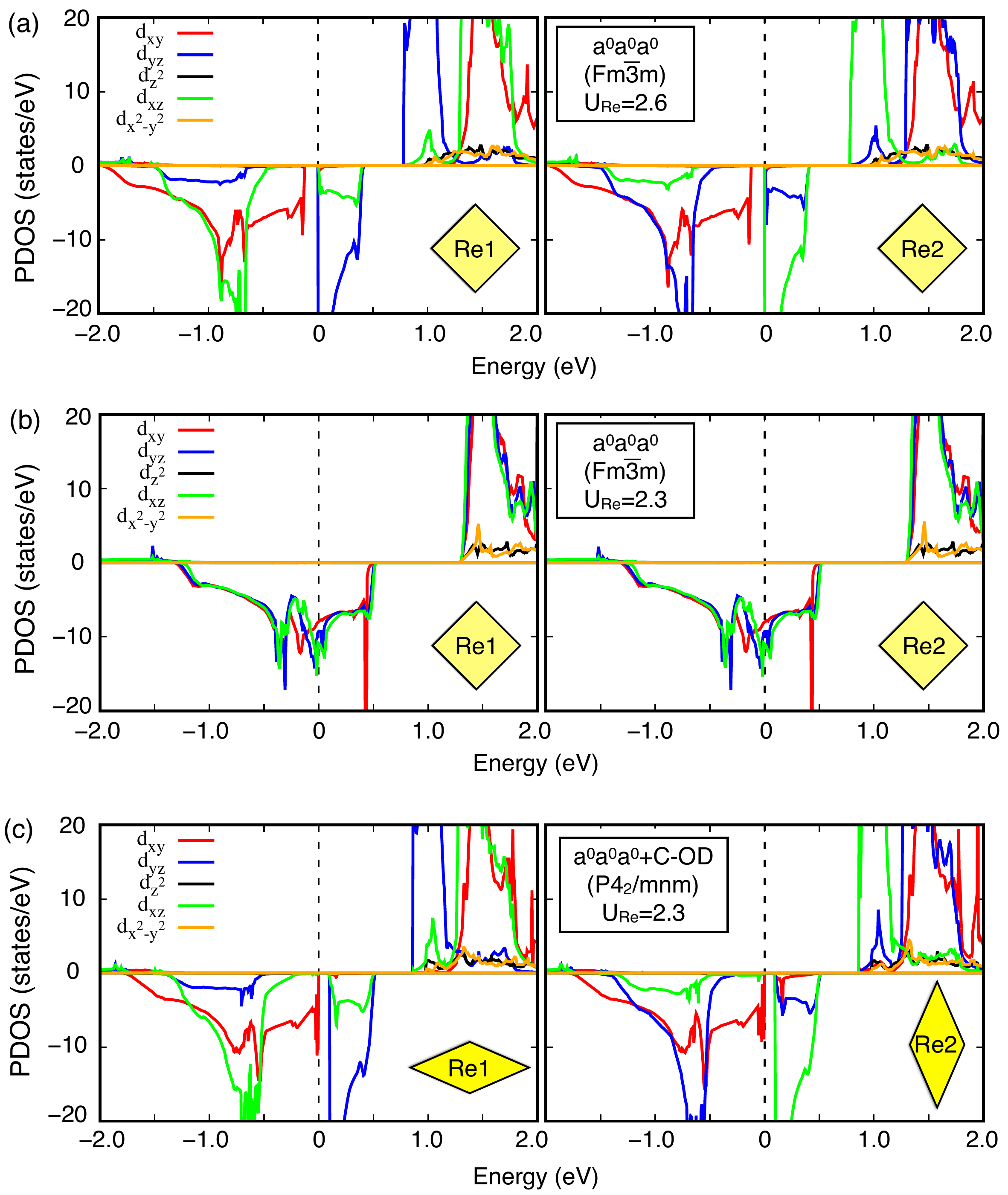}
  \caption{   Re projected density of states (PDOS) for Sr$_2$CrReO$_6$ within DFT+$U$.
  (a) Fm$\bar 3$m crystal structure and  $U_{\textnormal{Re}}$=2.6, resulting in an orbitally ordered insulator.
  (b) Fm$\bar 3$m crystal structure and  $U_{\textnormal{Re}}$=2.3, resulting in a metal.
  (c) P4$_2$/mnm crystal structure (ie. $a^{0}a^{0}a^{0}$ tilt, with C-OD) and $U_{\textnormal{Re}}$=2.3, 
  resulting in an orbitally ordered insulator.
  $U_{\textnormal{Cr}}$=0 in all cases.
}
\label{dos-COD}
\end{center}
\end{figure}

\begin{figure}
\begin{center}
\includegraphics[width=0.5\textwidth, angle=0]{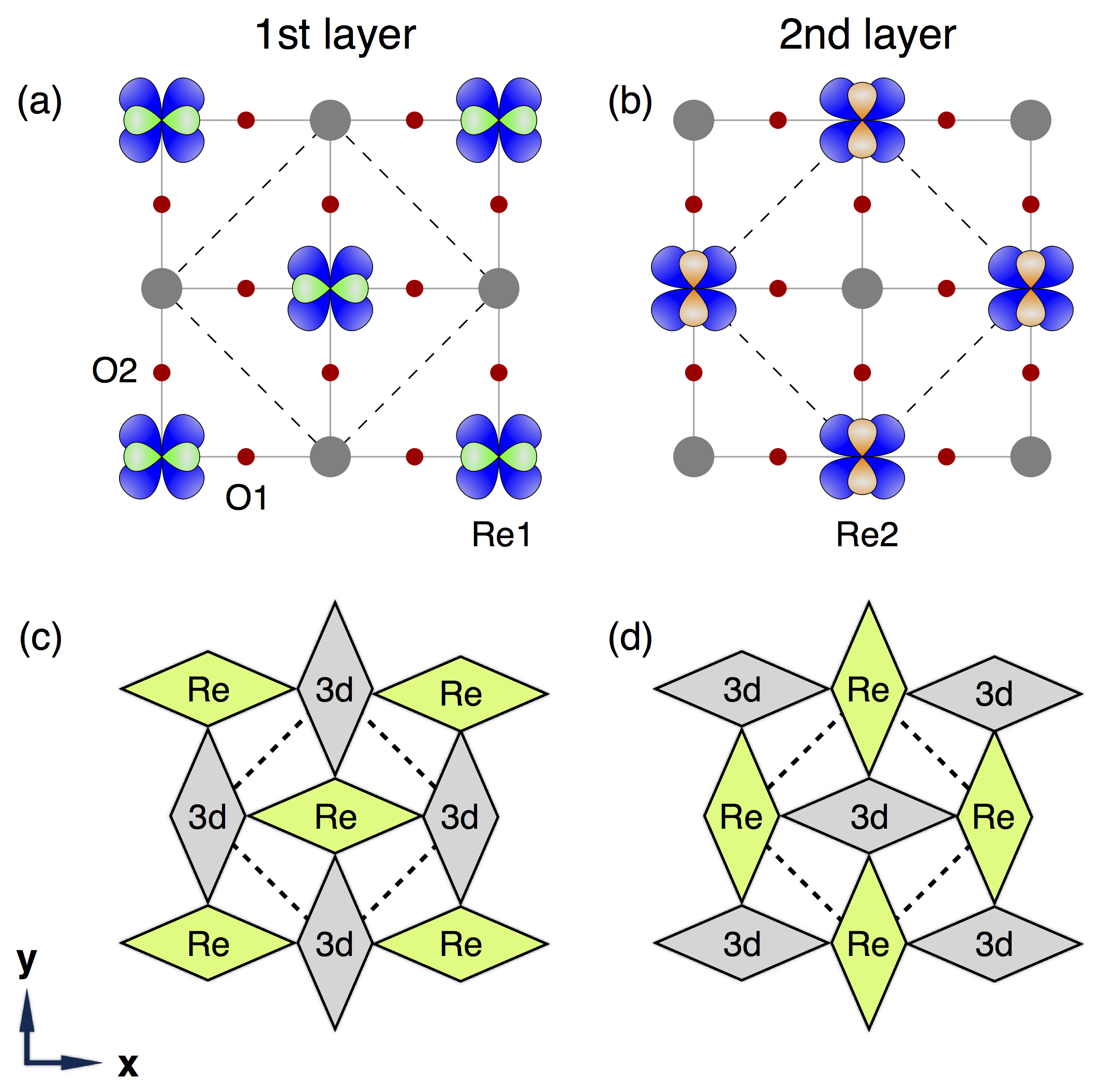}
\caption{   Schematic diagrams of the C-type orbital ordering for the 1st and 2nd layer along
the c-axis of the conventional cubic cell, shown in panels (a) and (b), respectively.  Red dots
correspond to oxygen atoms while grey dots correspond to $B$ atoms (ie. Cr or Fe). The pictured
$t_{2g}$ orbitals are located on Re sites ($d_{xy}$ is blue, $d_{xz}$ is green, $d_{yz}$ is
orange).  Panels  (c) and (d) depict a schematic of the C-OD type structural distortion,
whereby each local octahedron is distorted in a positive or negative $E_g^{(0)}$ distortion.
Positive $E_g^{(0)}$ corresponds to the elongation along the y-direction and contraction along
the x-direction (see Ref.  \cite{Marianetti2001224304} for formula).  Dotted lines correspond
to the unit cell.  }
\label{COD_and_OO}
\end{center}
\end{figure}

We begin by discussing the nominal charge states of the transition metals, the basic energy
scales,  and the common mechanism of the metal-insulator transition in these compounds; which
is a C-type antiferro orbital ordering. A perfect cubic structure ($Fm\bar{3}m$) is first
considered, in the absence of SOC (which will be addressed at the end of this discussion).
Given the Re double perovskite $A_2B$ReO$_6$ ($A$=Sr,Ca, $B$=Fe,Cr), nominal charge counting
dictates that the transition metal pair $B$Re must collectively donate 8 electrons to the
oxygen (given that A$_2$ donates 4 electrons), and it is energetically favorable (as shown
below) to have Re$^{5+}$ ($d^2$) and B$^{3+}$ (Cr$\rightarrow d^3$ and Fe $\rightarrow d^5$ )
in a high spin configuration. The Re spin couples antiferromagnetically to the $B$ spin via
superexchange, yielding a ferrimagnetic state.  Given that the nominally $d^5$ Fe has a half
filled shell when fully polarized, and that the nominally filled $d^3$ Cr has a half-filled
$t_{2g}$-based shell when fully polarized, none of these compounds would be expected to have Fe
or Cr states at the Fermi energy when strongly polarized.  Given that Re is in a $d^2$
configuration, group theory dictates that the system will be metallic with majority spin Re
states present at the Fermi energy within band theory. 

These naive expectations are clearly realized in DFT calculations (ie. $U$=0), as illustrated
in Sr$_2$CrReO$_6$ and Sr$_2$FeReO$_6$ using the $Fm\bar{3}m$ structure (see Figure
\ref{electronic_struct}).  It is useful to compare the Re states crossing the Fermi energy,
which are substantially narrower for Sr$_2$CrReO$_6$ as compared to Sr$_2$FeReO$_6$.
Relatedly, the Cr states hybridize less with and are further from the Re states as compared to
the case of Fe. The net result is that the Cr based compounds will have a smaller effective Re
bandwidth, and therefore stronger electronic correlations which result in a higher propensity
to form an insulating state.

At the level of DFT+$U$, or any static theory for that matter, one can only obtain an insulator
from the fully spin-polarized scenario outlined above via an additional spontaneously broken
symmetry, which could be driven either via the on-site Re Coulomb repulsion
$U_{\textnormal{Re}}$, structural distortions (which includes effects of electron-phonon
coupling), or combinations thereof. As we will detail in the remainder of the paper, structural
distortions alone (i.e. if $U_{\textnormal{Re}}$=0) cannot drive an insulating state in any of
the four Re-based materials studied. Therefore, a non-zero $U_{\textnormal{Re}}$ is a
\emph{necessary condition} to drive the insulating state, but the minimum required value of
$U_{\textnormal{Re}}$ will be influenced by the details of the structural distortions; in
addition to the on-site $U$ of the 3$d$ transition metal and the SOC.

In order to illustrate the points of the preceding paragraph, we show that
DFT+$U$ calculations (with the only nonzero $U$ being
$U_{\textnormal{Re}}$=2.6) for Sr$_2$CrReO$_6$ with the nuclei frozen in the
$Fm\bar{3}m$ structure results in a spontaneously broken symmetry of the
electrons where the Re orbitals order and result in an insulating state (see
Figure \ref{dos-COD}, panel a). 
We investigated ordered states consistent with $\bm{q_\textnormal{fcc}}=(0,0,0)$, $\bm{q_\textnormal{fcc}}=(0,0,\frac{1}{2})$, 
$\bm{q_\textnormal{fcc}}=(0,\frac{1}{2},\frac{1}{2})$, and $\bm{q_\textnormal{fcc}}=(\frac{1}{2},\frac{1}{2},\frac{1}{2})$
(where $q$ is a fractional coordinate of the reciprocal lattice vectors constructed from  the primitive FCC DP lattice vectors; see Fig. \ref{fcc_lattice}); 
resulting in a ground state of $\bm{q_\textnormal{fcc}}=(0,\frac{1}{2},\frac{1}{2})$ (ie. C-type ordering).
Specifically, 
a Re $d_{xy}$ orbital is occupied on every site and there is a C-type
alternation between $d_{xz}$ and $d_{yz}$ (see schematic in Figure \ref{COD_and_OO}, panels a and b).
This C-type antiferro orbital ordering (denoted as C-OO) is generic among this $A_2B$ReO$_6$ family.
We will demonstrate that other orderings are possible and even favorable under certain conditions. 
For example, for small values of $U_{\textnormal{Re}}$, the orbitals order in a ferro fashion (denoted F-OO), 
whereby the $d_{xy}$ and either the $d_{xz}$ or $d_{yz}$ is
occupied at every Re site. For intermediate values of $U_{\textnormal{Re}}$, a ferri version of the C-OO ordering (denoted FI-OO) is found, 
though it is destroyed by octahedral tilts.
These detailed scenarios are explored in Section \ref{sec:electronic_struct}.

\begin{figure}
\begin{center}
\includegraphics[width=0.5\textwidth, angle=0]{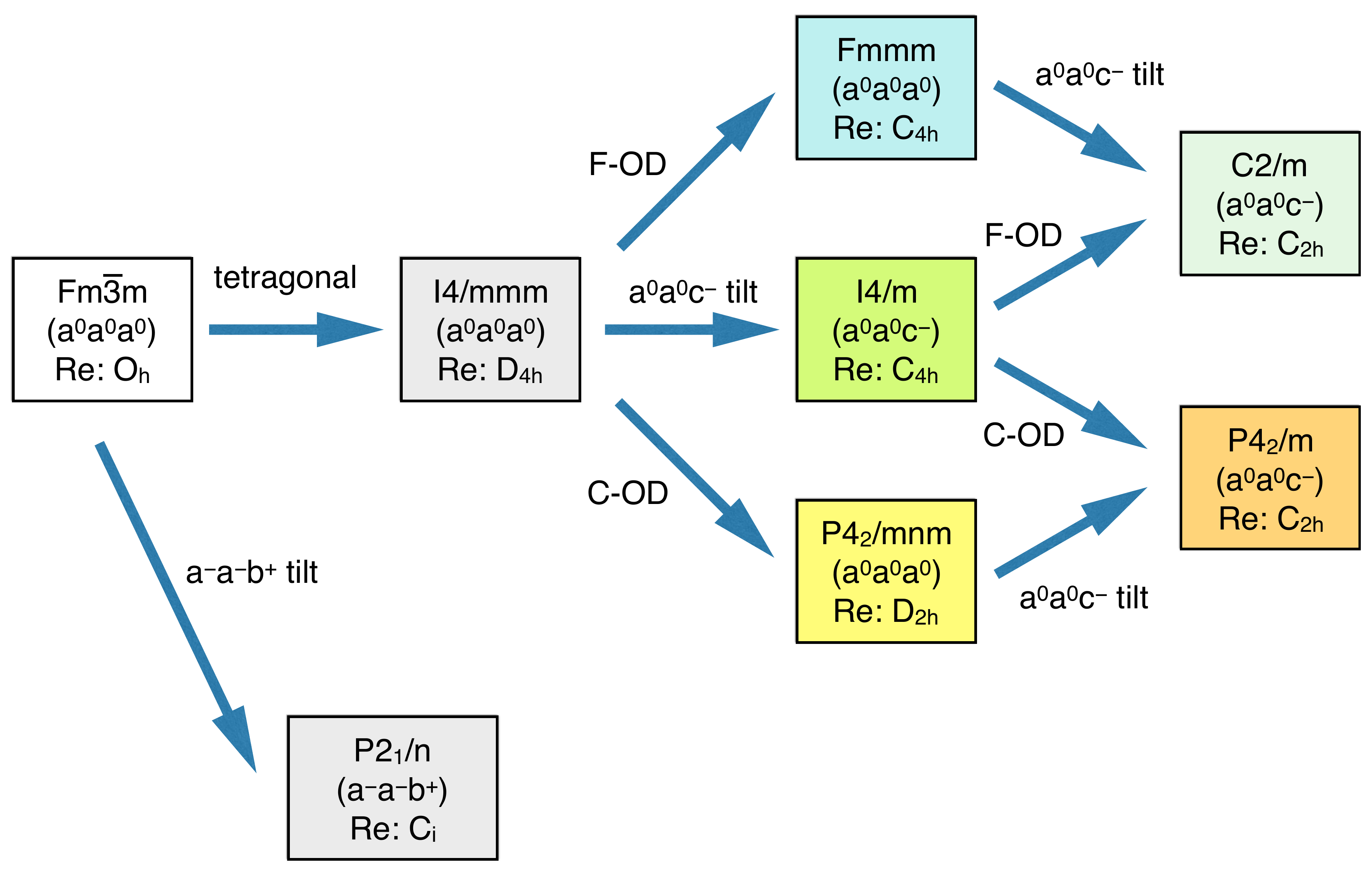}
  \caption{   Hierarchy of the $A_{2}BB'$O$_6$ double perovskite space groups connected by various
  distortions, including octahedral tilts, F-OD, and C-OD. The point group
  symmetry of the Re site is listed for all structures, along with the octahedral tilt pattern.  
  Ca-based systems follow the $a^{-}a^{-}b^{+}$ arrow, while Sr-based systems proceed along the
  tetragonal arrow.
}
\label{symmetry_schematic}
\end{center}
\end{figure}

We now turn to the importance of structural distortions, such as the $E_g$ octahedral
distortions which are induced by the orbital ordering.  We first remain in the
$Fm\bar{3}m$ structure and lower the value of $U_{\textnormal{Re}}$ to 2.3eV,
demonstrating that the orbital ordering is destroyed and the gap is closed (see Figure
\ref{dos-COD}, panel b).  Subsequently, we allow any internal relaxations of the ions
consistent with $\bm{q_\textnormal{fcc}}=(0,0,0)$ or
$\bm{q_\textnormal{fcc}}=(0,\frac{1}{2},\frac{1}{2})$, demonstrating that an $E_g^{(0)}$
octahedral distortion with C-type wavevector (denoted C-OD)  condenses (see Figure
\ref{COD_and_OO} (c)/(d) for schematic); lowering the structural symmetry from
$Fm\bar{3}m$ to $P4_2/mnm$ (see symmetry lineage in Figure \ref{symmetry_schematic}) and
allowing the  C-OO to occur at $U_{\textnormal{Re}}$=2.3eV (see Figure \ref{dos-COD},
panel c).  This demonstrates how the C-OD can be an essential ingredient for realizing the
orbitally ordered insulating state, by influencing the critical value of
$U_{\textnormal{Re}}$ for the transition. Incidentally, it should be noted that when the
orbital ordering changes, the structural distortion changes as expected. For example,
ferro orbital ordering (i.e. F-OO) will lead to a ferro octahedral distortion (i.e. F-OD).

The above analysis proves that it is reasonable to characterize the insulating state as an orbitally ordered
state, despite the fact that the C-OD structural distortion could play a critical
role in moving the MIT phase boundaries to smaller values of $U_{\textnormal{Re}}$. 
We will demonstrate that this renormalization of the critical $U_{\textnormal{Re}}$ via the C-OD 
allows a common value of $U_{\textnormal{Re}}$ to realize the insulating in Sr$_2$CrReO$_6$, while retaining
a metallic state in Sr$_2$FeReO$_6$; and we predict that  
the orbitally ordered state can persist in the near absence of the C-OD in Ca$_2$CrReO$_6$ where electronic correlations are strongest. 
Given that the C-OD does not occur in the absence $U_{\textnormal{Re}}$, we refrain from
characterizing this as a Jahn-Teller effect, or pseudo Jahn-Teller effect in the case were the C-OO/C-OD is not
a spontaneously broken symmetry, which could have been a primary driving force given the orbital degeneracy (or near degeneracy)
present in these systems.

Another generic consideration is octahedral tilting, which will influence both the C-OO
and the C-OD.  The $a^{-}a^{-}b^{+}$ tilt pattern of the Ca-based systems is a relatively
large energy scale and therefore the tilts in these system exist independently of orbital
ordering and or the C-OD. Alternatively, the $a^{0}a^{0}c^{-}$ tilt pattern of the
Sr-based systems is a much weaker energy scale, and therefore it may be somewhat coupled
to the orbital ordering and the concomitant C-OD. These statements will be investigated in
detail below (see Section \ref{sec:electronic_struct}), where we find that the differences
of Sr/Ca are dominant over those of Fe/Cr in terms of setting the effective Re bandwidth;
which results in a ordering of Sr$_2$FeReO$_6$, Sr$_2$CrReO$_6$, Ca$_2$FeReO$_6$,
Ca$_2$CrReO$_6$ (smallest to largest effective Re bandwidth or electronic correlations).
For example, the resulting Re-bandwidths are 1.84, 1.70, 1.50, and 1.35 eV, respectively
(using $U_{\textnormal{Re}}$=0, $U_{\textnormal{Fe}}$=4, and $U_{\textnormal{Cr}}$=2.5).

Furthermore, the $a^{0}a^{0}c^{-}$ tilt pattern may be isolated from the C-OO/C-OD as they
break symmetry in a distinct manner (see symmetry lineage in Figure
\ref{symmetry_schematic}).  Therefore, the C-OO/C-OD will be a spontaneously broken
symmetry in the Sr based systems (should it occur).  Alternatively, the $a^{-}a^{-}b^{+}$
tilt pattern already has a sufficiently low symmetry such that the C-OO/C-OD is not a
spontaneously broken symmetry. Therefore, the C-OD cannot strictly be a signature of
orbital ordering in the case of the $a^{-}a^{-}b^{+}$ tilt pattern.  However, experiment
dictates that the magnitude of the C-OD is a useful metric given the discontinuous
structural phase transition at 140K between two crystal structures of the same space group
($P2_{1}/n$, no. 14-2), whereby the magnitude of the C-OD changes discontinuously; and the
variant switches from C-OD$^+$ to C-OD$^-$.

\begin{figure*}
\begin{center}
\includegraphics[width=0.9\textwidth, angle=0]{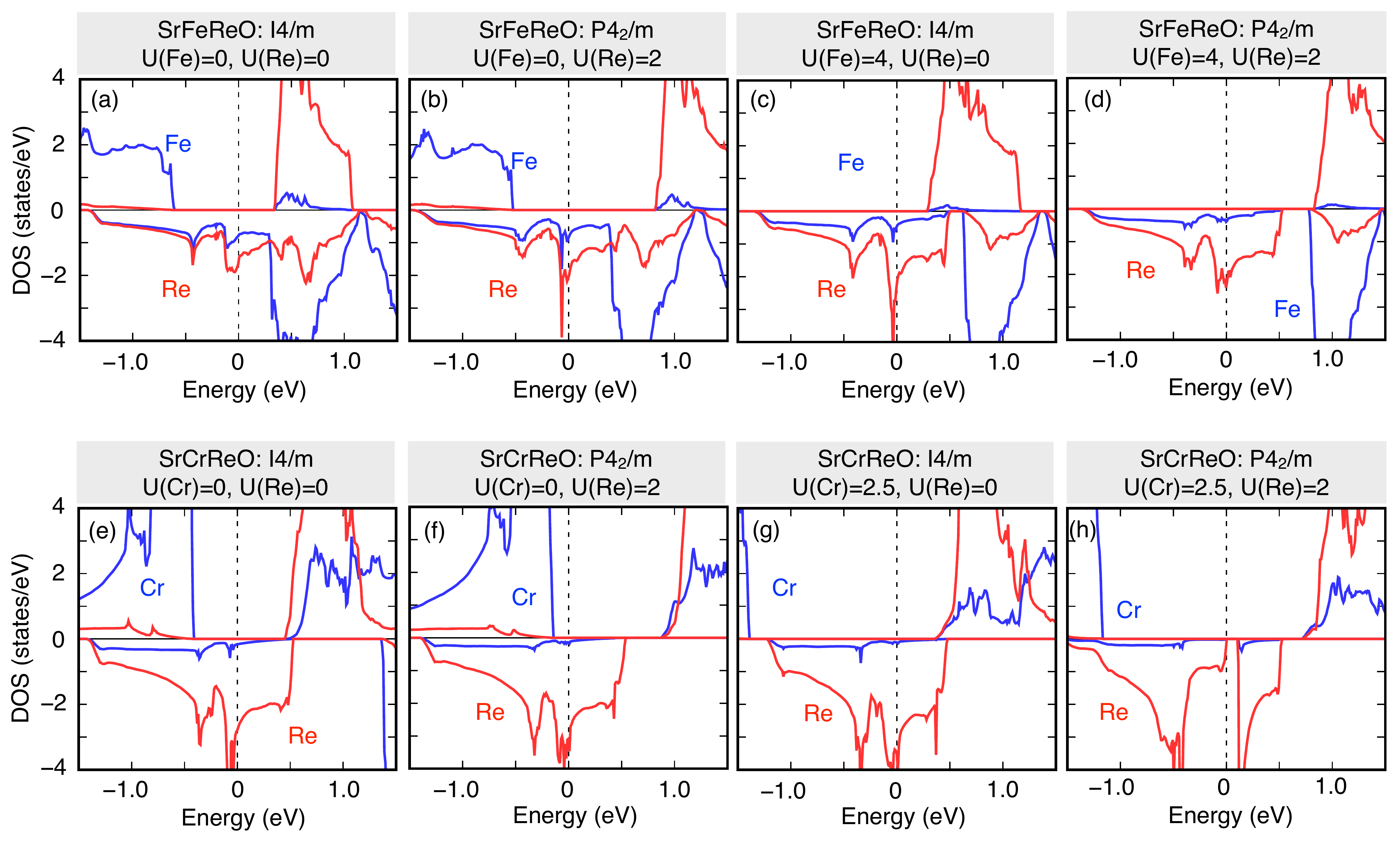}
\caption{   Projected density of states of Sr$_2$CrReO$_6$ and Sr$_2$FeReO$_6$, 
 projected on Fe (blue), Cr (blue), and Re (red) $d$ orbitals.
 Values of $U$ and space group (obtained from relaxing internal coordinates) are indicated in each panel. 
}
\label{hybrid}
\end{center}
\end{figure*}

Another generic consideration is the effect of the on-site Coulomb repulsion $U$ for the
$3d$ transition metals, which do not nominally have states at low energies given the half
filled (sub)shell of (Cr) Fe. However, in reality there is a non-trivial amount of $3d$
states at low energies due to hybridization, more so for Fe than for Cr, and this
determines the effective Re-$d$ bandwidth.  While $U_{\textnormal{Re}}$ can drive orbital
ordering even in the absence of $U_{\textnormal{Fe}}$/$U_{\textnormal{Cr}}$, as previously
illustrated above (see Fig. \ref{dos-COD}), we will demonstrate the quantitative influence
of $U_{\textnormal{Fe}}$/$U_{\textnormal{Cr}}$ in renormalizing the critical value of
$U_{\textnormal{Re}}$ for orbital ordering.  First, considering $U_{\textnormal{Re}}$=0,
one can clearly see an unmixing of 3$d$ states from the Re-$d$ states as
$U_{\textnormal{Fe}}$/$U_{\textnormal{Cr}}$ is applied, further narrowing the effective
Re-$d$ bandwidth (compare panels $a\leftrightarrow c$ and $e\leftrightarrow g$ in Fig.
\ref{hybrid}). 
%XX
%In Sr$_2$CrReO$_6$, spin-down Re-$t_{2g}$ bandwidths of $I4/mmm$ and $I4/m$ phases within GGA are 1.95 and 1.93 eV, respectively.
%If $U_{\textnormal{Cr}}$=2.5 is applied, bandwidths decreased by $\sim$0.2eV; become 1.75 and 1.70 eV for $I4/mmm$ and $I4/m$, respectively.

This effect is more dramatic in the case of Fe, which started with a larger degree of
hybridization.  Focussing on the Fe compound, we see that applying $U_{\textnormal{Re}}$=2
does not drive the orbitally ordered insulator even when $U_{\textnormal{Fe}}$=4, and thus
the system remains metallic despite the diminished Re-$d$ bandwidth. Alternatively, when
applying $U_{\textnormal{Re}}$=2 to the Cr compound, the addition of
$U_{\textnormal{Cr}}$=2.5 is sufficient to move the critical value of
$U_{\textnormal{Re}}$ below 2eV, and an orbitally ordered insulator is obtained. This
demonstrates that, while indirect, the on-site $U$ for the 3$d$ transition metal can play
a critical role. Interestingly, $U_{\textnormal{Cr}}$ also turns out to be critical for
stabilizing the experimentally observed $a^{0}a^{0}c^{-}$ tilt pattern in Sr$_2$CrReO$_6$
(see Section \ref{subsubsec_srcr}).

\begin{comment}
\begin{table}
\label{bandwidth}
\begin{ruledtabular}
\begin{center}
  \caption{ Re bandwidths (eV) of Sr$_2$CrReO$_6$. }
\renewcommand{\arraystretch}{1.3}
\begin{tabular}{c c c c c}
	& \multicolumn{2}{c}{$U_{\textnormal{Cr}}$=0} & \multicolumn{2}{c}{$U_{\textnormal{Cr}}$=2.5} 	 \\
	\cline{2-3} \cline{4-5}
	& $U_{\textnormal{Re}}$=0 & $U_{\textnormal{Re}}$=2 & $U_{\textnormal{Re}}$=0 & $U_{\textnormal{Re}}$=2 \\
\hline
$I4/mmm$ 	&  1.95	& 1.98	& 1.75	&1.83 \\
$P4_{2}/mnm$ &  1.95	& 2.02	& 1.74	&1.87 \\	
$I4/m$		&  1.93	& 1.93	& 1.70	&1.79 \\	
$P4_{2}/m$	&  1.93	& 2.02	& 1.70	& 2.11 \\	
\end{tabular}
\end{center}
\end{ruledtabular}
\end{table}
\end{comment}

Yet another generic consideration is the spin-orbit coupling. We demonstrate the SOC is a
relatively small perturbation in this system by comparing the Re states near the Fermi
energy for the cubic reference structure computed using GGA (ie. $U$=0) with and without
SOC (see Fig. \ref{electronic_struct}, panels c and d). As shown, the DOS only exhibits
small changes upon introducing SOC. Indeed, we will demonstrate the SOC can shift the
phase boundary of the C-OO/C-OD by small amounts, and this can be very relevant in the
Ca-based systems (including a strong magnetization direction dependence in
Ca$_2$FeReO$_6$, see Section \ref{sec:SOC}).

Finally, we discuss how temperature will drive the insulator to metal transition and the
structural transition associated with the C-OD. For the most part, we will only address
ground state properties in this study, as finite temperatures will be beyond our current
scope; though some of our analysis will shed light on what may occur. As outlined above,
the insulating ground state in this family of materials is driven by C-type orbital
ordering on the Re sites, though two main factors will influence the critical value of
$U_{\textnormal{Re}}$: the C-OD and octahedral tilting. One can imagine several different
scenarios which could play out depending on the energy scales. First, the temperature of
the electrons could disorder the C-OO. Given that our DFT+U calculations predict that this
C-OO induced insulator is Slater-like (i.e.  the gap closes given ferro and other orbital
orderings, see Section \ref{subsubsec_srfe} and \ref{subsubsec_srcr}), the material will
become metallic upon disordering the orbitals. Given that weak nature of the
electron-phonon coupling (i.e. the C-OD cannot condense without an on-site
$U_{\textnormal{Re}}$), this means that the C-OD would disorder along with the orbitals. 

A different scenario can be envisioned at an opposite extreme, whereby the energy scale
for orbital ordering is very large and we can neglect the electronic temperature and only
consider the phonons. In this case, temperature could disorder the C-OD and or the
octahedral tilts which would substantially increase the critical value of
$U_{\textnormal{Re}}$, driving the system into a metallic state.  We will entertain this
latter scenario (see Section \ref{sec:electronic_struct}, and Figs. \ref{CaFeReO-pdos} in
particular), though it does not appear consistent with our preferred values of $U$ when
including SOC unless there is a reorientation of the magnetization direction as seen in
experiment (see Section \ref{sec:SOC}, and Fig. \ref{CaFeReO-pdos-soc}).  In reality, it
is possible that all ingredients may be needed in order to properly capture the MIT and
structural transitions from first-principles, and our paper will lay the groundwork for
future study.

\subsection{Crystal and electronic structures}
\label{sec:electronic_struct}
Here we compute the crystal and electronic structure of $A_2B$ReO$_6$ ($A$=Sr, Ca and
$B$=Cr,Fe), exploring a range of Hubbard $U$s for all transition metals.  We approach the
four materials in order of increasing strength of electronic correlations:
Sr$_2$FeReO$_6$, Sr$_2$CrReO$_6$, Ca$_2$FeReO$_6$, Ca$_2$CrReO$_6$. We will address
orbital ordering, axial octahedral distortions, octahedral tilt pattern, the presence of a
band gap, and relative structural energetics.

\subsubsection{Sr$_{2}$FeReO$_{6}$ }
\label{subsubsec_srfe}

\begin{table}
\begin{ruledtabular}
\begin{center}
\caption{ Relative energy difference (in meV per formula unit) of the $a^{0}a^{0}c^{-}$
structure minus the $a^{0}a^{0}a^{0}$ structure for Sr$_2$FeReO$_6$ within both GGA and
GGA+$U$. The resulting space groups are indicated.  We consider full structural
relaxations, in addition to only relaxing internal coordinates.  }
\label{SrFeReO}
\renewcommand{\arraystretch}{1.3}
\begin{tabular}{c c c c c}
  \multicolumn{1}{c}{$U$ (eV)} & symmetry	& inter. rel. 	& full rel.  \\
\hline
$U_{\textnormal{Fe}}$=0,  $U_{\textnormal{Re}}$=0	& $I4/m-I4/mmm$		& $-$56	& $-$65 \\
$U_{\textnormal{Fe}}$=4,	$U_{\textnormal{Re}}$=0	& $I4/m-I4/mmm$		& $-$44	& $-$51 \\
$U_{\textnormal{Fe}}$=0,  $U_{\textnormal{Re}}$=2	& $P4_2/m-P4_2/mnm$	& $-$55	& $-$60 \\
$U_{\textnormal{Fe}}$=4,	$U_{\textnormal{Re}}$=2	& $P4_2/m-P4_2/mnm$	& $-$48	& $-$53 
\end{tabular}
\end{center}
\end{ruledtabular}
\end{table}

Experimentally, Sr$_2$FeReO$_6$ is found to be a metal with an $a^{0}a^{0}c^{-}$
octahedral tilt pattern and $I4/m$ symmetry (see Section \ref{lit_review}).  Given that Sr
will have a smaller propensity to drive octahedral tilts relative to Ca, and that in
Figure \ref{electronic_struct} we showed that Re has a larger effective bandwidth in Fe
based systems as opposed to Cr based systems, it is easy to understand why Sr$_2$FeReO$_6$
is the only metal among the four compounds considered. 

\begin{figure}
\begin{center}
\includegraphics[width=0.49\textwidth, angle=0]{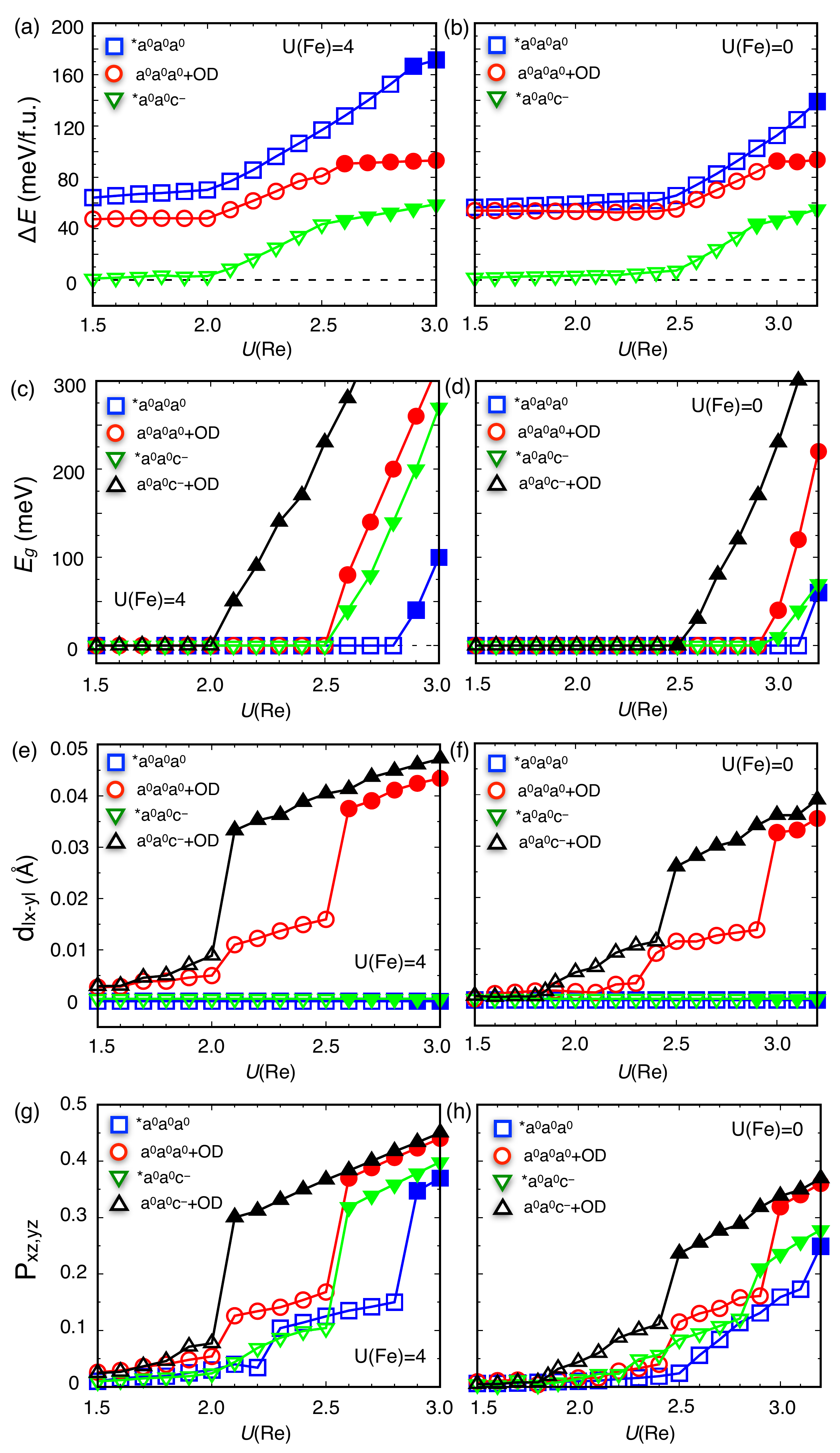}
\caption{(a),(b)   Relative energy of Sr$_2$FeReO$_6$ in several structures  with
  respect to the ground state (i.e. $a^{0}a^{0}c^{-}$+OD).  (c),(d)  Electronic band gaps
  of different phases of Sr$_2$FeReO$_6$.  (e),(f) Octahedral distortion (OD) amplitude
  $d_{|x-y|} $ of the ReO$_6$ octahedron.  (g),(h) Orbital polarization
  $P\left(d_{xz},d_{yz}\right)$ (see eq. \ref{eq:orbpol}) for Re.  Panels (a), (c), (e),
  and (g) correspond to $U_{\textnormal{Fe}}$=4, while panels  (b), (d), (f), and (h)
  correspond to  $U_{\textnormal{Fe}}$=0.  Filled and empty points stand for the
  insulating and metallic phases, respectively.  }
\label{SrFeRe-JT}
\end{center}
\end{figure}

Here we explore the interplay of octahedral tilts, octahedral distortions, and the Hubbard
$U$ in detail (see Fig. \ref{SrFeRe-JT}); including at least six different crystal
structures (i.e. all structures in Fig. \ref{symmetry_schematic} except $Fm\bar 3 m$ and
$P2_1/n$).  We will use the acronym OD (i.e. octahedral distortion) to generically refer to
any spatial ordering of $E_g^{(0)}$ octahedral distortions ($E_g^{(0)}$ is shown
schematically in Fig. \ref{COD_and_OO}, panels c, d and mathematically defined in Ref.
\cite{Marianetti2001224304}), such as C-OD for C-type ordering, F-OD for ferro ordering,
etc; and the same nomenclature will be used for the orbital ordering (i.e. OO generically
refers to C-OO, F-OO, etc.).  In the higher symmetry structures which lack an OD (i.e.
$I4/m$ and  $I4/mmm$), the Hubbard $U$ may cause the  electrons to spontaneously break
space group symmetry despite the fact that we will prevent the nuclei from breaking
symmetry; allowing us to disentangle different effects.  This is achieved by using a
reference crystal structure obtained from relaxing with $U_{\textnormal{Re}}$=0 and then
retaining this structure for $U_{\textnormal{Re}}>$0 (this process is repeated for
different values of $U_{\textnormal{Fe}}/U_{\textnormal{Cr}}$). Anytime a reference
structure is employed, it will be indicated using an asterisk.  Given that the OO/OD is a
spontaneously broken symmetry for $I4/m$, we could have created a reference structure
simply by enforcing space group symmetry, but this is not possible in the Ca-based systems
where the OO/OD is not a spontaneously broken symmetry; and we prefer to have a uniform
approach throughout.  

We note that in all cases we retain the small degree of
tetragonality in the lattice parameters, so there is technically always a very small
tetragonal distortion ($\sqrt{2}a$=7.865 and $c$=7.901 \cite{Re-Kato}).  Fully relaxing
the lattice parameters had a very small effect on the results in the test cases we
evaluated (see Table \ref{SrFeReO}).  In all panels, solid points indicate an insulator,
while hollow points indicate a metal.  

It should be noted that the structures with an OO (e.g. C-OO, F-OO, etc.) are merged into
the same line for brevity, despite the fact that they have different space groups (see
Fig. \ref{symmetry_schematic}). The C-OO can easily be distinguished as it is always
insulating in this compound (it is only favorable at larger values of
$U_{\textnormal{Re}}$), and the F-OO is always metallic (it is only favorable at smaller
values of $U_{\textnormal{Re}}$). The same statements clearly follow for C-OD and F-OD,
given that the orbital ordering is what causes the structural distortion.  Interestingly,
we will show that there is a different state which can occur at intermediate values of
$U_{\textnormal{Re}}$ in the region between the F-OO/F-OD and the C-OO/C-OD, and this is a
ferrimagnetic orbital ordering (FI-OO) and corresponding octahedral distortion (FI-OD);
though the smaller magnitude OO/OD within the FI-OO/FI-OD is always nearly zero. These
three regimes, F-OO/F-OD, FI-OO/FI-OD, and C-OO/C-OD,  are easy to identify due to kinks
in the curves, as we shall point out. The FI-OD will prove not to be important given that
it tends to lose a competition with octahedral tilting.  For each structure, we present
the relative energy  $\Delta E$ (i.e. the energy of a reference structure with respect to
the ground state, panels a and b), the band gap (panel b,c), the amplitude of the OD
(denoted $d_{|x-y|}$, see panel d,e), and the magnitude of the OO defined as the orbital
polarization (panels f and g): 
\begin{equation}
  P_{xz,yz}
=\frac{1}{N_{\tau}}\sum_{\tau}{\frac{|n_{d_{yz}}^\tau-n_{d_{xz}}^\tau|}{n_{d_{yz}}^\tau+n_{d_{xz}}^\tau}}
\label{eq:orbpol}
\end{equation}
where $n_{d}^\tau$ is the occupancy of a given
minority spin $d$ orbital,
$\tau$ labels a Re site in the unit cell, and $N_{\tau}$ is the number of Re atoms in the unit cell.

We first focus on the left column of panels in Fig. \ref{SrFeRe-JT} (i.e. a, c, e, and g),
where $U_{\textnormal{Fe}}$=4, though nearly qualitative behavior is independent of
$U_{\textnormal{Fe}}$; the few small differences will be noted as they arise.  Focussing
on the blue curves corresponding to the  $*a^{0}a^{0}a^{0}$ structure (where the nuclei
are constrained to space group $I4/mmm$), we see that  $d_{|x-y|}$ is zero, as it must be
when the nuclei are confined to this space group. Despite this fact, $P_{xz,yz}$  reveals
a small symmetry breaking of the electronic density for $U_{\textnormal{Re}}\le$2.2 (see
panel g) where F-OO is found; and this sharply transitions to a new plateau for  2.3$\le
U_{\textnormal{Re}}\le$2.7 where FI-OO is found; and finally there is a sharp transition
to the C-OO insulating state for $U_{\textnormal{Re}}\ge$2.8. Therefore, the MIT occurs at
approximately  $U_{\textnormal{Re}}$=2.8 in this scenario.  Inspecting the relative
energy, $\Delta E$ is roughly constant up until approximately $U_{\textnormal{Re}}$=2,
whereafter $\Delta E$  increases linearly due to the fact that the ground state structure
has formed the C-OO/C-OD.

Allowing the C-OD to condense, but still in the absence of tilts, will shift the orbital
ordering to lower values of $U_{\textnormal{Re}}$; and this is illustrated in the red
curves labeled $a^{0}a^{0}a^{0}$+OD (space group $Fmmm$ and $P4_2/mnm$ for the F-OD and
C-OD, respectively). A jump in the value of the OD amplitude $d_{|x-y|}$ can be seen
occurring concomitantly with the orbital polarization. Clearly, the C-OD cooperates with
the C-OO, allowing the latter to form at smaller values of $U_{\textnormal{Re}}$ and
saturate at larger values.  Here $\Delta E$ has two clear kinks in the slope, given that
the curve begins as roughly constant, then changes to linear when the ground state forms
the C-OO/C-OD, and then becomes constant once again when the C-OO/C-OD forms in this
$a^{0}a^{0}a^{0}$+OD structure.

We can now explore the results where we allow $a^{0}a^{0}c^{-}$ tilts, but not the OD (ie.
nuclei are frozen in $I4/m$ space group, see green curves labeled $*a^{0}a^{0}c^{-}$). The
tilts also reduce the threshold $U_{\textnormal{Re}}$ needed to drive the C-OO insulating
state as compared to the $*a^{0}a^{0}a^{0}$ reference structure. Serendipitously, this
reduction is roughly the same as the $a^{0}a^{0}a^{0}$+OD reference structure; though we
see that when comparing the energetics of these two cases, $*a^{0}a^{0}c^{-}$ is favorable
up to the largest $U_{\textnormal{Re}}$  considered (see panel a).  It should be noted
that the ferri FI-OO state is not realized in this case (ie. F-OO transitions directly to
C-OO).

Finally, we can allow both $a^{0}a^{0}c^{-}$ tilts and the OD (ie. space group $P4_2/m$,
see black curves labeled $a^{0}a^{0}c^{-}$+OD), which cooperate to strongly reduce the
threshold for the C-OO insulating state to $U_{\textnormal{Re}}$=2.1.  Interestingly, this
appears to occur because the tilts have a preference for converting the FI-OD to the C-OD
(see panels e and g), which appears reasonable given that the tilt pattern of the Re
alternates in the z-direction with the same phase as the C-OD. All of the same generic
trends can be observed in the right column where $U_{\textnormal{Fe}}$=0, though all
transitions are shifted to higher values of $U_{\textnormal{Re}}$ as is expected for a
larger effective Re bandwidth.

Given that Sr$_{2}$FeReO$_{6}$ is metallic in experiment, and that we expect
$U_{\textnormal{Fe}}$=4 to be a reasonable value, we would infer that
$U_{\textnormal{Re}}\le$2.0 in order to be consistent with experiment (see Section
\ref{U_of_Re} for a more detailed discussion).  In the region $U_{\textnormal{Re}}\le$2.0,
the energy differences are nearly constant, and it is worth noting that the predicted
energy gain for octahedral tilting is reasonable given the experimentally observed
transition from $I4/m\rightarrow Fm\bar 3 m$ at $T=490K$.

It is also useful to determine the effect of $U$ on the magnetic moment of the transition
metal sites in addition to the number of electrons ($N_d$) in the correlated manifold (see
Table \ref{magtable}).  The Fe and Re moments are 3.65 (4.09) and 0.82 (1.34) $\mu_B$,
respectively, within GGA (GGA+$U$).  The number of $d$ electrons decreases by roughly 0.15
electrons for Fe as $U$ is turned on, reflecting the unmixing the Fe; while the changes in
Re are more modest.

\begin{table}
\begin{ruledtabular}
\begin{center}
\caption{Magnetic moment $M$ ($\mu_B$) and $N_{d}$  of Cr/Fe and Re atoms in Re-DPs within GGA and GGA+$U$,
  where $U_{\textnormal{Cr}}$=2.5, $U_{\textnormal{Fe}}$=4, and $U_{\textnormal{Re}}$=2. }
\label{magtable}
\renewcommand{\arraystretch}{1.3}
\begin{tabular}{c c c c c}
& \multicolumn{2}{c}{$M$(Cr/Fe)}	& \multicolumn{2}{c}{$M$(Re)} \\
\cline{2-3} \cline{4-5}
& GGA & GGA+$U$ & GGA & GGA+$U$ \\
 \hline
Sr$_{2}$FeReO$_{6}$ 	& 3.647	& 4.086	& 0.815	& 1.340 \\	
Sr$_{2}$CrReO$_{6}$	& 2.239	& 2.647 	& 1.068	& 1.532 \\
Ca$_{2}$FeReO$_{6}$	& 3.593	& 4.094	& 0.804	& 1.411 \\
Ca$_{2}$CrReO$_{6}$ 	& 2.329	& 2.689	& 1.135	& 1.581 \\
& \multicolumn{2}{c}{$N_{d}$(Cr/Fe)}	& \multicolumn{2}{c}{$N_{d}$(Re)} \\
\cline{2-3} \cline{4-5}
& GGA & GGA+$U$ & GGA & GGA+$U$ \\
 \hline
Sr$_{2}$FeReO$_{6}$ 	& 5.896	& 5.743	& 4.178	& 4.140 \\	
Sr$_{2}$CrReO$_{6}$	& 4.131	& 4.075 	& 4.251	& 4.210 \\
Ca$_{2}$FeReO$_{6}$	& 5.884	& 5.717	& 4.171	& 4.152 \\
Ca$_{2}$CrReO$_{6}$ 	& 4.107	& 4.031	& 4.229	& 4.194 \\
\end{tabular}
\end{center}
\end{ruledtabular}
\end{table}

\subsubsection{Sr$_{2}$CrReO$_{6}$}
\label{subsubsec_srcr}

As we discussed in the literature review (see Section \ref{lit_review}), several
experiments suggested that Sr$_2$CrReO$_6$ is metallic with space group $I4/m$ (demanding
that $d_{|x-y|} $=0)\cite{Re-Kato,Teresa-SrCrReO,Majewski,Michalik}.  However, recently
Hauser \emph{et al.} proposed that a fully-ordered Sr$_2$CrReO$_6$ film on the STO
substrate is in fact a semiconductor with $E_\textnormal{gap}$=0.21 eV
\cite{Hauser-SrCrRe}, and further suggested that the previously reported metallicity of
Sr$_2$CrReO$_6$ may be due to oxygen vacancies \cite{Hauser-Vo}.  Our calculations lend
support to the observations of Hauser \emph{et al.}, showing that the C-OO/C-OD can induce
an insulating state for reasonable values of $U_{\textnormal{Re}}$.

\begin{figure}
\begin{center}
\includegraphics[width=0.5\textwidth, angle=0]{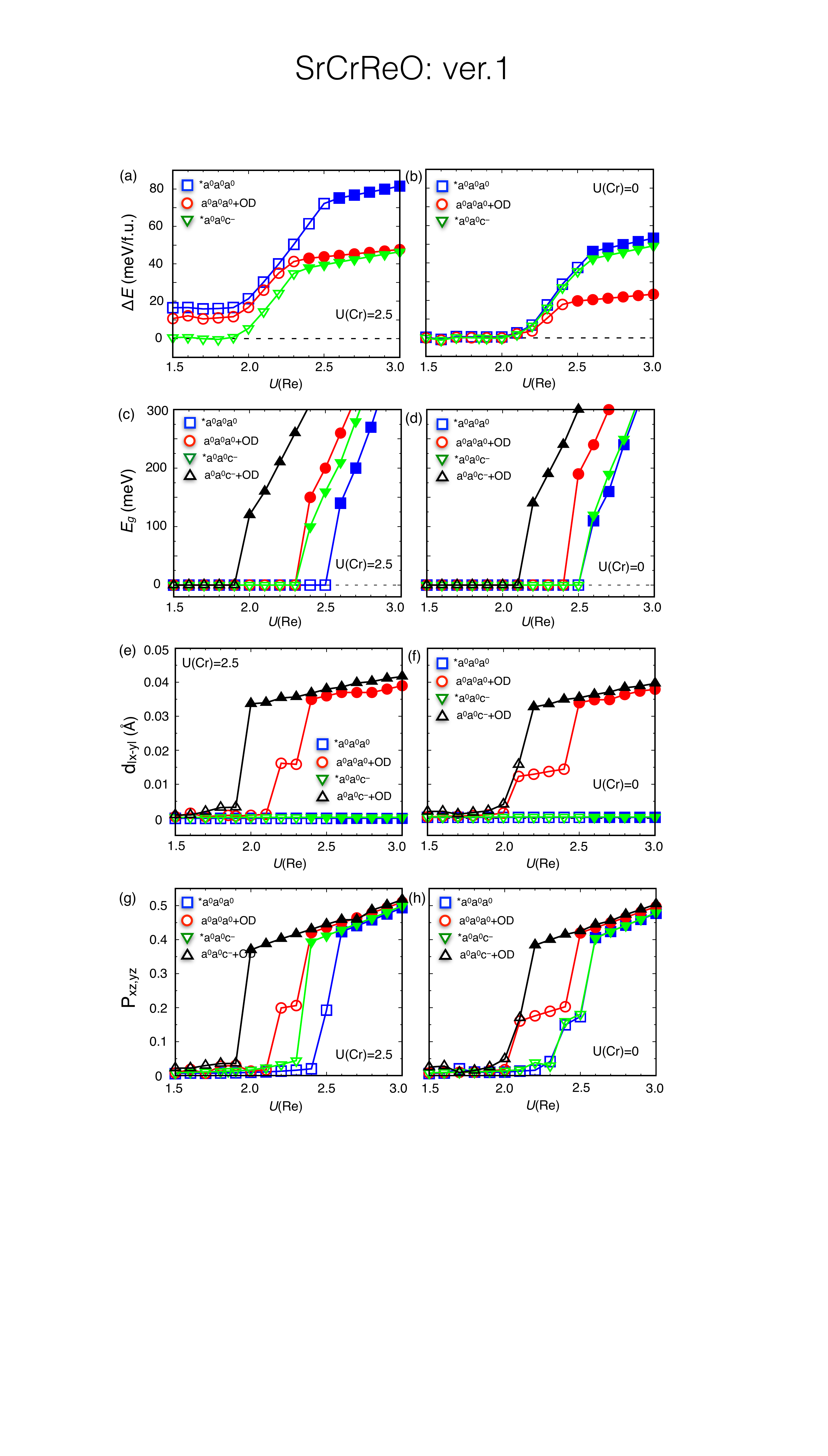}
\caption{(a),(b)   Relative energy of Sr$_2$CrReO$_6$ in several structures  with respect
 to the ground state (ie. $a^{0}a^{0}c^{-}$+OD).  (c),(d)  Electronic band gaps of
 different phases of Sr$_2$CrReO$_6$.  (e),(f) Octahedral distortion (OD) amplitude
 $d_{|x-y|} $ of the ReO$_6$ octahedron.  (g),(h) Orbital polarization
 $P\left(d_{xz},d_{yz}\right)$ (see eq. \ref{eq:orbpol}) for Re.  Panels (a), (c), (e),
 and (g) correspond to $U_{\textnormal{Cr}}$=2.5, while panels  (b), (d), (f), and (h)
 correspond to  $U_{\textnormal{Cr}}$=0.  Filled and empty points stand for the insulating
 and metallic phases, respectively.
}
\label{SrCrRe-JT}
\end{center}
\end{figure}

Here we perform the same analysis for Sr$_{2}$CrReO$_{6}$ as in the previous section for
Sr$_{2}$FeReO$_{6}$, demonstrating the same generic behavior; but different quantitative
thresholds (see Fig. \ref{SrCrRe-JT}).  The main notable difference observed in
Sr$_{2}$CrReO$_{6}$ as compared to Sr$_{2}$FeReO$_{6}$ is the energy scale for octahedral
tilting, where $U_{\textnormal{Cr}}$ plays a role in stabilizing the tilt pattern. For
example, when $U_{\textnormal{Cr}}$=0 octahedral tilting is either unstable or stabilized
by less than 1meV, depending on whether or not one relaxes lattice parameters in addition
to internal coordinates (see Table \ref{SrCrReO}). However, applying a non-zero
$U_{\textnormal{Cr}}$ results in a small stabilization energy for $a^{0}a^{0}c^{-}$
tilting, and this effect only depends weakly on $U_{\textnormal{Re}}$ prior to the onset
of the C-OO (i.e. $U_{\textnormal{Re}}< 2$; see Fig. \ref{SrCrRe-JT}, panels a and b).
Clearly, a non-zero $U_{\textnormal{Cr}}$ is essential to obtaining an energy scale for
octahedral tilting which is consistent with a tilt transition of $T=260$K (see Table
\ref{str_temp}).  Otherwise, all of the same trends from Sr$_{2}$FeReO$_{6}$ can be seen
in Sr$_{2}$CrReO$_{6}$. If we then take a value of $U_{\textnormal{Cr}}$=2.5, an
insulating state can only be achieved if $U_{\textnormal{Re}}\gtrapprox $2  for the ground
state structure $P4_2/m$ (i.e. $a^{0}a^{0}c^{-}$+C-OD).

Given our preferred values of $U$ (i.e. $U_{\textnormal{Cr}}$=2.5 and
$U_{\textnormal{Re}}$=2), Sr$_{2}$CrReO$_{6}$ is insulating as in experiment. However,
given these values of $U$, the C-OD is a necessary condition for realizing the insulating
state (i.e. compare the black and green curves in Fig. \ref{SrCrRe-JT}, panel c), and the
C-OD is only energetically favorable by 5.5 meV (i.e. green curve in panel a). Therefore,
if thermal fluctuations of the phonons were to disorder the C-OD, the system would be
driven through the MIT. The system could remain insulating in the absence of the C-OD if
$U_{\textnormal{Re}}\gtrapprox 2.4$, but then Sr$_{2}$FeReO$_{6}$ would be insulating with
a C-OD stabilized by 34.4 meV for $U_{\textnormal{Re}} = 2.4$; inconsistent with
experiment.  Therefore, the C-OD should condense at sufficiently low temperatures in
experiment and the space group should be measured to be $P4_2/m$ instead of $I4/m$ given
sufficiently clean samples. Later we demonstrate that SOC introduces quantitative changes,
but the same general conclusion holds. Future experiments can test this prediction. 

Given that the experimental insulating state was realized via growth on STO, it is
worthwhile to determine the influence of imposing the STO lattice parameter (a=3.905);
which is $\sim$0.04\% of compressive strain compared to the optimized lattice parameter
within GGA+$U$ (with $U_{\textnormal{Cr}}$=2.5 and $U_{\textnormal{Re}}$=2).  We find that
this strain has only a small effect on energy differences, resulting in a difference of
$-$17.8 meV for $P4_2/m-P4_2/mnm$, as compared to $-$18.5 meV for the bulk case in Table
~\ref{SrCrReO}; and therefore we do not believe the substrate has any substantial effect.

The magnetic moments and number of electrons as a function of  $U$ are summarized in Table
\ref{magtable}.  The Cr and Re moments are 2.24 (2.65) and 1.07 (1.53) $\mu_B$,
respectively, within GGA (GGA+U).  However, note that the total moment is constant
(1$\mu_B$/f.u.) within both GGA and GGA+$U$.  The number of $d$ electrons decreases by
$\sim$0.06 for Cr as $U$ is turned on, which is almost half of the change of $N_d$ of Fe
in Sr$_2$FeReO$_6$.  Smaller change of $N_d$(Cr) also reflects the weaker Cr-Re
hybridization.

\begin{table}
\begin{ruledtabular}
\begin{center}
\caption{  Relative energy difference (in meV per formula unit) of the $a^{0}a^{0}c^{-}$
structure minus $a^{0}a^{0}a^{0}$ for Sr$_2$CrReO$_6$ within both the GGA and GGA+$U$
calculations, and their crystal symmetries.  We consider full structural relaxations, in
addition to only relaxing internal coordinates.
}
\label{SrCrReO}
\renewcommand{\arraystretch}{1.3}
\begin{tabular}{c c c c c}
  \multicolumn{1}{c}{$U$(eV)} & symmetry & inter. rel.	& full rel.  \\
\hline
$U_{\textnormal{Cr}}$=0,  $U_{\textnormal{Re}}$=0		  & $I4/m-I4/mmm$		& 0.07		& $-$0.06 \\
$U_{\textnormal{Cr}}$=2.5, $U_{\textnormal{Re}}$=0		& $I4/m-I4/mmm$		& $-$4.90		& $-$7.96 \\
$U_{\textnormal{Cr}}$=3.5, $U_{\textnormal{Re}}$=0		& $I4/m-I4/mmm$		& $-$11.90	& $-$13.89 \\
$U_{\textnormal{Cr}}$=0, $U_{\textnormal{Re}}$=2		  & $P4_2/m-P4_2/mnm$	& $-$0.33		& $-$0.48 \\
$U_{\textnormal{Cr}}$=2.5, $U_{\textnormal{Re}}$=2		& $P4_2/m-P4_2/mnm$	& $-$18.47	& $-$20.45 \\
$U_{\textnormal{Cr}}$=3.5, $U_{\textnormal{Re}}$=2		& $P4_2/m-P4_2/mnm$	& $-$29.67	& $-$28.98 \\
\end{tabular}
\end{center}
\end{ruledtabular}
\end{table}

\subsubsection{Ca$_{2}$FeReO$_{6}$}
\label{subsubsec_cafe}

\begin{table}
\begin{ruledtabular}
\begin{center}
\caption{ Nonzero octahedral modes of ReO$_6$ in Ca$_{2}$FeReO$_{6}$, for one of the two
symmetry equivalent Re atoms in the unit cell.  The mathematical definition of each mode
can be found in Figure 4 of Ref. \cite{Marianetti2001224304}.  The local coordinate system
is chosen by having zero rotation modes (i.e., amplitude of $T_{1g}$ modes are zero).  The
amplitudes for the other Re-O octahedron in the corresponding local coordinate system can
be obtained by inverting the sign of $E_{g}^{(0)}$, and swapping the values of
$T_{2g}^{(1)}$ and $T_{2g}^{(2)}$.  It should be noted that the low and high temperature
experimental structures are different C-OD variants.
}
\label{CaFeReO-Re_mode}
\renewcommand{\arraystretch}{1.3}
\begin{tabular}{c c c c c c c}
		& \multicolumn{2}{c}{exp \cite{Oikawa}} 	& \multicolumn{3}{c}{GGA}  \\
\cline{2-3} \cline{4-6}
modes	& 7K 		& 300K				& GGA	& +$U$	& +$U$+SOC \\
\hline
$A_{1g}$		& 4.7834			& 4.7780		& 4.7655		& 4.8077		& 4.8146        \\
$E_{g}^{(0)}$		& $-$0.0137		& $-$0.0046	& $-$0.0017		& $-$0.0335	& $-$0.0260        \\
$E_{g}^{(1)}$		& $-$0.0126		& $-$0.0196	& $-$0.0066	& $-$0.0171	& $-$0.0012        \\
$T_{2g}^{(0)}$	& 0.0005			& 0.0179		& 0.0109		& 0.0022		& 0.0078     \\
$T_{2g}^{(1)}$	& $-$0.0423		& 0.0192		& $-$0.0293	& $-$0.0678	& $-$0.0582        \\
$T_{2g}^{(2)}$	& 0.0225			& $-$0.0255	& 0.0205		& 0.0256		& 0.0271     \\
\end{tabular}
\end{center}
\end{ruledtabular}
\end{table}

\begin{figure}
\includegraphics[width=0.45\textwidth, angle=0]{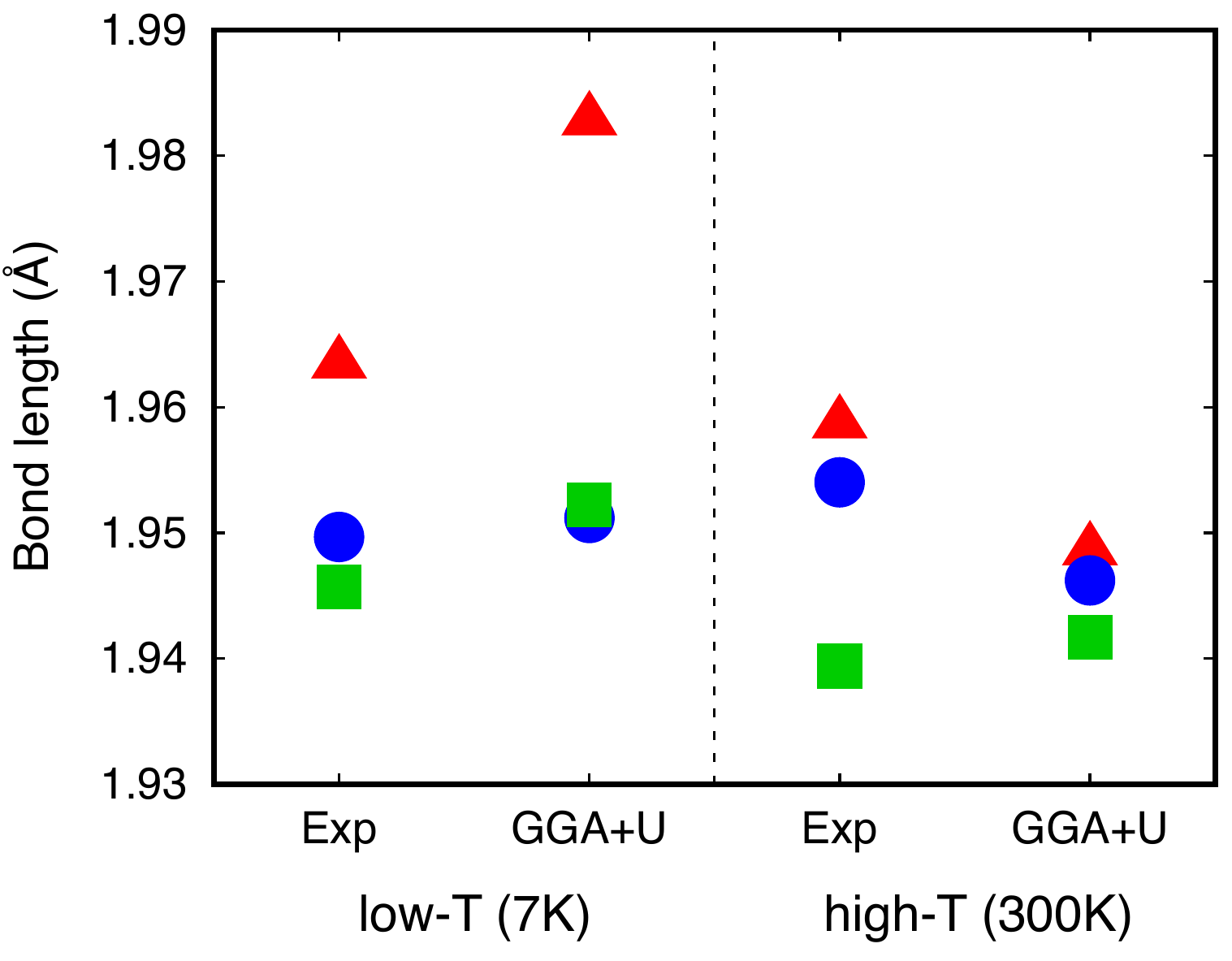}
\caption{   Re1--O bond lengths of Ca$_2$FeReO$_6$ for the experimental structures at
high- and low-temperature\cite{Oikawa}, and from the relaxed structures within GGA and
GGA+$U$.  Blue circle, red triangle, and green box stand for Re1--O1, Re1--O2, and
Re1--O3, respectively, where Re1--O3 is the out-of-plane Re--O bond. The Re2--O bond
lengths are given by Re2--O1=Re1--O2, Re2--O2=Re1--O1, and Re2--O3=Re1--O3.  Given our
unit cell conventions (see Fig. \ref{COD_orientation}), the experimental paper plots
Re1--O in the low temperature phase and Re2--O in the high temperature phase.
}
\label{CaFeReO-bonds}
\end{figure}

We now move on to the case of Ca$_{2}$FeReO$_{6}$, which has the lower symmetry space
group $P2_{1}/n$ ($a^-a^-b^+$ tilt, see Fig. \ref{FeRe-str}) and is measured to be an
insulator with a 50meV energy gap at low temperature (see Section \ref{lit_review} for a
detailed review).  Given the smaller size of Ca relative to Sr, the tilts in both Ca-based
materials are large in magnitude (see Table \ref{CaFeReO-Re_mode} for octahedral mode
amplitudes) and retained up to the highest temperatures that have been studied in
experiment (i.e. 300K and 550K for Ca$_{2}$CrReO$_{6}$ and Ca$_{2}$FeReO$_{6}$,
respectively).  For example, two in-plane and one out-of-plane $\angle$Fe-O-Re are 151.2,
151.8, and 152.4$^\circ$ at 7K, and both DFT and  DFT+$U$ accurately capture the large
magnitude of the octahedral tilts: $\angle$Fe-O-Re are 149.9, 151.1, 150.5 using DFT;
149.7, 150.0, and 150.8$^\circ$ using DFT+$U$ ($U_{\textnormal{Fe}}$=4 and
$U_{\textnormal{Re}}$=2).  Furthermore, these large tilts substantially reduce the
effective Re bandwidth, resulting in a smaller critical value of $U_{\textnormal{Re}}$
needed to drive the C-OO induced insulating state, as we will now illustrate.

In Sec. \ref{lit_review}, we briefly discussed the structures of Ca$_{2}$FeReO$_{6}$
obtained at low and high temperatures \cite{Oikawa}, as summarized in Fig.
\ref{CaFeReO-bonds}.  According to experiment, there is an appreciable  C-OD amplitude
below the phase transition (e.g. C-OD$^+$, $d_{|x-y|}=0.014\AA$ at $T$=7K), and it is
highly suppressed and swapped to the alternate variant above the transition temperature of
$T=140K$  (e.g. C-OD$^-$, $d_{|x-y|}=0.005\AA$ at $T$=300K).  It should be emphasized that
the C-OD is not a spontaneously broken symmetry in this structure, in contrast to the Sr
case (see symmetry lineage in Fig. \ref{symmetry_schematic}).

\begin{figure}
\includegraphics[width=0.5\textwidth, angle=0]{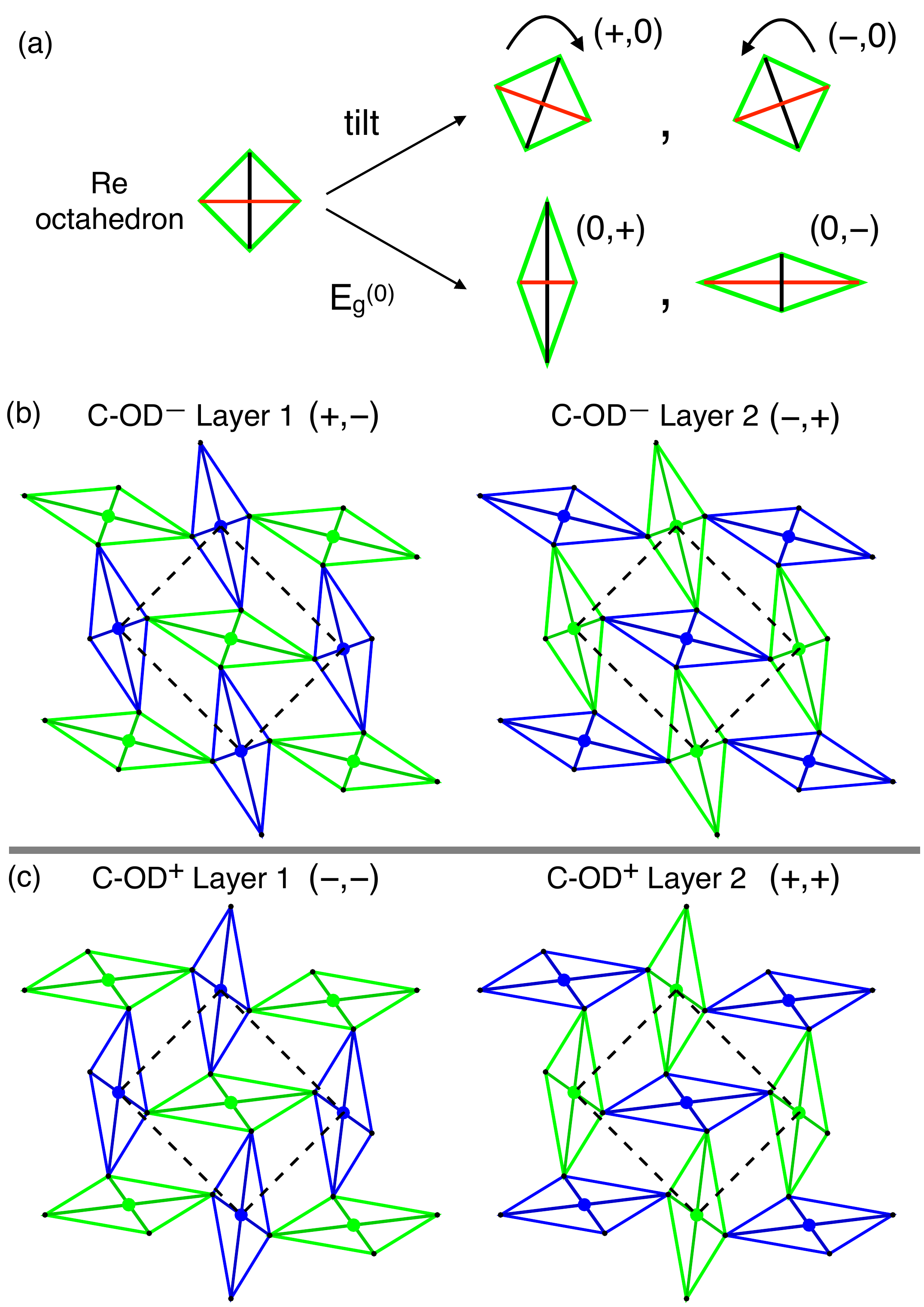}
\caption{  (a) Signs of the tilt mode (clockwise being positive) and $E_g^{(0)}$ mode ($y$
  elongation being positive) of Re-O octahedron.  Black and red lines correspond to Re-O
  bonds. Quantities in parenthesis, give the sign of the tilt and $E_g^{(0)}$ amplitude,
  respectively.  (b), (c) Two dimensional schematic of the  two possible orientations of
  C-OD: C-OD$^-$ and C-OD$^+$.  Black dots represent oxygen, green dots represent Re, and
  blue dots represent $B$ atom (Fe or Cr).  The dashed rectangle is the unit cell, where
  $a<b$.  The coordinate system is the same as depicted in Fig. \ref{COD_and_OO}.  Each
  schematic is defined by three numbers: the tetragonality of the unit cell, the
  octahedral tilt amplitude, and the amplitude of the $E_g^{(0)}$ mode.  Panels (a) and
  (b) (i.e. C-OD$^-$ and C-OD$^+$, respectively) only differ in the sign of the octahedral
  tilt amplitude.  The amplitudes of the distortions are exaggerated to showcase the
  difference between C-OD$^+$ and C-OD$^-$.  }
\label{COD_orientation}
\end{figure}

We now elaborate on the fact that there are two types of C-OD within the monoclinic
$P2_1/n$ structure (see schematic in Fig. \ref{COD_orientation}).  We will use a notation
of  C-OD$^+$ to denote the ordering where a given  Re-O octahedron has the same sign for
the $E_g^{(0)}$ mode (defined in the unrotated coordinate system) and the rotation mode
(i.e. both modes positive or both modes negative); whereas C-OD$^-$ indicates opposite
signs.  Structures below and above the MIT exhibit  C-OD$^+$ and C-OD$^-$, respectively.
Note that the C-OD$^+$ and C-OD$^-$ are distinguishable only in the monoclinic (i.e.,
$a\neq b$) double perovskites, whereas they are symmetry equivalent in the tetragonal
double perovskites (e.g. in the Sr-based systems).

\begin{figure}
\includegraphics[width=0.5\textwidth, angle=0]{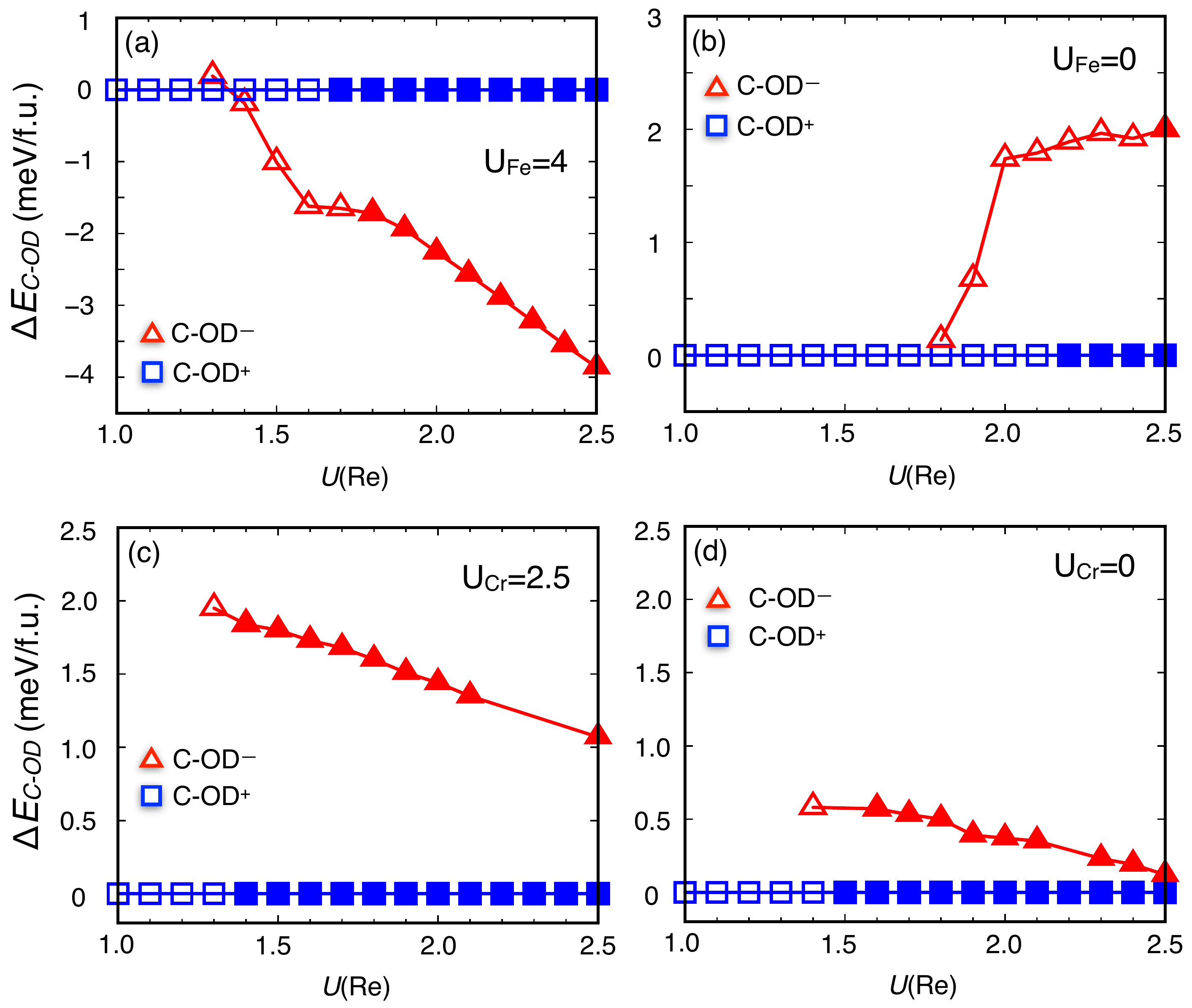}
\caption{ Energies for C-OD$^+$ (blue square) and C-OD$^-$ (red triangle) relative 
  to C-OD$^+$ as a function of $U_{\textnormal{Re}}$.
  (a),(b) Ca$_{2}$FeReO$_{6}$ with $U_{\textnormal{Fe}}$=4 and  $U_{\textnormal{Fe}}$=0, respectively. 
  (c),(d) Ca$_{2}$CrReO$_{6}$ with $U_{\textnormal{Cr}}$=2.5 and  $U_{\textnormal{Cr}}$=0, respectively. 
Empty and filled points stand for metallic and insulating phases, respectively.
}
\label{COD_energy}
\end{figure}

GGA results in C-OD$^+$,  and C-OD$^-$ is not even metastable (i.e. it relaxes back to
C-OD$^+$); though the C-OD$^+$ amplitude is negligibly small (i.e., $d_{|x-y|}<0.001\AA$).
The energies of C-OD$^+$ and C-OD$^-$ become distinct as $U_{\textnormal{Re}}$ increases,
while the relative stability also depends on $U_{\textnormal{Fe}}$.  As depicted in Fig.
\ref{COD_energy}(a), C-OD$^+$ switches to C-OD$^-$ at $U_{\textnormal{Re}}$=1.4 when
$U_{\textnormal{Fe}}$=4, and the energy difference increases as a function of
$U_{\textnormal{Re}}$.  When $U_{\textnormal{Fe}}$=0, as shown in Fig.
\ref{COD_energy}(b),  C-OD$^+$ is always favorable, and its stability increases in the
range $U_{\textnormal{Re}}$=1.8$-$2.5eV.  Since the energy difference between two
different orderings is very small, we simply follow C-OD$^+$ (i.e. low temperature
orientation) whenever applying DFT+$U$.  In terms of the C-OD amplitudes, GGA+$U$ and GGA
agree more closely for the low-$T$ and high-$T$ structures, respectively (see Figure
\ref{CaFeReO-bonds}), though GGA+$U$ overestimates and GGA underestimates $d_{|x-y|} $.

We now perform the same analysis as for the Sr-based systems, except that the untilted
structure does not need to be considered given its large energy scale.  In the Sr-based
systems, we considered high symmetry reference structures, where we allowed the electrons
to spontaneously break symmetry but prevented the structure from doing so by fixing it at
the relaxed $U_{\textnormal{Re}}$=0 structure (though non-zero
$U_{\textnormal{Cr}}$/$U_{\textnormal{Fe}}$ was included in creating a relaxed reference
structure). The same recipe can be followed in the Ca-based cases, despite the fact that
the $U_{\textnormal{Re}}$=0 structure has an identical space group symmetry, and this
reference structure will be referred to as $*a^-a^-b^+$; where the asterisk indicates that
this a reference structure where we have effectively removed the C-OD which is induced by
orbital ordering.  Comparison to the reference structure will give insight into the
importance of the C-OD in realizing the C-OO.  Additionally, we will also study the
unrelaxed experimentally measured structures from $T$=120K and $T$=160K, which straddle
the $T$=140K phase transition.  Due to the strong octahedral tilting, only the C-OO/C-OD
is found in the Ca-based results, as opposed the Sr-based systems where ferro and ferri
OO/OD's are observed.

\begin{figure}
\includegraphics[width=0.485\textwidth, angle=0]{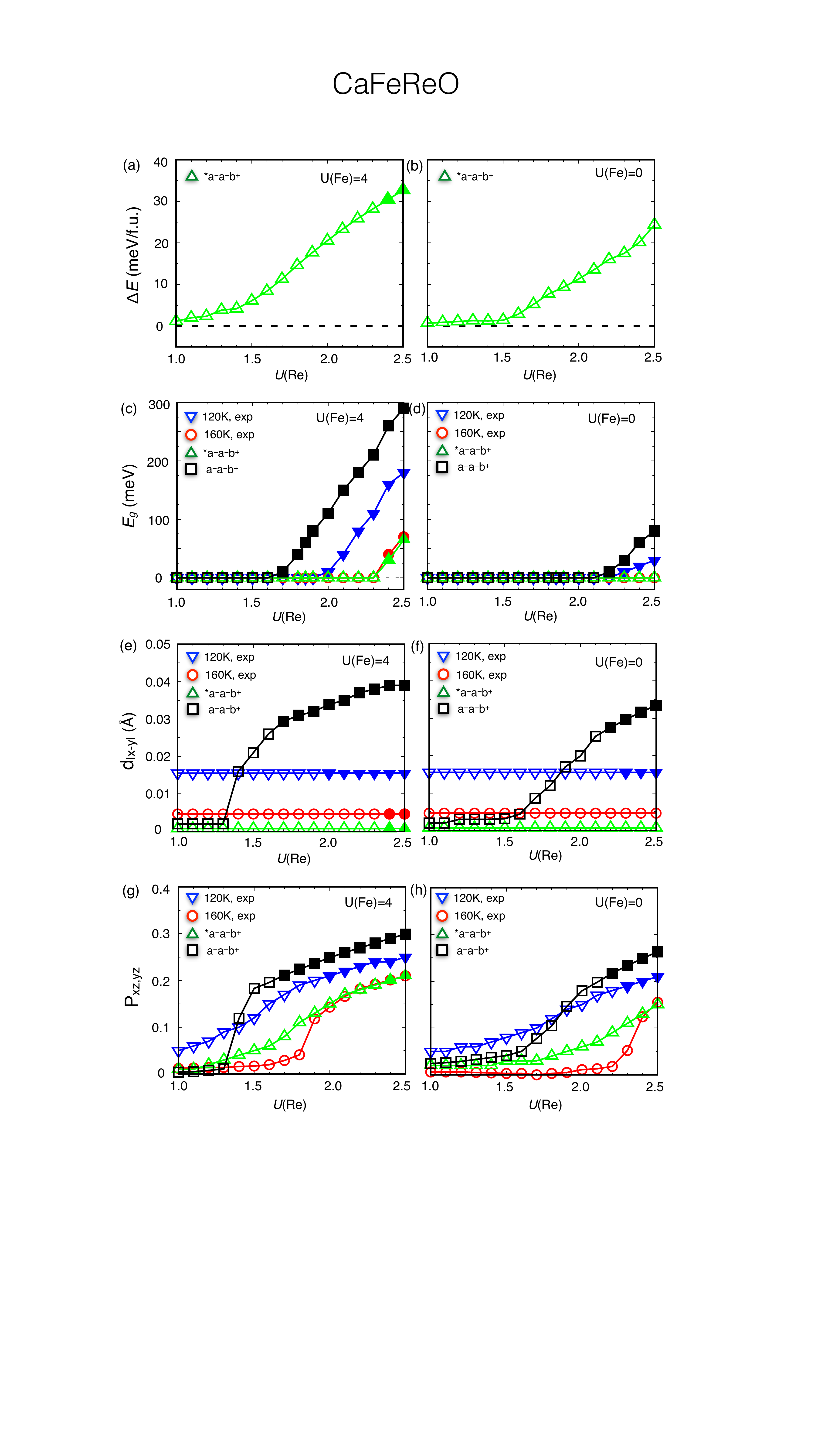}
\caption{(a),(b)   Relative energy of Ca$_2$FeReO$_6$ in the reference structure
$*a^{-}a^{-}b^{+}$ (where the C-OD amplitude is suppressed, see text)  with respect to the
ground state (ie. $a^{-}a^{-}b^{+}$).  (c),(d)  Electronic band gaps of different phases.
(e),(f) Octahedral distortion (OD) amplitude $d_{|x-y|} $ of the ReO$_6$ octahedron.
(g),(h) Orbital polarization $P\left(d_{xz},d_{yz}\right)$ (see eq. \ref{eq:orbpol}) for
Re.  Panels (a), (c), (e), and (g) correspond to $U_{\textnormal{Fe}}$=4, while panels
(b), (d), (f), and (h) correspond to  $U_{\textnormal{Fe}}$=0.  Filled and empty points
stand for the insulating and metallic phases, respectively.  The frozen experimental
structures at 120K and 160K\cite{Oikawa}, which bound the phase transition, are included
for comparison.
}
\label{CaFeRe-JT}
\end{figure}

\begin{figure}
\includegraphics[width=0.45\textwidth, angle=0]{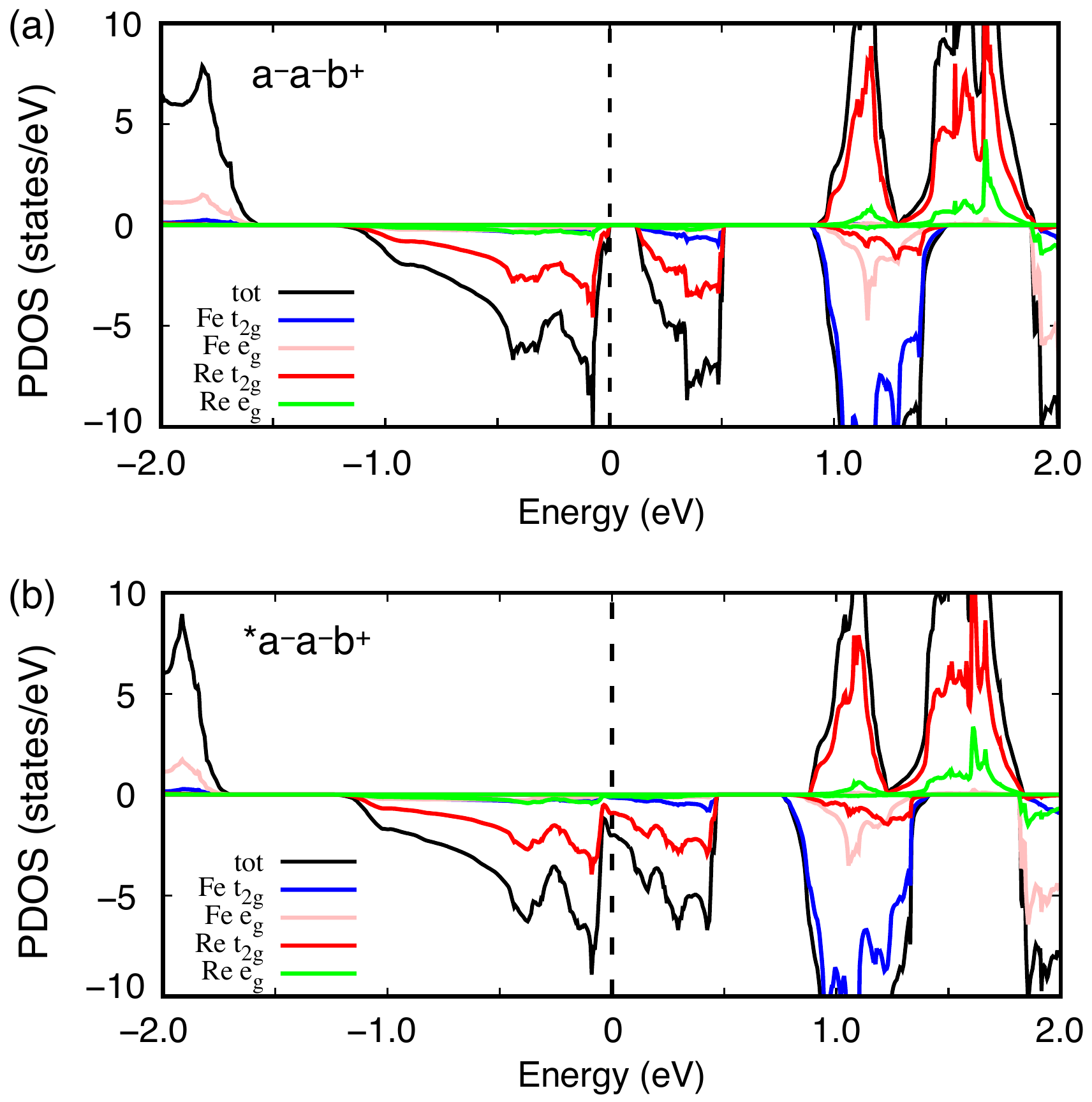}
\caption{   DFT+$U$ projected density of states of Ca$_2$FeReO$_6$  (a) in the ground
state structure ($a^{-}a^{-}b^{+}$) (b) in the reference structure $*a^{-}a^{-}b^{+}$.
The Fermi energy is set to be zero;   $U_{\textnormal{Re}}$=2.0 and
$U_{\textnormal{Fe}}$=4. 
}
\label{CaFeReO-pdos}
\end{figure}

We begin by focussing on the reference structure $*a^-a^-b^+$, depicted by a green curve,
where the C-OD amplitude is negligibly small irrespective of $U_{\textnormal{Fe}}$ (see
Fig. \ref{CaFeRe-JT}, panels e and f).  Increasing $U_{\textnormal{Re}}$ causes the
orbital polarization to increase, and an insulating state (solid point) is eventually
realized at $U_{\textnormal{Re}}$=2.4 for the case of $U_{\textnormal{Fe}}$=4 (see Fig.
\ref{CaFeRe-JT}, panels c, e, and g).  For $U_{\textnormal{Fe}}$=4, the relative energy
$\Delta E$ increases rather slowly for $U_{\textnormal{Re}}\lessapprox$1.4, and the slope
increases thereafter due to the fact that the ground state experiences the C-OO/C-OD at
$U_{\textnormal{Re}}\approx 1.4$.  As in Sr-based systems, turning off the $U$ on the 3$d$
transition metal shifts the metal-insulator phase boundary to larger values of
$U_{\textnormal{Re}}$, and an insulating state is not achieved for
$U_{\textnormal{Re}}\le$2.5 if $U_{\textnormal{Fe}}$=0 (panels d, f, and h).  Hereafter we
focus our discussion on $U_{\textnormal{Fe}}$=4, as all the same qualitative trends hold
upon decreasing $U_{\textnormal{Fe}}$.  The experimental $T$=160K structure (depicted by a
red curve) shows relatively small differences as compared to the $*a^-a^-b^+$ reference
structure, with the band gap being quantitatively similar.

We now move on to the fully relaxed structure, where C-OD amplitude is allowed to relax as
$U_{\textnormal{Re}}$ is increased (depicted as black curve).  In this discussion, we
focus on $U_{\textnormal{Fe}}$=4.  For $U_{\textnormal{Re}}\le$1.3, both the orbital
polarization (i.e. $P_{xz,yz}$) and the C-OD amplitude (i.e. $d_{|x-y|}$) are very small
with a weak $U_{\textnormal{Re}}$ dependence; comparable in magnitude to the reference
structure. Once $U_{\textnormal{Re}}>$1.3, there is a sharp increase in the C-OO/C-OD
amplitude, and the system becomes an insulator for $U_{\textnormal{Re}}\ge$1.7. Therefore,
the cooperation of the C-OO and C-OD greatly reduces the critical $U_{\textnormal{Re}}$
needed to drive the insulating state, from a value of $U_{\textnormal{Re}}$=2.4 in the
reference $*a^-a^-b^+$ structure down to a value of 1.7; which is the same trend as the
case of the Sr-based systems.  It is interesting to compare the relaxed C-OD amplitude to
that of the $T$=120K experimental structure, depicted as a blue curve. In the relaxed
structure, the smallest value of $U_{\textnormal{Re}}$ which has an insulating state is
1.7, and already the C-OD amplitude is nearly twice that of the experimental $T$=120K
structure.  However, later we demonstrate that including SOC dampens the C-OD amplitude
(though not enough to agree with experiment, see Section \ref{sec:SOC}, Fig.
\ref{Ca-soc}).  The $T$=120K and $T$=160K experimental structures produce a critical
$U_{\textnormal{Re}}$ of 2.0 and 2.4 for the C-OO/C-OD, respectively, which is still
appreciably different.

Given the substantial renormalization of the critical $U_{\textnormal{Re}}$ between the
$*a^-a^-b^+$ reference structure and the fully relaxed structure, and analogously between
the two experimental structures,  it is interesting to consider the possibility of the
anharmonic phonon free energy being the primary driving force of the MIT as a function of
temperature. In this scenario, the structural transition is driven by the phonon free
energy, and the resulting change in the structure is sufficient to renormalize the
critical value of $U_{\textnormal{Re}}$ and drive the system through the MIT. 

Using our
prescribed values of $U_{\textnormal{Re}}$=2.0 and  $U_{\textnormal{Fe}}$=4 (given that we
are not yet using spin-orbit coupling), we plot the site/orbital projected electronic
density-of-states for the $*a^-a^-b^+$ reference structure and the relaxed $a^-a^-b^+$
structure (see Fig. \ref{CaFeReO-pdos}). As shown, the result is a metal for the
$*a^-a^-b^+$ structure and an insulator for the $a^-a^-b^+$ structure, with the latter
having a gap of 110meV; slightly larger than relatively small experimental gap of 50meV.
While a greater value of $U_{\textnormal{Re}}$ would yield an insulator in the
$*a^-a^-b^+$ structure, this sort of tuning is discouraged by the fact that
Sr$_{2}$FeReO$_{6}$ would wrongly be driven into a C-OO/C-OD insulating state in
contradiction with experiment (assuming a common value of $U_{\textnormal{Re}}$ is
utilized).  Future work will determine if this phonon driven scenario is dominant, as
opposed to the other extreme where temperature disorders the electrons (see Section
\ref{sec:general_aspects} for further discussion of these scenarios).

\subsubsection{Ca$_{2}$CrReO$_{6}$}

\begin{figure}
\includegraphics[width=0.495\textwidth, angle=0]{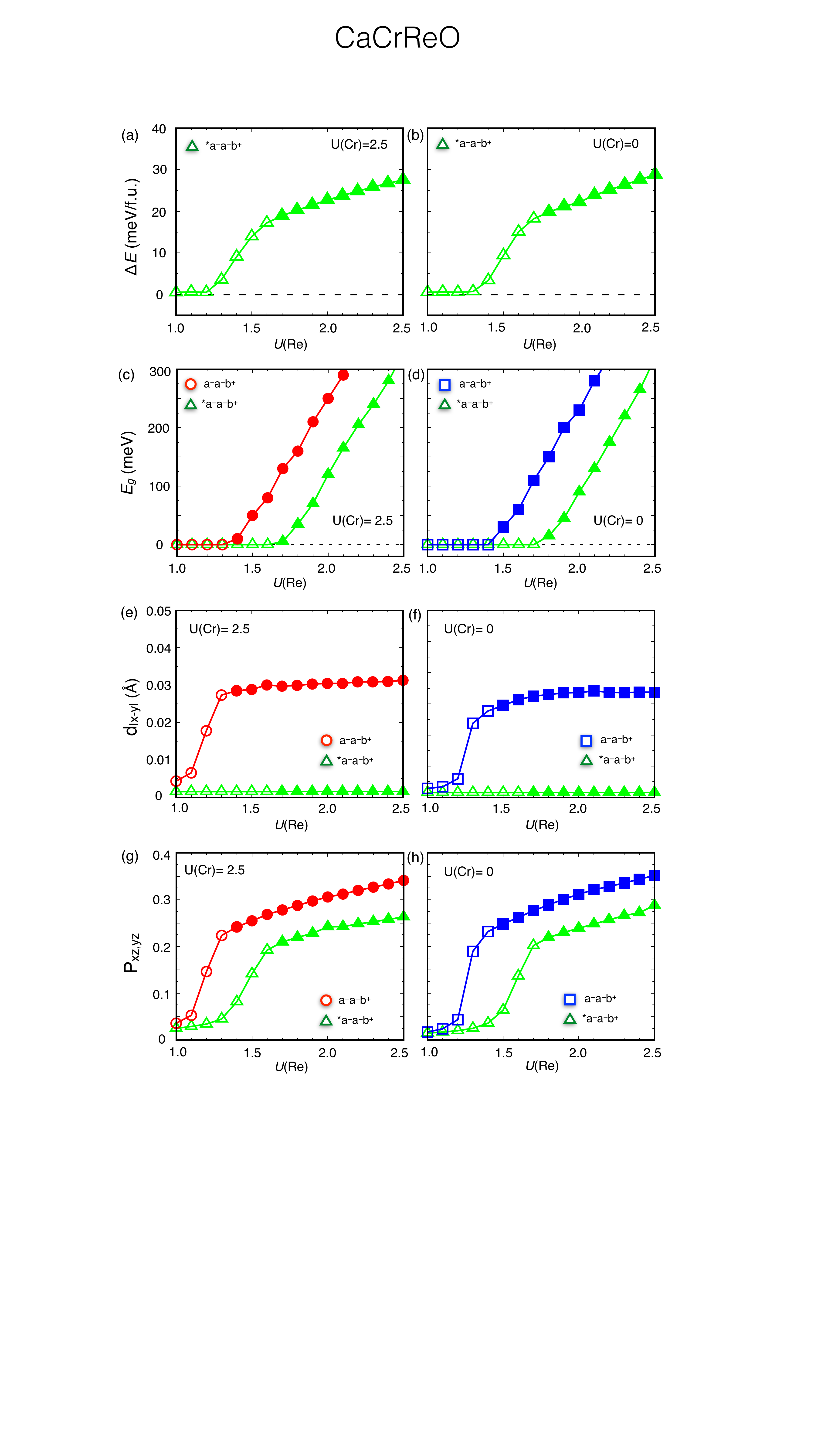}
  \caption{(a),(b)   Relative energy of Ca$_2$CrReO$_6$ in the reference structure
  $*a^{-}a^{-}b^{+}$ (where the C-OD amplitude is suppressed, see text)  with respect to
  the ground state (i.e. $a^{-}a^{-}b^{+}$).  (c),(d)  Electronic band gaps of different
  phases.  (e),(f) Octahedral distortion (OD) amplitude $d_{|x-y|} $ of the ReO$_6$
  octahedron.  (g),(h) Orbital polarization $P\left(d_{xz},d_{yz}\right)$ (see eq.
  \ref{eq:orbpol}) for Re.  Panels (a), (c), (e), and (g) correspond to
  $U_{\textnormal{Cr}}$=2.5, while panels  (b), (d), (f), and (h) correspond to
  $U_{\textnormal{Cr}}$=0.  Filled and empty points stand for the insulating and metallic
  phases, respectively.
}
\label{CaCrRe-JT}
\end{figure}

\begin{figure}
\includegraphics[width=0.45\textwidth, angle=0]{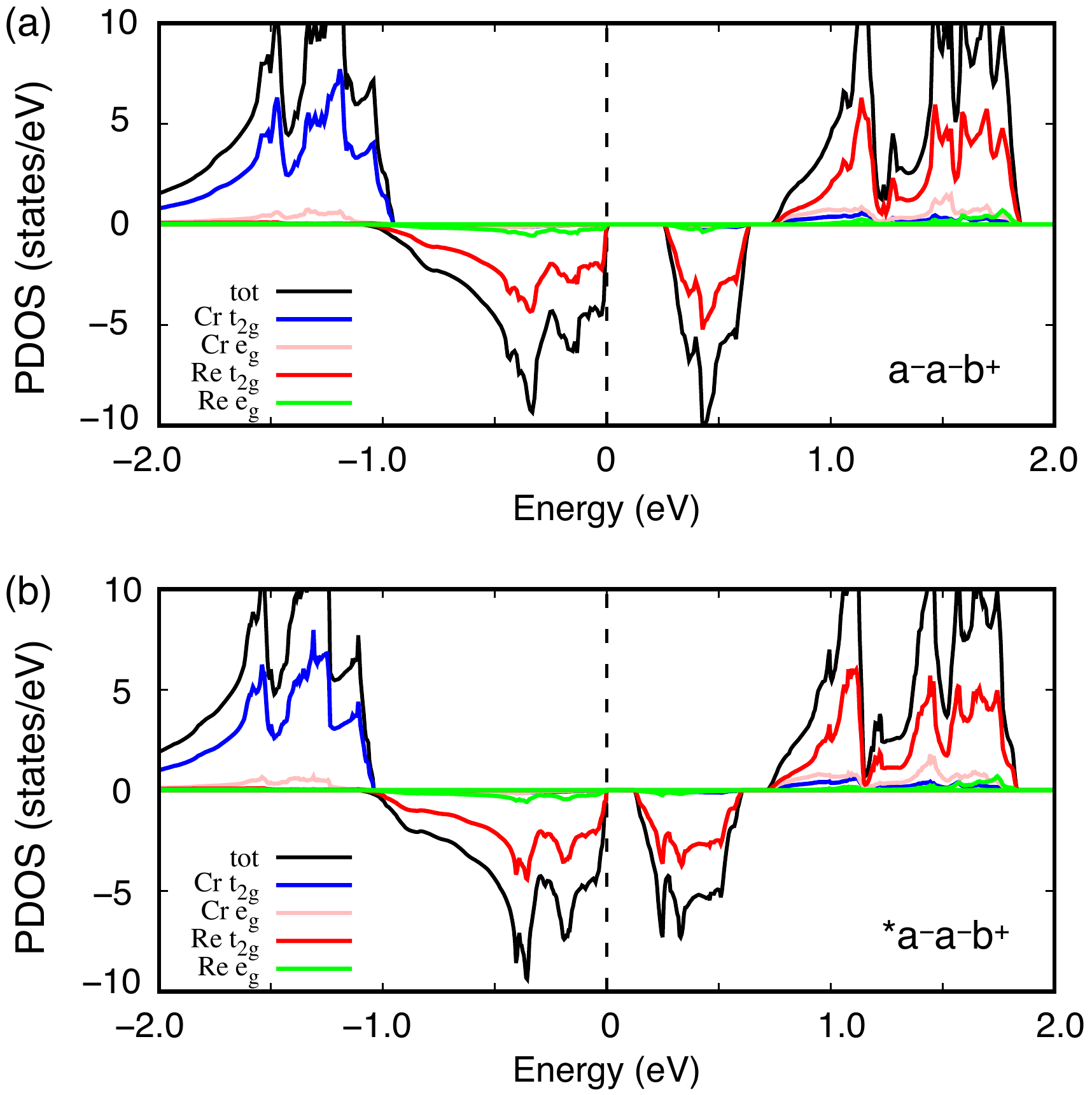}
\caption{   DFT+$U$ projected density of states of Ca$_2$CrReO$_6$  
  (a) in the ground state structure ($a^{-}a^{-}b^{+}$) (b) in the reference structure $*a^{-}a^{-}b^{+}$.
The Fermi energy is set to be zero;   
 $U_{\textnormal{Re}}$=2.0 and  $U_{\textnormal{Cr}}$=2.5. 
}
\label{CaCrReO-pdos}
\end{figure}

Similar to Ca$_{2}$FeReO$_{6}$, Ca$_{2}$CrReO$_{6}$ results in a  monoclinic structure
$P2_{1}/n$ ($a^-a^-b^+$ tilt, see Fig. \ref{FeRe-str}) with an insulating ground state.
Both DFT and  DFT+$U$ reasonably capture the large magnitude of the octahedral tilts: the
two in-plane and one out-of-plane $\angle$Cr-O-Re are 153.8, 154.1, 154.9 using DFT;
151.7, 151.0 and 152.7$^\circ$ using DFT+$U$ ($U_{\textnormal{Cr}}$=2.5 and
$U_{\textnormal{Re}}$=2); and  153.1, 154.3, and 155.0$^\circ$  as measured at $T$=300K in
experiment\cite{Re-Kato}.  In terms of the C-OD amplitude, the experimental value of
$d_{|x-y|}$ at 300K reported by Kato \emph{et al.} is 0.003 \AA, which is smaller than
$d_{|x-y|} =0.005\AA$ of Ca$_{2}$FeReO$_{6}$ at the same temperature\cite{Re-Kato}; and
this suggests that the C-OD has been disordered at 300K, yet the transport still suggests
an insulating state.  Unfortunately, the low temperature values of $d_{|x-y|} $ have not
yet been measured, but we will demonstrate that a large C-OD amplitude is expected just as
in the case of Ca$_{2}$FeReO$_{6}$. 

Just as in the case of Ca$_{2}$FeReO$_{6}$, the C-OD may form in either the C-OD$^+$ or
C-OD$^-$ variant.  Unlike Ca$_{2}$FeReO$_{6}$, C-OD$^+$ ordering is  more stable over a
broad range of $U_{\textnormal{Re}}$, as depicted in Figs. \ref{COD_energy}.  The energy
difference between the C-OD variants are relatively small as compared to the case of
Ca$_{2}$FeReO$_{6}$, which might be due to the smaller difference between the respective
$a$ and $b$ lattice parameters.  More specifically, $b-a$  is  0.070$\AA$ in
Ca$_{2}$FeReO$_{6}$, while  $b-a$ is 0.026$\AA$ in Ca$_{2}$CrReO$_{6}$.  In both cases,
the energy difference between C-OD$^+$/C-OD$^-$ is well within the error of DFT+$U$.  As
in the case of Ca$_{2}$FeReO$_{6}$, here we only present the results of C-OD$^+$ ordering.

We now perform the same analysis as in the case of Ca$_{2}$FeReO$_{6}$, computing the
orbital polarization, C-OD amplitude, band gap, and relative energy of the ground state
structure $a^-a^-b^+$ and the reference structure $*a^-a^-b^+$ as a function of $U$ (see
Fig. \ref{CaCrRe-JT}). The same trends are observed as compared to Ca$_{2}$FeReO$_{6}$,
with the only differences being quantitative changes due to the smaller effective Re
bandwidth in the Cr-based systems.  Interestingly, the C-OD amplitude rapidly saturates
after its onset, and the relative energy difference $\Delta E$ shows three distinct
regions.  The third region, corresponding to $U_{\textnormal{Re}}>$1.6 and
$U_{\textnormal{Cr}}$=2.5, corresponds to the formation of the C-OO in the  $*a^-a^-b^+$
reference structure, whereby the energy penalty of $U_{\textnormal{Re}}$ in the
$*a^-a^-b^+$  structure is reduced via polarization. This region could not be clearly seen
in the Ca$_{2}$FeReO$_{6}$ case given that the corresponding transition occurs just
preceding the maximum value of $U_{\textnormal{Re}}$ in the plot, and the magnitude of the
effect should be smaller given the larger effective Re bandwidth.  

Most importantly, the
critical threshold of $U_{\textnormal{Re}}$ for driving the MIT is strongly reduced,
requiring only $U_{\textnormal{Re}}=1.4$ in the relaxed structure (with
$U_{\textnormal{Cr}}=2.5$); and a similar renormalization occurs in the reference
structure $*a^-a^-b^+$ which now only needs $U_{\textnormal{Re}}=1.7$ to achieve an
insulating state.  This has interesting implications, as the critical
$U_{\textnormal{Re}}$ is now sufficiently small in the reference structure that the
insulating state may survive in the absence of any appreciable C-OD amplitude. If we
assume our preferential values of $U_{\textnormal{Re}}=2$ and $U_{\textnormal{Cr}}=2.5$,
we see that both the relaxed structure and the reference structure are insulators (see
Fig. \ref{CaCrReO-pdos} for projected DOS). This result is consistent with the
experimental measurements on Ca$_{2}$CrReO$_{6}$ which find no appreciable C-OD amplitude,
as in our reference structure, yet still measure an insulating
state\cite{Re-Kato,Kato-SrCrReO}; though further experiments are clearly needed in this
system before drawing conclusions. 

One could argue that choosing a smaller value of $U_{\textnormal{Re}}$ could yield the
same behavior as Ca$_{2}$FeReO$_{6}$, where the loss of the C-OD amplitude destroys the
C-OO and results in an metallic state, but this sort of tuning would be forbidden by the
fact that $U_{\textnormal{Re}}\ge$2.0 is needed to obtain the experimentally observed
insulating state in Sr$_{2}$CrReO$_{6}$.  Therefore, Ca$_{2}$CrReO$_{6}$ could be a
concise example where orbital ordering can clearly be observed in the (near) absence of a
concomitant structural distortion (i.e. at a temperature where the C-OD is suppressed but
the C-OO survives).

\subsection{Effect of spin-orbit coupling}
\label{sec:SOC}

\begin{figure}
\includegraphics[width=0.5\textwidth, angle=0]{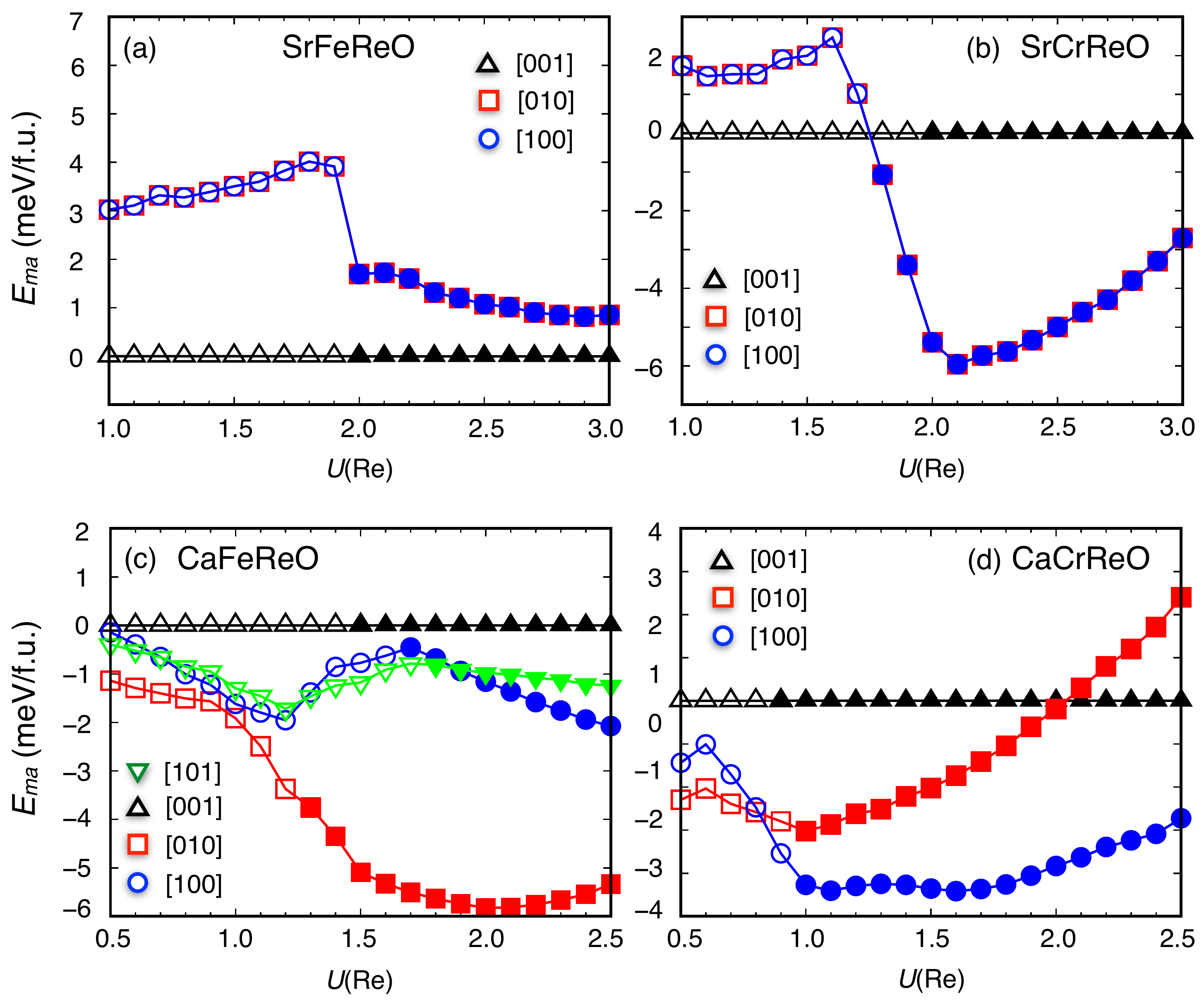}
 \caption{   Magnetic anisotropy energies as a function of $U_{\textnormal{Re}}$ for (a)
 Sr$_{2}$FeReO$_{6}$, (b) Sr$_{2}$CrReO$_{6}$, (c) Ca$_{2}$FeReO$_{6}$, and (d)
 Ca$_{2}$CrReO$_{6}$.  Energies of the magnetization direction  [001] ($c$-axis) is set to
 zero; $U_{\textnormal{Fe}}$=4 and $U_{\textnormal{Cr}}$=2.5.  Empty and filled points
 stand for metallic and insulating phases, respectively.
}
\label{aniso}
\end{figure}

The strength of the spin-orbit coupling ($\lambda$) can be up to 0.5eV in the 5$d$
transition metal oxides, which is non-negligible when compared to $U$ and the bandwidth.
In the better known example of the irridates, the $t_{2g}$ bandwidth is approximately 1eV,
and thus a spin-orbit coupling of $\lambda$=0.3-0.5eV plays an important role in realizing
the insulating state \cite{BHKim-2012,Watanabe,BJKim-2008}.  The effect of SOC in the Re
based DPs will be smaller than the irridates given that the $t_{2g}$ bandwidth of Re is
closer to 2eV and the strength of the SOC of Re will also be smaller due to the smaller
atomic number of Re.  For example, our comparison of the Re-projected DOS with and without
SOC in the Sr-based systems demonstrated changes of approximately 0.2eV (see Fig.
\ref{electronic_struct}, panels c and  d).  
While SOC does not qualitatively change any major trends, the small quantitative changes can be relevant; 
as we will demonstrate.  
%XX
%Specifically, changes are larger in the Ca-based systems due to the smaller $t_{2g}$ bandwidth of Re 
%(1.50eV and 1.35eV for Ca$_{2}$FeReO$_{6}$ and Ca$_{2}$CrReO$_{6}$, respectively, 
%when $U_{\textnormal{Fe}}$=4,  $U_{\textnormal{Cr}}$=2.5 and $U_{\textnormal{Re}}$=0; see Sec. \ref{sec:general_aspects})
In this section, we will explore the magnetic anisotropy energy as a function of
$U_{\textnormal{Re}}$, in addition to repeating our previous analysis of the orbital
polarization, the OD amplitude, band gap, and relative energetics. Here we will only
consider $U_{\textnormal{Fe}}$=4 and $U_{\textnormal{Cr}}$=2.5.

We begin by considering the magnetic anisotropy energy ($E_{ma}$) as summarized in Fig.
\ref{aniso}.  We define $E_{ma}$ as the relative energy (per Re) of a given magnetic
orientation with respect to the energy of the [001] orientation (e.g. $E_{ma}[010]=E[010]-E[001]$).  
The magnetic orientation is particularly important since the threshold
of $U_{\textnormal{Re}}$ for the C-OO/C-OD depends on the magnetic orientation, and shifts
as large as 0.4eV can observed for Ca$_{2}$FeReO$_{6}$.

For Sr$_{2}$FeReO$_{6}$, the magnetization along [001] is most stable in our calculations,
as shown in Fig \ref{aniso} panel (a), whereas magnetic moments are aligned in $ab$-plane
in the experiment at 298K \cite{Nakamura}.  This appears to be a discrepancy, though we
only explored [100] and [010] directions within the $ab$-plane, so it is possible that
some other direction within the plane is lower. Also, our calculations are at $T$=0, while
the experiments were done at $T$=298K. Otherwise, this could serve as an interesting
failure of the method (albeit for a very small energy scale).  Nonetheless,
Sr$_{2}$FeReO$_{6}$ is metallic with $U_{\textnormal{Re}}<$2.0 in all orientations that we
explored.

For Sr$_{2}$CrReO$_{6}$, the magnetization along the [100] and [010] directions are
equivalent, as shown in Fig. \ref{aniso} panel (b).  Interestingly, [001] is more stable
for small $U_{\textnormal{Re}}$, but then this trend is reversed once the system goes
through the C-OO/C-OD and there is a magnetic easy $ab$ plane for
$U_{\textnormal{Re}}\geq$1.8. Given our preferred values of $U_{\textnormal{Re}}$=1.9 and
$U_{\textnormal{Cr}}$=2.5 (see Section \ref{U_of_Re}), DFT+$U$ results in an easy $ab$
plane.  Recent experiment by Lucy \emph{et al.} showed that a Sr$_{2}$CrReO$_{6}$ film on
SrTiO$_3$ and (LaAlO$_3$)$_{0.3}$(Sr$_2$AlTaO$_6$)$_{0.7}$, corresponding to  0.09\% and
1.04\% of compressive strains, results in a magnetic easy axis within the $ab$ plane at
both low (20K) and high $T$ (300K) \cite{Lucy-SrCrRe,Lucy_SrCrReO_2015}.

For Ca$_{2}$FeReO$_{6}$, Rietveld refinement determined that the magnetization easy axis
below $T_{MIT}$ is the $b$-axis (ie. [010]), while above $T_{MIT}$ the magnetization easy
axis changes \cite{Granado,Oikawa}; though there is not yet consensus on the direction.
Granado \emph{et al.} suggested that Fe and Re moments lie on the $ac$-plane, where the
magnetization angle from the $a$ axis is 55$^\circ$ (close to [101]) \cite{Granado},
whereas Oikawa \emph{et al.} showed that [001] is the easy axis \cite{Oikawa}. We will
explore [100], [010], [001], and [101] in the ground state structure, while primarily
focussing on [001] in the $*a^{-}a^{-}b^{+}$ reference structure; though with the latter
we investigate a few scenarios using [101].

By using the experimental atomic coordinates and LDA+$U$ calculations
($U_{\textnormal{Re}}$=3 and $J_{\textnormal{Re}}$=0.7), Antonov \emph{et al.} showed that
[010] is the easy axis and [001] is lower in energy than [100], for both low $T$ and high
$T$ experimental structures \cite{Ernst}.  Gong \emph{et al.} found the same result using
the mBJ potential\cite{Gong}, despite the fact that they were using the GGA relaxed
structure which more closely resembles the experimental structure above the phase
transition.  We also found the same ordering, which proved to be independent of the value
of  $U_{\textnormal{Re}}$, even when crossing the C-OO/C-OD transition (see Fig.
\ref{aniso}, panel c).  Given that above the MIT Granado \emph{et al.} found [101] to be
the easy-axis, we also explore this direction; demonstrating that it is very similar to
[100].  Interestingly, the magnetic orientation can have an appreciable effect on the
onset of C-OO/C-OD.

For Ca$_{2}$CrReO$_{6}$, we are not aware of any experimental data on the magnetic easy
axis.  From an mBJ study with GGA-relaxed structure, Gong \emph{et al.} reported that
[010] is the easy axis, and $E_{ma}[001] > E_{ma}[100] $\cite{Gong}.  Alternatively, our
GGA+$U$+SOC calculations suggest that [100] is the easy axis for 
$U_{\textnormal{Re}} \geq 0.9$ (see Fig. \ref{aniso}, panel (d)).  Given our preferred values of
$U_{\textnormal{Re}}$=1.9 and $U_{\textnormal{Cr}}$=2.5 (see Section \ref{U_of_Re}), we
would expect an easy axis of [100] and that [010],[001]  are very close in energy.

\begin{figure*}
\begin{minipage}[c]{0.70\textwidth}
\includegraphics[width=\textwidth]{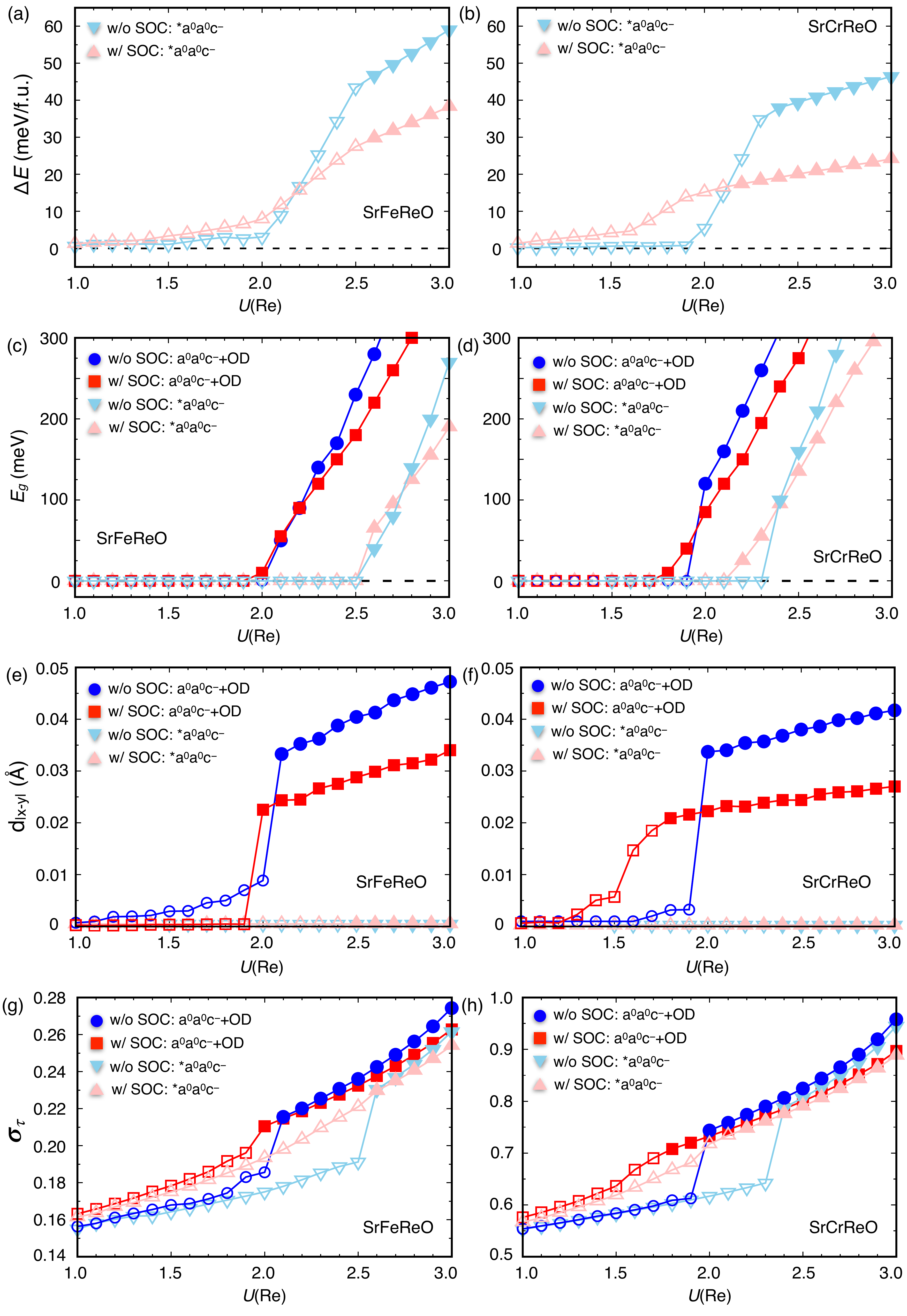}
\end{minipage}\hfill
\begin{minipage}[c]{0.25\textwidth}
 \caption{(a),(b)   Relative energy of Sr$_2$FeReO$_6$ and Sr$_2$CrReO$_6$ in  the
 reference structure $I4/m$ with respect to the ground state (i.e. $a^{0}a^{0}c^{-}$+OD);
 with and without spin-orbit coupling.  (c),(d)  Electronic band gaps of different phases.
 (e),(f) Octahedral distortion (OD) amplitude $d_{|x-y|} $ of the ReO$_6$ octahedron.
 (g),(h) Orbital polarization $\sigma_\tau$ (see eq. \ref{eq:sigma_tau}) for Re.  Panels
 (a), (c), (e), and (g) correspond to Sr$_2$FeReO$_6$ with $U_{\textnormal{Fe}}$=4, while
 panels  (b), (d), (f), and (h) correspond to Sr$_2$CrReO$_6$ with
 $U_{\textnormal{Cr}}$=2.5.  Filled and empty points stand for the insulating and metallic
 phases, respectively.  Magnetization is along the [001] and [100] for Sr$_{2}$FeReO$_{6}$
 and Sr$_{2}$CrReO$_{6}$, respectively.  }
\label{Sr-soc}
\end{minipage}
\end{figure*}

\begin{figure*}
\begin{minipage}[c]{0.70\textwidth}
\includegraphics[width=\textwidth]{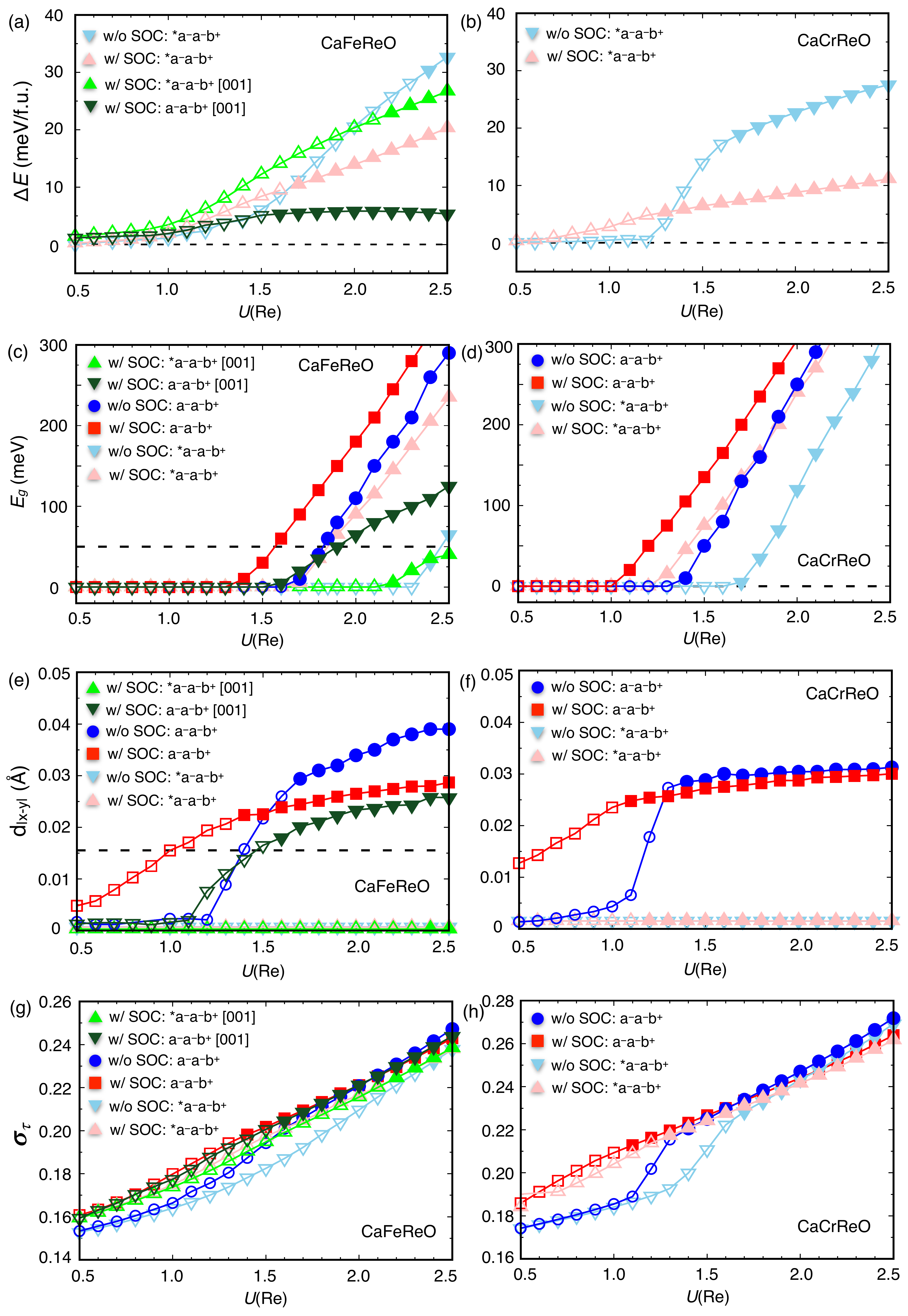}
\end{minipage}\hfill
\begin{minipage}[c]{0.25\textwidth}
 \caption{(a),(b)   Relative energy of Ca$_2$FeReO$_6$ and Ca$_2$CrReO$_6$ in  the
 reference structure $*a^{-}a^{-}b^{+}$ with respect to the ground state (i.e.
 $a^{-}a^{-}b^{+}$); with and without spin-orbit coupling.  (c),(d)  Electronic band gaps
 of different phases.  (e),(f) Octahedral distortion (OD) amplitude $d_{|x-y|} $ of the
 ReO$_6$ octahedron.  (g),(h) Orbital polarization $\sigma_\tau$ (see eq.
 \ref{eq:sigma_tau}) for Re.  Panels (a), (c), (e), and (g) correspond to Ca$_2$FeReO$_6$
 with $U_{\textnormal{Fe}}$=4, while panels  (b), (d), (f), and (h) correspond to
 Ca$_2$CrReO$_6$ with  $U_{\textnormal{Cr}}$=2.5.  Filled and empty points stand for the
 insulating and metallic phases, respectively.  Magnetization is along the [010] and [100]
 for Ca$_{2}$FeReO$_{6}$ and Ca$_{2}$CrReO$_{6}$, respectively.  Additionally, the [001]
 magnetization direction (experimentally observed for T]140K) is included for
 Ca$_2$FeReO$_6$ where indicated.
}
\label{Ca-soc}
\end{minipage}
\end{figure*}

Having established the easy axis for each material, we now repeat the previous analysis
probing the behavior as a function of the Hubbard $U$ but now  including SOC and the easy
axis axis as determined from DFT+$U$ (see Figs. \ref{Sr-soc}-\ref{Ca-soc}); and it should
be kept in mind that the predicted easy-axis for Sr$_{2}$FeReO$_{6}$ disagrees with
experiment.  Summarizing, we consider Sr$_{2}$FeReO$_{6}$ [001], Sr$_{2}$CrReO$_{6}$
[100], Ca$_{2}$FeReO$_{6}$ [010], and Ca$_{2}$CrReO$_{6}$ [100].  Given that SOC will
break the block diagonal structure of the single-particle density matrix in the spin
sector, it is useful to introduce a more general measure of orbital polarization rather
than the definition used in equation (\ref{eq:orbpol}); and we will utilize the standard
deviation of the Eigenvalues of the local single particle density matrix for the
correlated subspace, denoted $\sigma_\tau$ (this is a component of the DFT+$U$ energy
functional, see Ref. \cite{Isaacs2017045141} for a detailed derivation): 
\begin{equation}\label{eq:sigma_tau}
\sigma_{\tau}=\sqrt{\frac{\sum_{m}(n_{m}^{\tau }-\mu_{\tau})^2}{N_{\mathrm{orb}}}}
\end{equation}
and
\begin{equation}\label{eq:mu_tau}
\mu_{\tau}=\frac{\sum_{m}n_{m}^{\tau }}{N_{\mathrm{orb}}}  \ ,
\end{equation}
where $m$ labels an Eigenvalue of the single-particle density matrix for the correlated
subspace (ie. Eigenvalues of  the  $10\times 10$ single-particle density matrix for the
case of $d$ electrons), $\tau$ labels a Re site in the unit cell, and
$N_{\mathrm{orb}}$=10 for $d$-electrons. The orbital polarization is then defined to be
$\sigma_{\tau}$.

We begin with the Sr-based materials, Sr$_{2}$FeReO$_{6}$ and Sr$_{2}$CrReO$_{6}$,
characterizing the effect of the SOC for the relaxed structure $a^{0}a^{0}c^{-}$+OD (e.g.
$P4_{2}/m$ for $a^{0}a^{0}c^{-}$+C-OD, etc.) and the reference structure $I4/m$
($*a^{0}a^{0}c^{-}$)(see Fig. \ref{Sr-soc}). The previously presented results without SOC
are included to facilitate comparison, in addition to providing updated values for our new
metric of orbital polarization $\sigma_\tau$.  As expected, SOC is a relatively small
perturbation in all cases, though there are some interesting differences. We begin by
examining the orbital polarization for the reference structures $*a^{0}a^{0}c^{-}$  where
the C-OD amplitude is restricted to be zero (see panels g and h). For smaller values of
$U_{\textnormal{Re}}$, prior to the C-OO transition, SOC enhances the orbital polarization
at a given value of $U_{\textnormal{Re}}$ in the F-OO state (comparing lines with up and
down triangles). For Sr$_{2}$CrReO$_{6}$, the critical $U_{\textnormal{Re}}$ for the C-OO
transition is shifted down by about 0.2eV (compare lines with up and down triangles),
indicating the SOC is facilitating the onset of the C-OO and the resulting MIT.  This
renormalization of $U_{\textnormal{Re}}$ is much smaller for  Sr$_{2}$FeReO$_{6}$ and
cannot be seen at the resolution we have provided.  In both cases, the magnitude of the
orbital polarization beyond the C-OO transition is very similar with and without the SOC.

Allowing the C-OD to condense in the relaxed structures shows similar behavior (see red
and blue curves).  In both Sr$_{2}$FeReO$_{6}$ and Sr$_{2}$CrReO$_{6}$, SOC pushes the
onset of the C-OO/C-OD to smaller values of $U_{\textnormal{Re}}$; more substantially in
the case of Cr.  As a result, including SOC causes the gap to open at slightly smaller
values of $U_{\textnormal{Re}}$: approximately 0.1 less for Sr$_{2}$FeReO$_{6}$ and 0.2
less for Sr$_{2}$CrReO$_{6}$.  Notably, the C-OD amplitude for the metallic phase of
Sr$_{2}$FeReO$_{6}$ is dampened to zero, in agreement with experiment.  Somewhat
counterintuitively,  SOC results in  smaller C-OD amplitudes for values of
$U_{\textnormal{Re}}$ beyond the MIT, despite causing an earlier onset of the C-OD.  For
the relative energetics, in both compounds SOC decreases the stabilization energy of the
C-OD for $U_{\textnormal{Re}}\gtrapprox2.1$ (see panels a and b), consistent with the
reduced magnitude of the C-OD.  Given our preferred value of $U_{\textnormal{Re}}$=1.9 for
SOC, we find that Sr$_{2}$FeReO$_{6}$ is metallic with space group $I4/m$ (ie. no
condensation of OD), consistent with experiment; while Sr$_{2}$CrReO$_{6}$ is insulating
with a non-zero C-OD amplitude (i.e. space group $P4_2/m$), stabilized by roughly 14meV.

In the Ca-based systems, the effects of SOC are slightly more pronounced (see Fig. \ref{Ca-soc}),
which is likely associated with the smaller Re $t_{2g}$ bandwidth, 
but the trends are all the same as the Sr-based materials. We begin by
analyzing the orbital polarization in the reference structure $*a^{-}a^{-}b^{+}$, where
the C-OD has effectively been removed (see panels g and h, curves with pink-up and
blue-down triangles).  For small values of  $U_{\textnormal{Re}}$, SOC mildly enhances the
orbital polarization, but the differences diminish once both cases form the C-OO
insulator. However, SOC has a more dramatic effect in the Ca-based systems in terms of
shifting the C-OO induced MIT to smaller values of $U_{\textnormal{Re}}$, giving a
reduction of 0.7 and 0.4eV for the Fe-based and Cr-based material, respectively (see
panels c and d, curves with pink-up and blue-down triangles).  For the relaxed structures
(see red and blue curves), the C-OD is activated at much smaller values of
$U_{\textnormal{Re}}$ in both materials, more so for the case of Ca$_{2}$CrReO$_{6}$.
Furthermore, Ca$_{2}$FeReO$_{6}$ reaches a relatively smaller value of the C-OD amplitude
beyond the C-OO induced MIT, while Ca$_{2}$CrReO$_{6}$ saturates at roughly the same
value.  Given our preferred value of $U_{\textnormal{Re}}$=1.9, both Ca$_{2}$FeReO$_{6}$
and  Ca$_{2}$CrReO$_{6}$ are insulators with a appreciable C-OD amplitude, consistent with
known experiments (though the low temperature structural parameters of Ca$_{2}$CrReO$_{6}$
have not yet been measured). Furthermore, SOC has reduced the C-OD amplitude of
Ca$_{2}$FeReO$_{6}$, moving it closer to the experimental value (see panel e, red curve).

For Ca$_{2}$FeReO$_{6}$, we also investigate the behavior of the [001] magnetization
direction for both the reference structure $*a^{-}a^{-}b^{+}$ and the ground state
structure  $a^{-}a^{-}b^{+}$, which is essential given that experiment dictates [001] is
approximately the easy-axis above the MIT where the C-OD is suppressed. For
$a^{-}a^{-}b^{+}$, the [001] orientation is higher in energy than [010], with the
difference being enhanced as $U_{\textnormal{Re}}$ increases (see Fig. \ref{aniso}, panel
c, green curve).  Furthermore, for [001] the threshold value of $U_{\textnormal{Re}}$ for
the onset of the C-OO/C-OD is increased, and the magnitude of the band gap and C-OD
amplitude are diminished at a given value of $U_{\textnormal{Re}}$ (see Fig. \ref{Ca-soc},
panels c and e, dark green triangles).  More relevantly, the same trends are observed in
the reference structure  $*a^{-}a^{-}b^{+}$, but the effect is amplified (light green
triangles).  In particular, the critical value of $U_{\textnormal{Re}}$  for the C-OO/C-OD
dramatically increases from 1.8 to 2.2 eV as the magnetization switches from [010] to
[001] (compare pink and light green curves, respectively).  We also investigate the case
of [101] magnetization direction.  The overall features of [101] are similar to the case
of [001]  (not shown), except that the critical value of $U_{\textnormal{Re}}$ for the
C-OO/C-OD in the reference structure is increased to 2.4eV.

\begin{figure}
\includegraphics[width=0.45\textwidth, angle=0]{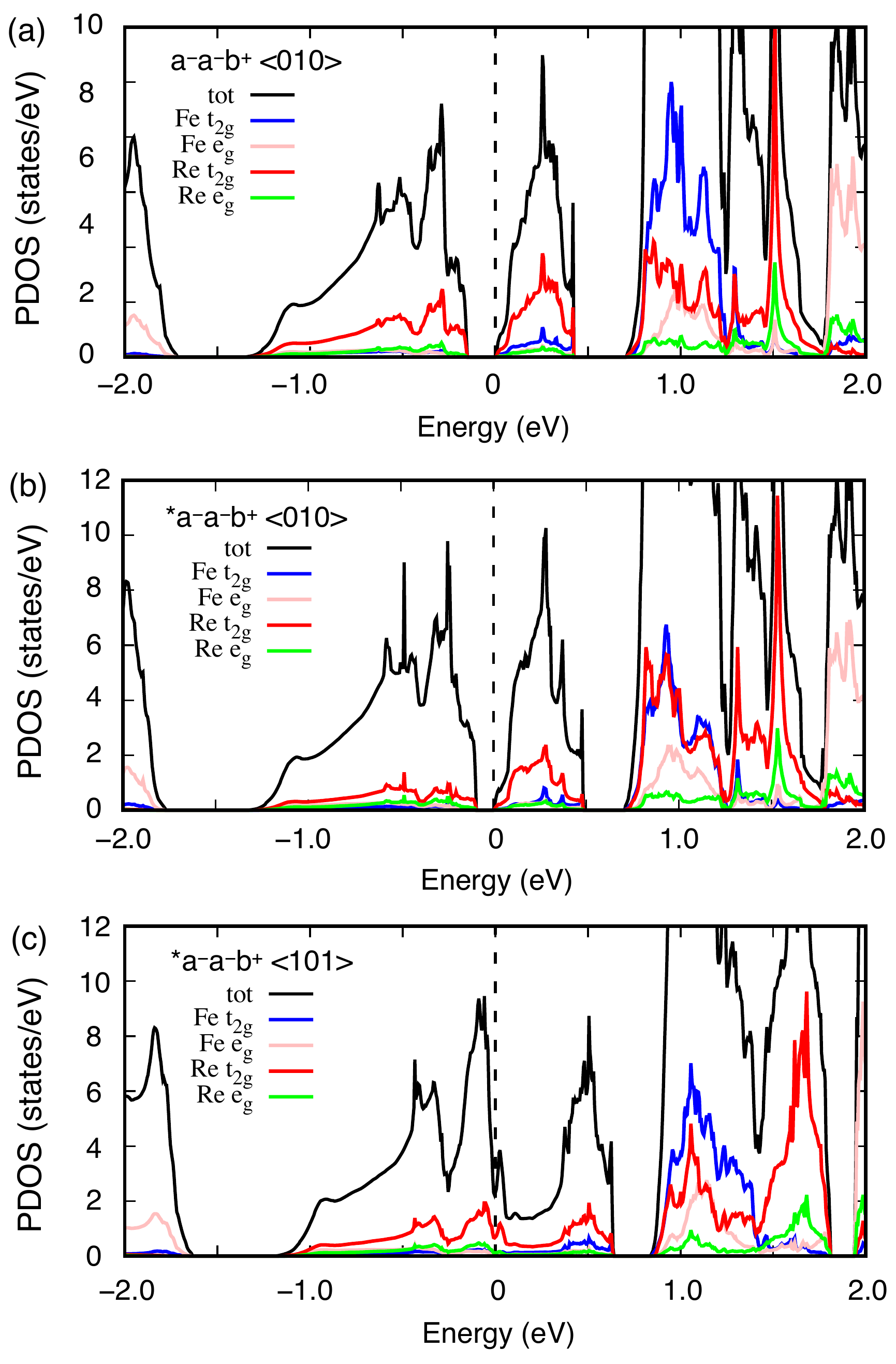}
\caption{   DFT+$U$+SOC projected density of states of Ca$_2$FeReO$_6$  
  (a) in the ground state structure ($a^{-}a^{-}b^{+}$) with the [010] magnetization direction,
  (b) in the reference structure ($*a^{-}a^{-}b^{+}$) with the [010] magnetization direction,
  (c) in the reference structure ($*a^{-}a^{-}b^{+}$) with the [001] magnetization direction.
The Fermi energy is set to be zero;   
 $U_{\textnormal{Re}}$=1.9 and  $U_{\textnormal{Fe}}$=4. 
}
\label{CaFeReO-pdos-soc}
\end{figure}

In Section \ref{subsubsec_cafe}, where SOC was not yet included, we elucidated the
possibility that a suppression of the C-OD (e.g. via thermal fluctuations) closes the band
gap via moving the critical value of $U_{\textnormal{Re}}$ beyond our expected value of
$U_{\textnormal{Re}}$=2.0 within GGA+$U$ (see Figure \ref{CaFeReO-pdos}). This could have
been a viable mechanism for the MIT, but SOC is strong enough to alter this scenario (see
Figure \ref{CaFeReO-pdos-soc}, panels a and b, using $U_{\textnormal{Re}}$=1.9). Given the
[010] magnetization direction, the gap is reduced in the reference structure, but it does
not close, unlike the case where SOC is not included. However, the experiments of Oikawa
\emph{et al.} dictate that [001] should be the easy axis of the high temperature
structure, in contradiction with DFT+$U$+SOC using our reference structure (though our
predicted energy difference is less than 6meV). If we consider the [001] direction in the
reference structure $*a^{-}a^{-}b^{+}$, we see that the gap has indeed closed (see Figure
\ref{CaFeReO-pdos-soc}, panel (c)); the gap also closes for the [101] direction.
Therefore, it is possible that the reorientation of the magnetization is important to the
MIT.

In summary, we see that for $U_{\textnormal{Re}}$=1.9, Sr$_{2}$FeReO$_{6}$ is a metal,
while the remaining systems are C-OO induced insulators. The general physics that was
deduced in the absence of SOC holds true with some small renormalizations of various
observables. Slightly reducing the value of $U_{\textnormal{Re}}$ allows for results which
are qualitatively consistent with experiment, with the caveat that the easy-axis of
Sr$_{2}$FeReO$_{6}$ disagrees with experiment.

%Orbital moments
Another interesting feature of SOC is the nonzero orbital moments of Re.  Spin and orbital
moments of Re within GGA+SOC and GGA+$U$+SOC with $U_{\textnormal{Re}}$=1.9 are summarized
in Table~\ref{orb_mom}.  The direction of Re orbital moment is opposite to the spin
moment, in agreement with the previous experiments
\cite{Sikora-FeRe,Winkler,Serrate_SrCrReO_2007} and GGA+SOC \cite{Jeng,Vaitheeswaran}.

\begin{table}
\begin{ruledtabular}
\begin{center}
\caption{Spin ($m_S$), orbital ($m_L$), and total ($M_{tot}$) moments of Re in DPs. Values are given in $\mu_B$/Re. }
\label{orb_mom}
\renewcommand{\arraystretch}{1.15}
\begin{tabular}{c c c c c c c}
     	& $m_S$   & $m_L$	& $M_{tot}$	& $\left| m_{L}/m_{S} \right|$	& method \\
	   \hline 
Sr$_2$FeReO$_6$    & $-$0.74  & 0.21  & $-$0.53  & 0.28 &   exp \cite{Winkler}  \\
                                   & $-$1.07  & 0.33  & $-$0.74  & 0.31 &   exp \cite{Sikora-FeRe}  \\
                                   & $-$0.85  & 0.23  & $-$0.62  & 0.27 &   GGA \cite{Jeng}  \\
                                   & $-$0.68  & 0.15  & $-$0.53  & 0.22 &   GGA \cite{Winkler}  \\
                                   
                                   & $-$0.73  & 0.18  & $-$0.55  & 0.25 &   LDA \cite{Ernst}  \\
                                   & $-$0.88  & 0.24  & $-$0.64  & 0.28 &   LDA+$U$ \cite{Ernst}  \\
                                   & $-$1.01  & 0.26  & $-$0.75  & 0.26 &   mBJ \cite{Guo2015}  \\
                                   
                                   & $-$0.76  & 0.16  & $-$0.61  & 0.20 &   GGA   \\
                                   & $-$1.22  & 0.42  & $-$0.83  & 0.34 &   GGA+$U$   \\
                                   
                 \hline                          
Sr$_2$CrReO$_6$      & $-$0.68  & 0.25  & $-$0.43  & 0.37 &   exp \cite{Majewski}  \\
            		           & $-$0.85  & 0.18  & $-$0.67  & 0.21 &   GGA \cite{Vaitheeswaran}  \\
                                    & $-$1.17  & 0.31  & $-$0.85  & 0.27 &   mBJ \cite{Guo2015}  \\		           
		           
	                            & $-$0.99  & 0.19  & $-$0.80  & 0.19 &   GGA  \\
	                            & $-$1.40  & 0.48  & $-$0.95  & 0.35 &   GGA+$U$ \\  
	                            
        \hline  
Ca$_2$FeReO$_6$   & $-$0.47  & 0.16  & $-$0.31  & 0.34 &   exp \cite{Winkler}  \\
				  & $-$1.15  & 0.39  & $-$0.76  & 0.34 &   exp \cite{Sikora-FeRe}  \\
 
                                   & $-$0.75  & 0.34  & $-$0.42  & 0.45 &   LDA \cite{Ernst}  \\
                                   & $-$1.11  & 0.66  & $-$0.45  & 0.60 &   LDA+$U$ \cite{Ernst}  \\
				 & $-$1.10  & 0.18  & $-$0.91  & 0.17 &   mBJ \cite{Gong}  \\
 
                                   & $-$0.76  & 0.17  & $-$0.58  & 0.23 &   GGA  \\
                                   & $-$1.30  & 0.43  & $-$0.87  & 0.33 &   GGA+$U$  \\
                                   
	 \hline   
Ca$_2$CrReO$_6$	  & $-$1.24  & 0.19  & $-$1.05  & 0.15 &   mBJ \cite{Gong}  \\
				  & $-$1.04  & 0.24  & $-$0.80  & 0.23 &   GGA  \\                          
                                   & $-$1.41  & 0.56  & $-$0.85  & 0.40 &   GGA+$U$  \\  
	                                                        
\end{tabular}
\end{center}
\end{ruledtabular}
\end{table}

As presented in Table ~\ref{orb_mom}, varying results have been measured for the magnitude
of spin and orbital moments by different groups.  However, the $\left| m_{L}/m_{S}
\right|$ values are more or less consistent \cite{Winkler,Sikora-FeRe} since this quantity
is not affected by possible uncertainties in the calculated number of holes
\cite{Winkler}, thus these values are better quantities to compare theory and experiments.
While GGA largely underestimate the experimental $\left| m_{L}/m_{S} \right|$ values,
GGA+$U$ gives a much better estimation for $\left| m_{L}/m_{S} \right|$.

\subsection{Optimal  $U$ values}
\label{U_of_Re}
Exploring a range of $U$ is a necessary burden for several reasons. First, the procedure
for constructing both the interactions and the double-counting correction is still an open
problem. Second, given that the DFT+$U$ method is equivalent to DFT+DMFT when the DMFT
impurity problem is solved within Hartree-Fock\cite{Kotliar2006865}, DFT+$U$ contains well
known errors which may be partially compensated by artificially renormalizing the $U$ to
smaller values.  Given that our most basic concern in this paper is to develop a
qualitative, and perhaps even  semi-quantitative, understanding of an entire family of
Re-based double perovskites, performing an empirical search for a single set of $U$'s
which can capture the physics of this family was essential.

In Sec \ref{sec:electronic_struct} and \ref{sec:SOC}, we have explored various observables
for a range of values of $U$.  Clearly, $U_{\textnormal{Re}}$ is the main influence, as it
is a necessary condition for driving the C-OO insulating state in the entire family of
materials, in addition to the C-OD.  However, we also demonstrated that the $U$ of the
$3d$ transition metal could play an important indirect role, via renormalizing the
critical value of $U_{\textnormal{Re}}$ for the C-OO/C-OD to smaller values. Also, for the
case of Sr$_{2}$CrReO$_{6}$, a nonzero $U_{\textnormal{Cr}}$ was important for properly
capturing the energetics of the $a^0a^0c^-$ tilt pattern.  For the 3$d$ transition metals,
we typically only explored $U$=0 and another value which is in line with expectations
based on previous literature or methods for computing $U$.  For Cr, we used
$U_{\textnormal{Cr}}$=2.5 eV,  which is similar to values used for CaCrO$_3$
\cite{Komarek} and Cr-related DPs ($U$=3 eV and $J$=0.87 eV) \cite{Jeng}.  For Fe, we
focus on $U_{\textnormal{Fe}}$=4 eV, as widely used elsewhere \cite{CaFeRe-Wu,Jeng,Jeon}.
Excessive tuning of $U_{\textnormal{Fe}}$ or $U_{\textnormal{Cr}}$ is not needed based on
our results, and the nonzero values that we evaluated were either necessary to capture a
given phenomena (i.e. the tilts in Sr$_{2}$CrReO$_{6}$), or were needed for a consistent
and reasonable value of $U_{\textnormal{Re}}$ (via the indirect influence of the
$U_{\textnormal{Fe}}$ or $U_{\textnormal{Cr}}$).  Therefore, $U_{\textnormal{Fe}}$=4 eV
and $U_{\textnormal{Cr}}$=2.5 eV are reasonable values to adopt, though a range of values
could likely give sufficient behavior. 

In the case of $U_{\textnormal{Re}}$, we explored a large number of values between 0-3.2eV.
The overall goal for selecting a set of $U$'s is to obtain the proper ground states in the entire family of materials,
which is nontrivial given that Sr$_{2}$FeReO$_{6}$ is metallic and the rest are insulators. While it is possible for
$U_{\textnormal{Re}}$ to have small changes due to differences in screening among the four materials, these differences
should be relatively small given the localized nature of the $d$ orbitals which comprise the correlated subspace; and therefore
we do seek a common value for all four compounds.
We conclude that $U_{\textnormal{Re}}$=2.0 and 1.9 are reasonable values
within GGA+$U$ and GGA+$U$+SOC, respectively, and these values will properly result in a metal for Sr$_{2}$FeReO$_{6}$ and insulators for the rest.
The predicted bandgap $E_\textnormal{gap}$ for Ca$_{2}$FeReO$_{6}$ (i.e. 105meV and 150meV within GGA+$U$ and GGA+$U$+SOC, respectively)
is somewhat larger than the experiment (i.e. 50meV), but this seems reasonable given the nature of approximations we are dealing with.
For Ca$_{2}$CrReO$_{6}$, we obtain $E_\textnormal{gap}$=250  and 270meV using GGA+$U$ and GGA+$U$+SOC, respectively (experimental gap is not known); while
$E_\textnormal{gap}$ of Sr$_{2}$CrReO$_{6}$ within GGA+$U$ and GGA+$U$+SOC is 120 and 40meV, respectively,
somewhat smaller than the experimental value of 200meV \cite{Hauser-SrCrRe}.

It is also interesting to compute  $U_{\textnormal{Re}}$ via the linear response approach \cite{Cococcioni}.
In Sr$_{2}$CrReO$_{6}$ and Ca$_{2}$CrReO$_{6}$, we obtained $U_{\textnormal{Re}}$=1.3 for both systems;
the calculation employed a supercell containing 8 Re atoms.
Therefore, linear response predicts a relatively small value for $U$,
consistent with 5$d$ electrons, but too small in order to be qualitatively correct: Sr$_{2}$CrReO$_{6}$ could not be an insulator
with such a small value.

\subsection{Future challenges for experiment}
\label{sec:exp_chal}
The central prediction of our work is that the minority spin Re $d_{xz}/d_{yz}$ orbitals
order in a  $\bm{q_\textnormal{fcc}}=\left( 0,\frac{1}{2},\frac{1}{2} \right)$ motif,
along with occupied minority spin Re $d_{xy}$ orbitals, in Sr$_2$CrReO$_6$,
Ca$_2$FeReO$_6$, and Ca$_2$CrReO$_6$. This section explores how this prediction may be
tested in experiment.  This orbital ordering results in a narrow gap insulator in our
calculations, consistent with the insulating states observed in experiment for these
compounds (see Section \ref{lit_review}).  However, more direct signatures of the orbital
ordering are desired.

Perhaps the most straightforward experiment is precisely resolving the crystal structure
of insulating Sr$_2$CrReO$_6$ at low temperatures.  Given that the C-OO breaks the
symmetry of the $I4/m$ space group, inducing the C-OD, experiment may be able to detect
the resulting $P4_2/m$ space group at low temperatures.  Such a measurement would serve as
a clear confirmation of our predicted orbital ordering.

Precisely resolving the bond lengths of Ca$_2$CrReO$_6$ at low temperatures would also be
beneficial. While the C-OO/C-OD is not a spontaneously broken symmetry in Ca$_2$CrReO$_6$,
an enhancement of $d_{|x-y|}$ is predicted in our calculations; similar to what has
already been experimentally observed in the case of  Ca$_2$FeReO$_6$.

Other experiments could possibly directly probe the orbital ordering, such as X-ray linear
dichroism.  Once again, Sr$_2$CrReO$_6$ may be the best test case given that the orbital
ordering is a spontaneously broken symmetry.

\section{Summary}
\label{sec:summary}
In summary, we investigate the electronic and structural properties of Re-based double
perovskites  $A_2B$ReO$_6$  ($A$=Sr, Ca and $B$=Cr, Fe) through density-functional theory
+ $U$ calculations, with and without spin-orbit coupling.  All four compounds share a
common low energy Hamiltonian, which is a relatively narrow Re $t_{2g}$ minority spin band
that results from strong antiferromagnetic coupling to filled 3$d$ majority spin shell (or
sub-shell) of the $B$ ion. Cr results in a narrower Re $t_{2g}$ bandwidth than Fe, while
Ca-induced tilts result in a narrower Re $t_{2g}$ bandwidth than Sr-induced tilts;
resulting in a ranking  of the Re $t_{2g}$ bandwidth as Sr$_2$FeReO$_6$, Sr$_2$CrReO$_6$,
Ca$_2$FeReO$_6$, and Ca$_2$CrReO$_6$ (from largest to smallest).  Spin orbit coupling is
demonstrated to be a relatively small perturbation, though it still can result in relevant
quantitative changes.

In general, we show that the on-site $U_{\textnormal{Re}}$ drives a C-type (i.e.
$\bm{q_\textnormal{fcc}}=\left( 0,\frac{1}{2},\frac{1}{2} \right)$ given the primitive
face-centered cubic unit cell of the double perovskite)  antiferro orbital ordering
(denoted C-OO) of the Re $d_{xz}/d_{yz}$ minority spin orbitals, along with minority
$d_{xy}$ being filled on each site, resulting in an insulating ground state.  This
insulator is Slater-like, in the sense that the C-type ordering is critical to opening a
band gap.  Interestingly, this C-OO can even occur in a cubic reference structure
($Fm\bar{3}m$) in the absence of any structural distortions for reasonable values of
$U_{\textnormal{Re}}$. Furthermore, allowing structural distortions demonstrates that this
C-OO is accompanied by a local $E_g$ structural distortion of the octahedra with C-type
ordering (denoted as C-OD); and it should be emphasized that $U_{\textnormal{Re}}$ is a
necessary condition for the C-OO/C-OD to occur. The C-OO/C-OD will be a spontaneously
broken symmetry for $a^{0}a^{0}c^{-}$-type tilt patterns as in the Sr based systems (i.e.
$I4/m \rightarrow P4_2/m$), whereas not for the $a^{-}a^{-}b^{+}$-type tilting pattern of
the Ca based systems (i.e. $P2_1/n \rightarrow P2_1/n$).

While $U_{\textnormal{Re}}$ is a necessary condition for obtaining an insulating state,
the presence of the C-OD will reduce the critical value of $U_{\textnormal{Re}}$ necessary
for driving the orbitally ordered insulating state; as will the $U$ on the 3$d$ transition
metal.  Furthermore, the C-OD is necessary for reducing the critical $U_{\textnormal{Re}}$
to a sufficiently small value such that Sr$_{2}$FeReO$_{6}$ remains metallic while
Sr$_{2}$CrReO$_{6}$ is insulating.  More specifically, using a single set of interaction
parameters (i.e.  $U_{\textnormal{Re}}=1.9eV$, $U_{\textnormal{Fe}}=4eV$,
$U_{\textnormal{Cr}}=2.5eV$, when using SOC), we show that Sr$_{2}$CrReO$_{6}$,
Ca$_{2}$CrReO$_{6}$, and Ca$_{2}$FeReO$_{6}$ are all insulators, while Sr$_{2}$FeReO$_{6}$
is a metal; consistent with most recent experiments. 

Previous experiments concluded that Sr$_{2}$CrReO$_{6}$ was half-metallic
\cite{Re-Kato,Kato-SrCrReO,Teresa-SrCrReO,Winkler,Asano-SrCrRe}, but recent experiments
showed that fully ordered films grown on an STO substrate are
insulating\cite{Hauser-SrCrRe,Hauser-Vo}.  We show that Sr$_{2}$CrReO$_{6}$ is indeed
insulating with $U_{\textnormal{Re}}=1.9eV$, so long as the structure is allowed to relax
and condense the C-OD.  Given that the C-OD is a spontaneously broken symmetry in this
case, the challenge for experimental verification will be resolving the $P4_{2}/m$ space
group at low temperatures instead of the higher symmetry $I4/m$ group.

While the C-OD is not a spontaneously broken symmetry in Ca$_{2}$FeReO$_{6}$, experiment
dictates that there is an unusual discontinuous phase transition at $T$=140K between two
structures with the same space group, $P2_{1}/n$; with the high temperature structure
being metallic and the low temperature structure being insulating.  The main structural
difference between the experimental structures is the C-OD amplitude: $d_{|x-y|}$ is 0.016
and 0.005\AA\ in the structures at 120K and 160K, respectively. Additionally, the C-OD
changes variants across the transition, going from C-OD$^+$ (120K) to C-OD$^-$ (160K).
The appreciable C-OD$^+$ amplitude measured in low temperature experiments is consistent
with our prediction of a large C-OD amplitude which is induced by the C-OO.  The same
trends are found in Ca$_{2}$CrReO$_{6}$, which has a narrower Re bandwidth and  results in
a more robust insulator with a larger band gap.  Predicting the transition temperature
from first-principles will be a great future challenge given that the temperature of the
electrons and the phonons may need to be treated on the same footing, all while accounting
for the spin-orbit coupling.

SOC is a small quantitative effect, though it can have relevant impact, such as lowering
the threshold value of $U_{\textnormal{Re}}$ for inducing the C-OO/C-OD in the Ca-based
compounds; even having a strong dependence on magnetization direction for
Ca$_{2}$FeReO$_{6}$.   GGA+$U$+SOC predicts the easy axis of Sr$_2$CrReO$_6$ and
Ca$_2$FeReO$_6$ to be \{100\} and [010], respectively, consistent with the experiment, and
also compares well to the experimental measurements of the magnitude of the orbital
moment. It should be emphasized that $U_{\textnormal{Re}}$, and the C-OO/C-OD which it
induces, is critical to obtaining the qualitatively correct easy-axis in Sr$_2$CrReO$_6$.
In the case Sr$_2$FeReO$_6$, GGA+$U$+SOC predicts a [001] easy axis, in disagreement with
one experiment which measured the easy axis to be in the $a-b$ plane.  Additionally, the
GGA+$U$+SOC predicted ratios of  obital/spin moment  $m_{L}/m_{s}$ are close to the
experimental values, whereas GGA+SOC largely underestimates them.

\section{Acknowledgments}
We thank to K. Oikawa and K. Park for helpful discussion.  This work was supported by the
grant DE-SC0016507 funded by the U.S. Department of Energy, Office of Science.  This
research used resources of the National Energy Research Scientific Computing Center, a DOE
Office of Science User Facility supported by the Office of Science of the U.S. Department
of Energy under Contract No. DE-AC02-05CH11231.

\bibliography{myref}

\end{document}